\documentclass[aps,pra,eqsecnum,twocolumn,amsmath,amssymb,10pt]{revtex4-2}

\usepackage{graphicx}
\usepackage{bm}
\usepackage[usenames,dvipsnames]{color}
\definecolor{altncolor}{rgb}{0,0,0.8}
\usepackage[colorlinks=true, linkcolor=green, anchorcolor=altncolor,
citecolor=altncolor, filecolor=altncolor, menucolor=altncolor,
urlcolor=altncolor]{hyperref}

\begin{document}

\title{Resonant optical bistability in support of enhanced Rydberg atom sensors}

\author{Peter B. Weichman}

\affiliation{FAST Labs$^{TM}$, BAE Systems, 600 District Avenue, Burlington, MA 01803}

\date{\today}

\begin{abstract}

Radio frequency antennas based on highly excited Rydberg atom vapors can in principle reach sensitivities beyond those of any conventional wire antenna, especially at lower frequencies where very long wires are needed to accommodate the growing wavelength. Conventional Rydberg sensors are based on individual atom response, with increased signal resolution relying on the $O(10^3)$ electric dipole moment enhancement, scaling as the square of the Rydberg state principal quantum number $N \sim 50$. However, despite more than thirty years of steady advances, beyond-classical signal sensitivity has yet to be demonstrated. More recently, the optical bistability effect, a many body nonequilibrium phase transition occurring at somewhat higher vapor densities, has been exploited for an order of magnitude or more increased sensitivity for some setups through tuning into the critical region. However, the results fall significantly short of those using more advanced ``conventional'' Rydberg sensor setups---which achieve even greater enhancement by exploiting resonant interaction between a pair of nearby Rydberg levels. This paper seeks to \emph{combine} the many body and resonant enhancement effects by extending the bistable phase analysis to include a pair of resonant Rydberg levels, supported by an exact treatment of the atom velocity thermal average. A dynamic linear response formalism is developed as well to explore the tradeoff between measurement sensitivity and finite bandwidth signals. We demonstrate regions of the phase diagram which could be exploited for record breaking, beyond-classical sensitivity. Of course, only a limited number of vapor parameters are under full experimental control, and experiments will be needed to quantitatively constrain the effective mean field interaction parameters appearing in the theory, and thereby define the accessible regions of the phase diagram.

\end{abstract}

\maketitle

\section{Introduction}
\label{sec:intro}

Major advances in radio frequency (RF) antenna technology through exploitation of quantum sensors based on Rydberg atom vapors have been pursued for several decades now \cite{HFI1990,FIM2005,RBTEY2011,CTSSAW2012,Sedlacek2012, NIST2014,SK2018,Michigan2019,NIST2019a,NIST2019b,Jing2020,Waterloo2021,MITRE2021,NIST2023,Romalis2024,Wu2024, Sandidge2024,Warsaw2024}. The key underlying physics is that by application of very precisely controlled laser and RF excitations, the atoms in these alkali vapors (e.g., $^{133}$Cs or $^{87}$Rb) can be placed in a highly excited quantum superposition state that is extremely sensitive to additional incident RF signals, in some cases potentially more sensitive than any classical, e.g., wire or loop, antenna.

Currently all setups are based on rather dilute room temperature vapors ($\rho_a \sim 10^{10}$  atoms/cm$^3$) in which each atom may be treated as responding individually to external signals. However, it was observed roughly a decade ago \cite{CRWAW2013,MLDGL2014} that by increasing the density ($\rho_a = 10^{11}$  atoms/cm$^3$ or higher, activated through increased temperature $\sim$\,45$^\circ$C) the resonant dipole interactions between the excited atoms can induce a non-equilibrium phase transition into an ordered, optically bistable state, somewhat analogous to magnetic alignment of spins in a ferromagnet, in which the atoms can now respond collectively. Recently this effect has been exploited for improved metrology and sensing \cite{Ding2022}, for example demonstrating several orders of magnitude increase in RF electric field sensitivity. A further order of magnitude improvement has been obtained by placing the vapor cell in an optical cavity \cite{Wang2023}, which acts to further sharpen the vapor probe laser electromagnetically-induced transparency (EIT) linewidth.

Interestingly, however, the observed RF electric field signal sensitivity (49 nV/cm/Hz$^{1/2}$) is actually short of what has been achieved (10 nV/cm/Hz$^{1/2}$ or better) through innovative setup architectures \cite{Romalis2024,Wu2024,Sandidge2024,Warsaw2024}, absent any bistability enhancement. This paper develops the basic theory and modeling advances required to treat the \emph{combined} bistable--resonant Rydberg pair enhanced sensing setup. This advance is far from trivial because there is now a $2 \times 2$ dipole interaction \emph{matrix}, accounting for interactions between the two different ``species'' of Rydberg atoms, generating a more complex multi-species nonequilibrium state and resulting phase diagram.

In addition, previous models \cite{CRWAW2013,MLDGL2014,Ding2022,Wang2023} have not accounted properly for thermal effects, especially in the form of rather strong motion-induced Doppler-detuning of the resonances. This effect has been treated only through an effective atom density $\rho_\mathrm{eff} \ll \rho_a$, suppressed by a factor of 100 or more, of the statistical population of sufficiently slow-moving atoms---say $v < 1$ m/s compared to a mean thermal velocity $v_\mathrm{th} > 100$ m/s. Although providing a qualitatively correct picture, quantitative experimental predictions, e.g., of improvement in incident RF field sensitivity, require proper treatment of the (Maxwell distribution) thermal average. In support of this, the theory developed here includes an exact analytic formulation of this average.

Proper comparison with experiment also requires careful attention to time-dependent effects. Thus, highest sensitivity generally occurs when the sensor is tuned precisely to a monotone incident signal frequency. However, many signals of interest have finite bandwidth, and operation near a phase transition introduces critical slowing down effects and accompanying strongly frequency-dependent responses. To address this quantitatively, a full dynamic linear response theory is developed, generalizing that for conventional Rydberg antenna setups \cite{W2024}. This theory is used to compute the full frequency-dependent sensor response, exhibiting precisely the significant tradeoff between sensitivity and bandwidth.

It should be noted that there can be very significant differences between the behavior predicted by mean field treatments, such as used here, and that observed near full fluctuation-driven phase transitions \cite{HH1977}. However, the mean field approximation for this system should actually be rather accurate due to the long-range dipole interactions between excited atoms. Thus, the interaction-induced detuning experienced by an atom, leading to the bistability effect, is due to the internal electric field generated by the laser-polarized vapor. This field is well known to be determined self-consistently by the full 3D macroscopic polarization density profile \cite{Jackson}, hence results from a full volume average rather than the rapidly fluctuating immediate neighborhood of a given atom. In particular, the bistable phase diagram should depend in general on the geometry of the excited vapor volume, which would be an interesting topic for future experimental investigation.

\subsection{Outline}
\label{sec:outline}

The outline of the remainder of this paper is as follows. The laser and RF-driven nonequilibrium dynamical model is developed in Sec.\ \ref{sec:modelform}, including atom motion-induced Doppler effects and Lindblad operator dissipation terms characterizing single atom spontaneous decay processes. The basic mean field approximation is developed (Sec.\ \ref{sec:mfapprox}), including a discussion of the underlying dipole polarization effect (Sec.\ \ref{sec:macroRypol}), and the implementation of the thermal average, producing closed-form self-consistency equations for the equation of state (Sec.\ \ref{sec:Dopplerthermave}) and for the more general RF signal dynamical response (Sec.\ \ref{sec:statdynlinresp}, and developed formally in App.\ \ref{app:dynlinrespformal}). The connection to the sensor output signal, namely through the vapor EIT effect derived from the (highly atomic state sensitive) probe laser beam transmission coefficient, is described as well (Sec.\ \ref{sec:probeEIT}).

\begin{figure*}

\includegraphics[width=2.23in,viewport = 10 10 410 310,clip]{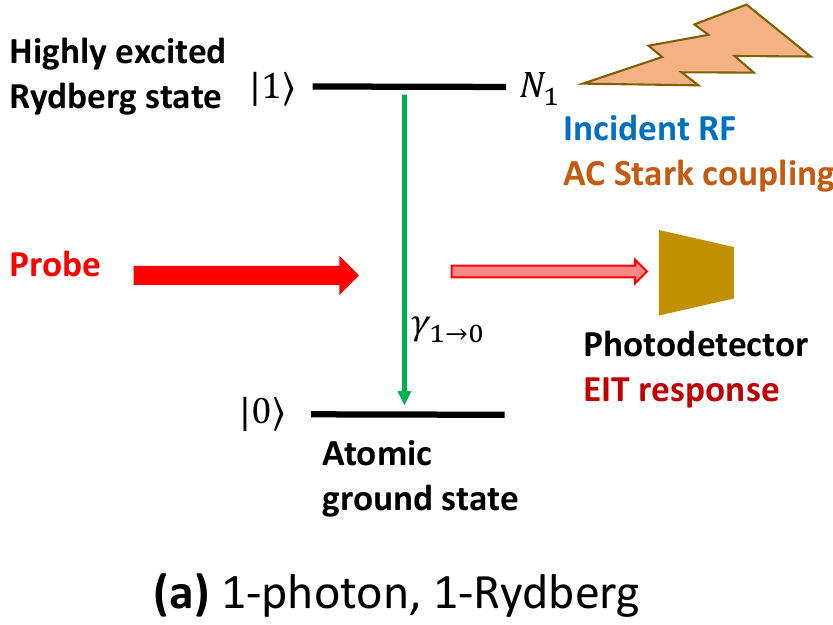} \quad
\includegraphics[width=2.23in,viewport = 10 10 410 360,clip]{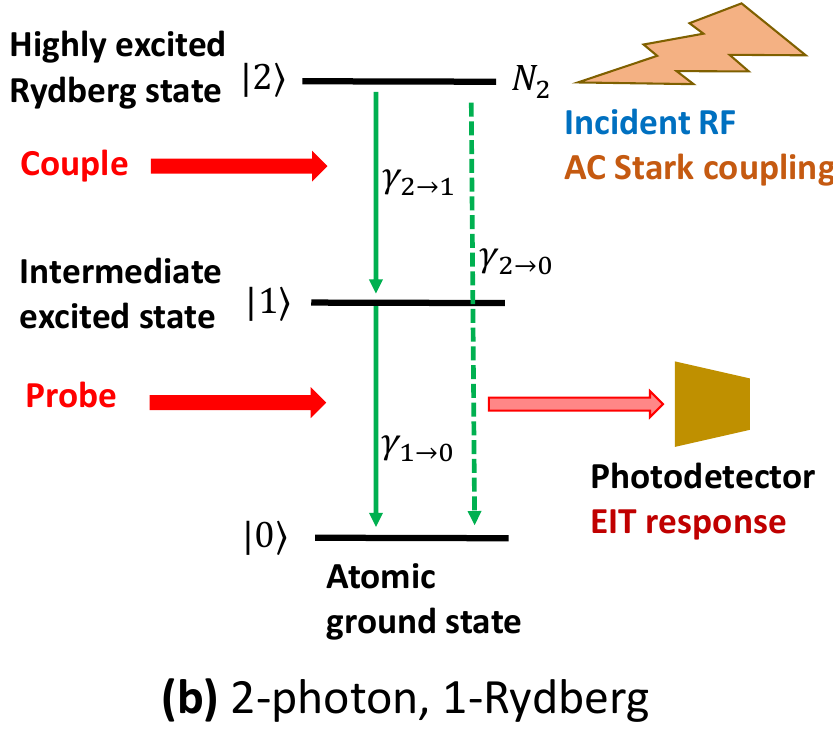} \quad
\includegraphics[width=2.23in,viewport = 10 10 410 370,clip]{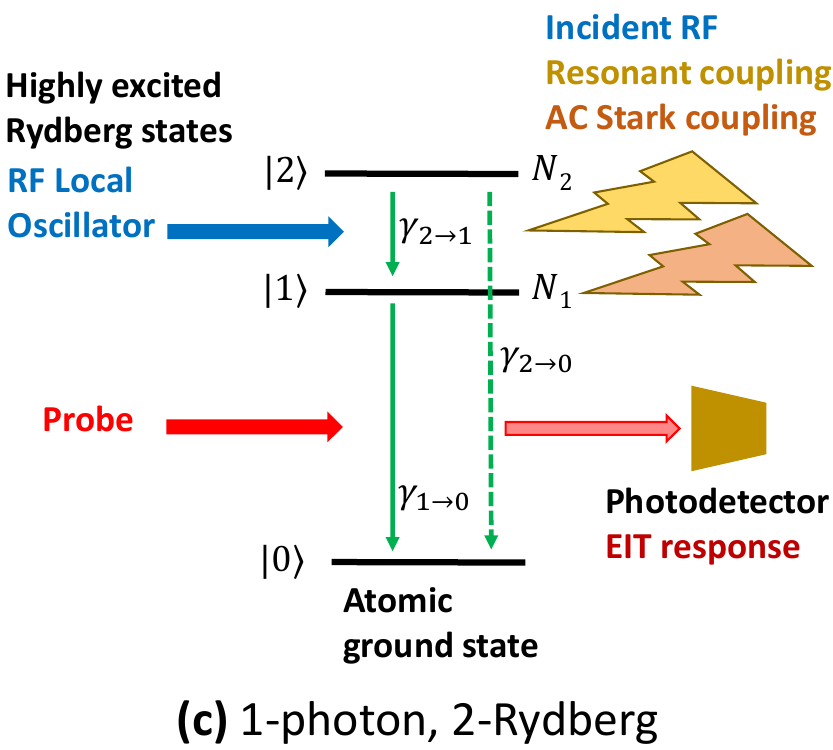}

\caption{Rydberg sensor measurement setups considered in this paper. (a) Simplest two-state setup \cite{CRWAW2013,MLDGL2014} in which the probe laser directly excites the atom from its ground state $|0\rangle$ to a Rydberg state $|1 \rangle$ characterized by large principal quantum number, typically $N_1 > 30$. Due to the correspondingly large electric dipole moment, the incident RF signal dominantly perturbs the Rydberg energy level $\varepsilon_1$ (AC Stark effect). (b) More complex two photon excitation setup in which the probe laser excites an intermediate (e.g., first excited) state $|1 \rangle$, followed by a coupling laser that excites the Rydberg state $|2 \rangle$ with large principal quantum number $N_2$. The RF coupling is again dominated by the Rydberg level AC Stark effect. The two (or more, in some setups) stages has practical advantages, including more accessible lasers and more convenient tunability. (c) The probe laser again directly excites a Rydberg level $|1\rangle$, but now a RF local oscillator (e.g., dielectric patterned on or near the vapor cell surface) then excites a second Rydberg level $|2 \rangle$, typically with quantum number difference $N_2 - N_1 \in  \{0, \pm 1\}$, so long as its frequency $\omega_\mathrm{LO}$ is tuned close to the frequency difference $\omega_{12} = (\varepsilon_2 - \varepsilon_1)/\hbar$. The incident RF signal, with setup selected so that its frequency $\omega_\mathrm{RF}$ is also close to $\omega_{12}$, now perturbs both the local oscillator (Rabi frequency perturbation) and the Rydberg energy levels (AC Stark effect). For small $|\omega_\mathrm{RF} - \omega_{12}|$ and $|\omega_\mathrm{RF} - \omega_\mathrm{LO}|$ the former heterodyne-type response strongly dominates, hence the intense interest in this setup \cite{NIST2019a,NIST2019b,Jing2020,Waterloo2021,MITRE2021,NIST2023,Romalis2024,Wu2024,Sandidge2024,Warsaw2024}. Intermediate state versions of this setup are very common as well, but for simplicity are not treated here. In all setups, the RF perturbation of the resulting tuned coherent two- or three-state superposition leads to a measurable change in the probe beam vapor transparency (EIT response). The various shown spontaneous decays rates (and other contributing forms of population loss) $\gamma_{m \to n}$ influence the measurement through associated fundamental linewidths.}

\label{fig:atomsetup}
\end{figure*}

The bistability theory is first specialized in Sec.\ \ref{sec:1ph1Rylevellim} to the previously explored single Rydberg level case depicted in Fig.\ \ref{fig:atomsetup}(a), but now fully including the Doppler-induced thermal average (Sec.\ \ref{sec:thermavemf}). In Sec.\ \ref{sec:critscale} the mean field equations are formulated in a convenient scaled form and the bistable phase diagram explored through some examples. The sensor signal detection is based on the AC Stark coupling to the Rydberg level energy which serves to perturb the probe laser detuning $\Delta$. Physical estimates for the bistability enhancement, based on choosing a near-critical operating point, are made based on experimentally motivated parameter choices. Contact is made with the EIT measurement in Sec.\ \ref{sec:probeEITsigsense} where an exact relation between the probe beam absorption rate and the excited Rydberg level population is derived (Sec.\ \ref{sec:probeEITsigsense}). The finite frequency dynamic linear response is investigated for this setup in Sec.\ \ref{sec:dynlinresp1}, and the degradation and broadening of the resonance peak with frequency is quantified in terms of the atomic decay and dephasing parameters.

In Sec.\ \ref{sec:2phmfsoln} these results are generalized to the two-photon, single Rydberg level setup depicted in Fig.\ \ref{fig:atomsetup}(b). This common setup, incorporating a second ``coupling'' laser (and sometimes three or more lasers), has significant practical experimental advantages, including cheaper, lower frequency lasers, more flexible tunability, and potentially greatly reduced Doppler broadening \cite{W2024}. The underlying physics of the bistable phase is unchanged, but the phase diagram is more complicated, with the possibility of reentrant bistable phases. The dynamic linear response, again based on the Rydberg level AC Stark coupling, is treated in Sec.\ \ref{sec:2ph1Rydlr}.

Finally, the full bistability theory of the resonant Rydberg level pair setup, depicted in Fig.\ \ref{fig:atomsetup}(c), is developed in Sec.\ \ref{sec:2Ryresonant}, including finite frequency dynamic linear response (Sec.\ \ref{sec:1ph2Rydlr}). Signal detection is now based on the far more sensitive tuned resonant coupling of the signal to this pair of levels, through the ``local oscillator'' (LO) Rabi frequency $\Omega_\mathrm{LO}$. The proposed super-enhancement is now obtained by combining this resonant enhancement with an appropriate choice of near-critical operating point in the phase diagram. Physical estimates are again made based on suggested experimental parameters. Multi-photon versions of this setup could be considered as well, but for simplicity we treat here only the single (probe) laser case.

The paper is concluded in Sec.\ \ref{sec:conclude}, including a discussion of future work and desired experimental inputs needed to more precisely quantify and constrain the theoretical results. Several appendices provide important calculational details underlying the presented results that would otherwise interrupt the narrative flow.

\section{Model formulation}
\label{sec:modelform}

\subsection{Experimental measurement setup and model Hamiltonian}
\label{sec:exptmeasure}

The two- and three-state measurement setups studied in this paper are depicted in Fig.\ \ref{fig:atomsetup}. The validity of ignoring all other atomic states relies on a very narrow linewidth probe laser that, for example, selects through a combination of frequency and polarization a particular pair of nearly degenerate hyperfine levels $F$ to define unique states $|0 \rangle$ and $|1 \rangle$. The initial ground state hyperfine levels, for example, are essentially equally thermally occupied. The coupling laser frequency $\omega_C$ [Fig.\ \ref{fig:atomsetup}(b)] or local oscillator frequency $\omega_\mathrm{LO}$ [Fig.\ \ref{fig:atomsetup}(c)] then selects a unique Rydberg state $|2 \rangle$ \cite{foot:mF}. In this section we will develop a three-state model using notation and physics appropriate to Fig.\ \ref{fig:atomsetup}(c). Setups (a) and (b) will then emerge as simpler special cases, treated in Secs.\ \ref{sec:1ph1Rylevellim} and \ref{sec:2phmfsoln}, respectively. Following this, full analysis of setup (c) will be described in Sec.\ \ref{sec:2Ryresonant}.

Given the three-state simplification, the model many body Hamiltonian (henceforth setting $\hbar = 1$) takes the form \cite{foot:corotate}
\begin{equation}
H = \sum_{l=1}^{N_a} \hat h_l + \frac{1}{2} \sum_{l \neq m} \hat v_{lm},
\label{2.1}
\end{equation}
where $N_a$ is the number of atoms. Using standard sign conventions, the individual atom contribution is defined by the $3 \times 3$ matrix
\begin{equation}
\hat h_l = -\left(\begin{array}{ccc}
0 & \frac{1}{2} \Omega_l & 0 \\
\frac{1}{2} \Omega_l^* & \Delta_l & \frac{1}{2} \Omega_{R,l} \\
0 & \frac{1}{2} \Omega_{R,l}^*    & \Delta_l + \Delta_{R,l}
\end{array} \right)
\label{2.2}
\end{equation}
in which $\Delta_l$, $\Delta_{R,l}$, are, respectively, the laser and ``local oscillator'' detuning defined by the (small) differences
\begin{equation}
\Delta_l = \omega_{L,l} - (\varepsilon_1 - \varepsilon_0), \ \
\Delta_{R,l} = \omega_{R,l} - (\varepsilon_2 - \varepsilon_1)
\label{2.3}
\end{equation}
between the illumination frequency and associated energy gap. The corresponding Rabi frequencies
\begin{equation}
\Omega_l = {\bf E}_P({\bf x}_l) \cdot {\bm \mu}_{10},\ \
\Omega_{R,l} = {\bf E}_\mathrm{LO}({\bf x}_l) \cdot {\bm \mu}_{21}
\label{2.4}
\end{equation}
define, respectively, the coupling of the ground state to the first Rydberg level, and first Rydberg level to the second, through the respective local electric fields ${\bf E}_P({\bf x}_l), {\bf E}_\mathrm{LO}({\bf x}_l)$ and transition dipole moments ${\bm \mu}_{10}, {\bm \mu}_{21}$. Specialization to setup (a) is obtained simply by setting $\Omega_{R,l} = 0$ so that state $|2 \rangle$ is not excited. Specialization to setup (b), in which $|1 \rangle$ is not a Rydberg level, amounts mainly to a change in nomenclature and to typical parameter numerical values, and will be clarified as needed below.

In most of what follows, $\Omega_l = \Omega$, $\Omega_{R,l} = \Omega_R$, $\Delta_{R,l} = \Delta_R$ will be taken uniform while
\begin{eqnarray}
\Delta_l &=& \Delta + {\bf k}_P \cdot {\bf v}_l,\ \ k_P = \frac{\omega_P}{c} = \frac{2\pi}{\lambda_P}
\nonumber \\
\Delta_{R,l} &=& \Delta_R + {\bf k}_\mathrm{LO} \cdot {\bf v}_l,\ \
k_\mathrm{LO} = \frac{\omega_\mathrm{LO}}{c} = \frac{2\pi}{\lambda_\mathrm{LO}} \ \ \ \ \ \
\label{2.5}
\end{eqnarray}
will be taken to have a background uniform values $\Delta$, $\Delta_R$ with a Doppler shift correction defined by the EM illumination wavevectors ${\bf k}_P$, ${\bf k}_\mathrm{LO}$ and atom velocity ${\bf v}_l$. For $\lambda_P \sim 1\ \mu$m and  $v \sim 100$ m/s one obtains Doppler shifts in the $\sim 100$ MHz range, which is extremely large compared to the few $\alt 1$ MHz typical atomic linewidths that will be encountered below. In the RF band, the LO transition Doppler shift will be $k_P/k_\mathrm{LO} \sim 10^5$ times smaller. In addition, ${\bf E}_\mathrm{LO}({\bf x})$ will generally have a complicated non-plane-wave distribution (unlike, e.g., a radar gun return signal) making both the magnitude and direction of ${\bf k}_\mathrm{LO}$ ill defined. For these reasons, the LO Doppler for setup (c) will typically be ignored. On the other hand, for setup (b) ${\bf k}_\mathrm{LO} \to {\bf k}_C$ is the coupling laser wavenumber and $k_P/k_C = O(1)$. Both Doppler contributions must then be accounted for in that application (Sec.\ \ref{sec:2phmfsoln}). If desired, the notational change $\omega_\mathrm{LO} \to \omega_C$, ${\bf E}_\mathrm{LO} \to {\bf E}_C$, $\Omega_{R,l} \to \Omega_{C,l}$, $\Delta_{R,l} \to \Delta_{C,l}$ may also be adopted, but this does not affect the general structure of the matrix (\ref{2.2}).

In the presence of an incident signal, for setup (c) the Rydberg level Rabi frequency will be time-dependent,
\begin{equation}
\Omega_R(t) = \Omega_R + e^{-i \Delta \omega_\mathrm{RF} t} \Omega^B_\mathrm{RF}(t)
\label{2.6}
\end{equation}
in which  $\Omega^B_\mathrm{RF}(t) = e^{i\omega_\mathrm{RF} t} \Omega_\mathrm{RF}(t)$ is the base-banded signal, with center frequency removed, and
\begin{equation}
\Delta \omega_\mathrm{RF} = \omega_\mathrm{RF} - \omega_\mathrm{LO}
\label{2.7}
\end{equation}
is the difference between the signal center and local oscillator frequencies. The highest signal sensitivity is expected to occur in the ``adiabatic limit'' with perfectly tuned local oscillator, $\Delta \omega_\mathrm{RF} \to 0$, and small bandwidth, $\Delta \omega^\mathrm{BW}_\mathrm{RF} \to 0$. The finite frequency degradation will be quantified in Secs.\ \ref{sec:dynlinresp1}, \ref{sec:2ph1Rydlr}, and \ref{sec:1ph2Rydlr} within dynamic linear response theory.

The pair interaction term takes the form
\begin{equation}
\hat v_{lm} = \sum_{\alpha,\beta \in \{1,2\}}
V^{\alpha\beta}_{lm} \hat n_l^\alpha \hat n_m^\beta
\label{2.8}
\end{equation}
in which
\begin{equation}
\hat n^\alpha_l = |\alpha \rangle_l \, {}_l\langle \alpha|
\label{2.9}
\end{equation}
are the projection operators onto states $\alpha = 0,1,2$ for atom $l$. Due to their strongly enhanced dipole moments, the (symmetric) interactions $V^{\alpha\beta}_{lm} = V^{\alpha\beta}_{ml} = V^{\beta\alpha}_{lm}$ are taken here as nonzero only for the Rydberg states: $\alpha = 1$ for setup (a), $\alpha = 2$ for setup (b), and both $\alpha,\beta = 1,2$ for setup (c). For the latter, since the two Rydberg states will generally have very similar dipole moments, the interactions may be expected to be only weakly dependent on $\alpha,\beta$. In a full model, $V_{lm}^{\alpha \beta} = V^{\alpha\beta}({\bf x}_l - {\bf x}_m)$ depends on atom relative positions (including dipolar anistropy) which would then need to be included as separate degrees of freedom. However, below we will adopt an approximation that takes advantage of the long-ranged character of the dipole, with each atom interacting simultaneously with many others, and replaces its effect by density-dependent ``mean field'' corrections to the detuning parameters.

\subsection{Quantum master equation}
\label{sec:qmastereq}

The nonequilibrium dynamics are treated within the semiclassical quantum master equation (QME) formalism in which the electromagnetic field is treated as a known classical applied field \cite{FIM2005}. The photon field is formally averaged over, resulting in an equation of motion
\begin{equation}
\partial_t \hat \rho = i [\hat \rho, H] + {\cal L}[\hat \rho]
\label{2.10}
\end{equation}
describing the evolution of the collective atomic degrees of freedom (many body, self-adjoint, positive definite) density matrix $\hat \rho$. The first term describes unitary evolution of an isolated atomic system. The second nonunitary dissipation term describes the effects of atomic decay processes. Recall that, as usual, this is a Schr\"odinger picture equation of motion so that the state projection operators (\ref{2.9}) are viewed as time-independent and we solve for the time varying amplitude of each state. The closure approximation leading to this equation is valid for strong illumination, and under the assumption that the photon interactions are very rapid on the time scale of the atom dynamics. This approximation fails to describe, for example, coherent light--atom entangled states in a cavity. However, it does describe, as will be seen below, dephasing of atom superposition states created by classically treated laser and microwave interactions, which sets limits on sensor capabilities.

If one formally diagonalizes $\hat \rho(t)$,
\begin{equation}
\hat \rho(t) = \sum_n p_n(t) |\psi_n(t) \rangle \langle \psi_n(t)|,
\label{2.11}
\end{equation}
in which $0 \leq p_n(t) \leq 1$ is the classical probability of finding the system in the atomic superposition state $|\psi_n(t) \rangle$. The probabilities must sum to unity, leading to the constraint
\begin{equation}
\mathrm{tr}[\hat \rho] = 1.
\label{2.12}
\end{equation}
The physical average of any operator $\hat O$ is given by
\begin{equation}
\langle \hat O \rangle = \mathrm{tr}[\hat \rho \hat O]
= \sum_n p_n \langle \psi_n|\hat O|\psi_n \rangle,
\label{2.13}
\end{equation}
which appropriately combines the quantum expectation value of $\hat O$ in the state $|\psi_n \rangle$ with the classical probability of finding the system in that state. It will turn out that the Rydberg sensor operation relies on each atom being in a carefully designed ``resonant'' coherent superposition of Rydberg and ground states in order that the perturbation of the Rydberg states by the incident field lead to the strong probe beam EIT response pictured in Fig.\ \ref{fig:atomsetup} \cite{FIM2005}. Consistency of the equation of motion (\ref{2.10}) with the probability conservation law (\ref{2.12}) requires that the dissipation operator satisfy
\begin{equation}
\mathrm{tr}[{\cal L}[\hat \rho]] = 0.
\label{2.14}
\end{equation}

In the model to be considered here, ${\cal L}[\hat \rho]$ is linear in $\hat \rho$ and parameterizes the various (non-unitary, single atom) decay and dephasing processes,
\begin{equation}
{\cal L}[\hat \rho] = \sum_\sigma \sum_l \left(L_{\sigma l} \hat \rho L^\dagger_{\sigma l}
- \frac{1}{2} \left\{L^\dagger_{\sigma l} L_{\sigma l}, \hat \rho \right\} \right),
\label{2.15}
\end{equation}
labeled here by the index $\sigma$. The condition (\ref{2.14}) follows immediately from the definition of the anticommutator and the cyclic property of the trace. Analogous to its classical master equation counterpart, the operator $L_{\alpha l}$ encodes irreversible transitions from one state to another (in this case limited to each individual atom---clearly this may be generalized), and the two terms in (\ref{2.15}) ensure that the rate of population gain of one state at any given time is compensated by an identical rate of population loss of another state.

Defining the transition operators
\begin{equation}
\hat \sigma_l^{\alpha \beta} = |\alpha \rangle_l \, {}_l\langle \beta|,\ \ \alpha \neq \beta,
\label{2.16}
\end{equation}
for the three state model treated here we consider the five processes defined by the operators
\begin{eqnarray}
L^\alpha_{1l} &=& \sqrt{\Gamma_\alpha} \hat \sigma_l^{0\alpha}
\nonumber \\
L^\alpha_{2l} &=& \sqrt{K_\alpha} \hat n_l^\alpha
\nonumber \\
L_{3l} &=& \sqrt{\Gamma_3} \hat \sigma_l^{12}
\label{2.17}
\end{eqnarray}
for $\alpha = 1,2$. We simplify the notation here from that appearing in Fig.\ \ref{fig:atomsetup} to $\Gamma_\alpha = \gamma_{\alpha \to 0}$ and  $\Gamma_3 = \gamma_{2 \rightarrow 1}$. The last line corresponds to the decay from the second Rydberg level to the first, whose rate is generally negligible compared to those to the ground state \cite{foot:Rydecay}. There can actually be no direct decay to ground from the second Rydberg level due to angular momentum selection rules. Thus, $\Gamma_2$ is an effective value accounting for multistep-step decays mediated by other low lying excited states and beam transit time effects.

\subsection{Mean field approximation}
\label{sec:mfapprox}

In order to obtain a more tractable model we next use the mean field approximation to further reduce the many body equation of motion (\ref{2.10}) to an effective single atom theory. Formally, one approximates the full density matrix by a direct product of individual atom density matrices
\begin{equation}
\hat \rho = \otimes_{l=1}^{N_a} \hat \rho_l.
\label{2.18}
\end{equation}
in which each $\hat \rho_l$ is a unit trace $3 \times 3$ matrix. The approximation is formally valid in the limit where the vapor is sufficiently dense and/or the interaction sufficiently long range (as in this case) that the interaction $V_{lm}$ involves many atoms, each experiencing the mean effect of many others. The equation of motion for any given $\hat \rho_k$ is obtained by averaging over all other degrees of freedom,
\begin{equation}
\partial_t \hat \rho_k = i[\hat \rho_k,\hat h_k]
+ \frac{i}{2} \sum_{l \neq m} \mathrm{tr}^{(k)}\{[\hat \rho, \hat v_{lm}]\}
+ {\cal L}_k[\hat \rho_k],
\label{2.19}
\end{equation}
in which $\mathrm{tr}^{(k)}\{ \cdot \}$ denotes a trace over all atoms except $k$. Since the dissipation term is a sum of single atom operators, one obtains
\begin{equation}
{\cal L}_k[\hat \rho_k] = \sum_{\alpha,\beta} \left(
L^\alpha_{\beta k} \hat \rho_k L^{\alpha \dagger}_{\beta k}
- \frac{1}{2} \left\{L^{\alpha \dagger}_{\beta k}
L^\alpha_{\beta k},\hat \rho_k \right\} \right),
\label{2.20}
\end{equation}
which is just the atom $k$ term in (\ref{2.15}), substituting individual decay process operators (\ref{2.17}) (in somewhat of an abuse of notation where the upper index $\alpha$ is understood to be suppressed when $\beta = 3$).

The result for the most general case, setup (c), of the interaction term follows in the form
\begin{eqnarray}
\mathrm{tr}^{(k)}\{[\hat \rho, \hat v_{lm}]\}
&=& \sum_{\alpha, \beta \in \{1,2\}}
V_{lm}^{\alpha\beta} \big(\delta_{lk} [\hat \rho_l,\hat n_l^\alpha] \langle \hat n_m^\beta \rangle
\nonumber \\
&&\hskip0.5in +\ \delta_{mk} \langle \hat n_l^\alpha \rangle [\hat \rho_m,\hat n_m^\beta] \big). \ \ \ \ \ \
\label{2.21}
\end{eqnarray}
Defining the interaction-induced detunings
\begin{equation}
V_{k\alpha} = -\sum_{\beta = 1}^2 \sum_{l (\neq k)}
V^{\alpha\beta}_{kl} \langle \hat n_l^\beta \rangle.
\label{2.22}
\end{equation}
which depend on the mean occupancy
\begin{equation}
\langle \hat n_l^\beta \rangle = \mathrm{tr}[\hat \rho_l \hat n_l^\beta]
= \rho_l^{\beta\beta}
\label{2.23}
\end{equation}
of the interacting atom Rydberg levels, the first two terms in (\ref{2.19}) may be combined to obtain the form
\begin{equation}
\partial_t \hat \rho_k = i[\hat \rho_k,\hat H_k] + {\cal L}_k[\hat \rho_k],
\label{2.24}
\end{equation}
with mean field effective single particle Hamiltonian
\begin{eqnarray}
\hat H_k &=& \hat h_k +  \sum_{\alpha=1}^2 V_{k\alpha} \hat n_\alpha
\label{2.25} \\
&=& - \left(\begin{array}{ccc}
0 & \frac{1}{2} \Omega_k & 0 \\
\frac{1}{2} \Omega_k^* & \Delta_k + V_{k1} & \frac{1}{2} \Omega_{R,k} \\
0 & \frac{1}{2} \Omega_{R,k}^*    & \Delta_k + \Delta_{R,k} + V_{k2}
\end{array} \right)
\nonumber
\end{eqnarray}

In general, the mean values $\langle \hat n_l^\beta \rangle$ in (\ref{2.22}) depend on $l$, e.g., via different atom velocities and inhomogeneities across the vapor cell, but again relying on a sufficiently dense vapor, we further approximate
\begin{equation}
V_{k\alpha} = -\sum_{\beta = 1}^2 V^{\alpha\beta}_\mathrm{eff} \bar n^\beta,
\label{2.26}
\end{equation}
independent of $k$, in which
\begin{equation}
\bar n^\beta = \frac{1}{N_a} \sum_{k=1}^{N_a} \langle \hat n^\beta_k \rangle
= \frac{1}{N_a} \sum_{k=1}^{N_a} \rho_k^{\beta\beta}
\label{2.27}
\end{equation}
is the total mean (thermally averaged) fraction of atoms in Rydberg level $\beta = 1,2$ and the mean potentials are given by
\begin{eqnarray}
V^{\alpha\beta}_\mathrm{eff} &=& \frac{1}{\bar n^\beta}
\left\langle \sum_{l (\neq k)} V_{kl}^{\alpha\beta} \right\rangle
\approx \rho_a V^{\alpha\beta}_\mathrm{tot}
\nonumber \\
V^{\alpha\beta}_\mathrm{tot} &\equiv& \frac{1}{\hbar}
\int d{\bf x} V^{\alpha\beta}({\bf x}),
\label{2.28}
\end{eqnarray}
in which $\rho_a = N_a/V_\mathrm{cell}$ is the total vapor number density. We temporarily restore $\hbar$ to make contact with the physical pair potentials $V^{\alpha\beta}$. These may be approximated as dipole potentials, regularized within the Rydberg state radius.

As described earlier, for the simpler cases one keeps only $V^{11}_\mathrm{eff} \neq 0$ (hence only $V_{k1} = V^{11}_\mathrm{eff} \bar n_1 \neq 0$) for setup (a) and only $V^{22}_\mathrm{eff} \neq 0$ (hence only $V_{k2} = V^{22}_\mathrm{eff} \bar n_2 \neq 0$) for setup (b).

\subsection{Macroscopic Rydberg polarization model}
\label{sec:macroRypol}

An exact form for the effective potential (\ref{2.28}), also validating the mean field approximation, may actually be derived if one treats the Rydberg atoms as classical dipoles and applies the macroscopic Maxwell equations. Thus, if one assumes that the effect of the laser excitation is to produce a smoothly varying polarization density
\begin{equation}
{\bf P}^\beta({\bf x}) = \bar \rho^\beta({\bf x})
\mu^\beta \hat {\bf p}^\beta({\bf x}),\ \
\bar \rho^\beta({\bf x}) \equiv \rho_a({\bf x}) \bar n^\beta({\bf x})
\label{2.29}
\end{equation}
in which
\begin{equation}
\mu^\beta = e r^\beta,\ \ r^\beta \approx N_\beta^2 a_B
\label{2.30}
\end{equation}
is the Rydberg state electric dipole moment approximated by the hydrogen atom orbital radius $N_\beta^2 a_B$ (for principal quantum number $N_\beta$) where $a_B = 0.5292$\,\AA\ is the Bohr radius. For generality the mean density $\bar \rho^\beta({\bf x})$ of atoms excited to Rydberg state $\beta$ and the associated dipole orientation unit vector $\hat {\bf p}^\beta({\bf x})$ (determined by the laser and local oscillator polarization) are permitted to vary smoothly as well. The induced macroscopic electric field ${\bf E}^\beta({\bf x})$ is obtained by solving the electrostatic equation \cite{Jackson}
\begin{equation}
\nabla^2 \Phi^\beta = \frac{1}{\epsilon_0}
\nabla \cdot {\bf P}^\beta,\ \ {\bf E}^\beta = -\nabla \Phi^\beta,
\label{2.31}
\end{equation}
constituting the equivalent continuum method for treating dipole interactions. This field produces the local Stark effect detuning contribution
\begin{equation}
V_\alpha({\bf x}) = \frac{1}{\hbar} \sum_{\beta = 1}^2
\mu^\alpha \hat {\bf p}^\alpha({\bf x}) \cdot {\bf E}^\beta({\bf x})
\label{2.32}
\end{equation}
to Rydberg level $\alpha$.

For uniform polarization in a spherical vapor cell one obtains the result \cite{Jackson}
\begin{equation}
{\bf E}^\beta = -\frac{1}{3\epsilon_0} {\bf P}^\beta,
\label{2.33}
\end{equation}
also uniform within the cell, allowing one to identify
\begin{eqnarray}
V_\mathrm{eff}^{\alpha\beta}
&=& -\frac{\mu^\alpha \mu^\beta}{3 \epsilon_0 \hbar}
\rho_a \hat {\bf p}^\alpha \cdot \hat {\bf p}^\beta
\nonumber \\
&=& -\frac{8\pi}{3} \frac{\varepsilon_H}{\hbar}
\frac{a_B r^\alpha r^\beta}{R_a^3}
\hat {\bf p}^\alpha \cdot \hat {\bf p}^\beta,
\label{2.34}
\end{eqnarray}
in which $\varepsilon_H = e^2/8\pi \epsilon_0 a_B = 13.6057\ \mathrm{eV} = 2\pi(3290\ \mathrm{THz})$ is the hydrogen ground state binding energy and $R_a = \rho_a^{-1/3} \gg r^{\alpha,\beta}$ is the mean separation between Rydberg atoms, assumed large for a low density vapor.

As a physical estimate, for quantum number $N_\beta = 50$ one obtains $r^\beta = 0.133\ \mu$m. Defining the mean Rydberg atom separations $R_\alpha = (\bar \rho^\alpha)^{-1/3} = R_a (\bar n^\alpha)^{-1/3}$, and estimating $r^{\alpha,\beta}/R_{\alpha,\beta} = 0.01$ (corresponding to $\bar \rho^{\alpha,\beta} = 4.25 \times 10^8$ cm$^{-3}$; e.g., atomic density $\rho_a = 4.25 \times 10^{10}$ cm$^{-3}$, Rydberg fractions $\bar n^{\alpha,\beta} = 10^{-2}$, and $\hat {\bf p}^\alpha \cdot \hat {\bf p}^\beta = 1$), one estimates the terms in (\ref{2.26})
\begin{equation}
\frac{1}{2\pi} V^{\alpha\beta}_\mathrm{eff} \bar n^\beta \approx 11\ \mathrm{MHz},
\label{2.35}
\end{equation}
a value that we will see later is comparable to a number of other physical parameters appearing in the Hamiltonian (\ref{2.25}).

A key observation is that since (\ref{2.32}) and (\ref{2.34}) are macroscopic in origin they are insensitive to atomic scale fluctuations and completely dominated by the average effect of distant atoms---the mean field approximation is indeed exact. In particular, the detuning potentials are sensitive to the shape of the vapor cell, making the optically bistable phase diagram sensitive to it as well.

\subsection{Probe laser EM-induced transparency}
\label{sec:probeEIT}

The EIT response is based on a measurement of the variation of the probe beam transmitted power $P_T = \hbar \omega_P N_T$ with input signal, where $N_T$ is the detected photon count. We focus here on the mean (thermally averaged) transmitted fraction
\begin{equation}
{\cal P}^\mathrm{th}_\mathrm{EIT} = \frac{P_T}{P_0}
= \frac{\Omega(L_\mathrm{cell})^2}{\Omega(0)^2}
\label{2.36}
\end{equation}
in which $P_0$ is the incident power into the vapor and $\Omega(s)$, $0 \leq s \leq L_\mathrm{cell}$, is the probe beam Rabi frequency vs.\ distance $s$ into the vapor cell of length $L_\mathrm{cell}$. Neglecting inhomogeneity across the beam, this quantity may be derived from the one-body density matrix in the form \cite{FIM2005}
\begin{equation}
{\cal P}^\mathrm{th}_\mathrm{EIT} = e^{-\alpha_P R_P(L_\mathrm{cell})}
\label{2.37}
\end{equation}
with exponential decay argument
\begin{equation}
R_P(L_\mathrm{cell}) = \int_0^{L_\mathrm{cell}} ds\,
\mathrm{Im}\left[\frac{\rho^{10}(s)}{\Omega(s)} \right]
\label{2.38}
\end{equation}
and coefficient
\begin{equation}
\alpha_P = \frac{2 k_P \rho_a |{\bm \mu}_{10}|^2}{\epsilon_0 \hbar}.
\label{2.39}
\end{equation}
As clarified further below, one should interpret $\hat \rho_l \to \hat \rho(s)$ as relabeling atoms through their positions, recognizing that the thermal average will depend only on the latter.

Note that $\mathrm{Re}[\rho^{10}/\Omega]$ corresponds to an index of refraction, perhaps detectable via an alternative interference measurement. Note also that absorption is a consequence of decay processes: $\mathrm{Im}[\hat \rho/\Omega]$ is nonzero only due to the presence of ${\cal L}[\hat \rho]$, mainly through $\gamma_{1 \to 0}$. It follows that spectral properties of $H$ dominate the resonant behavior, but the Lindblad term ${\cal L}[\hat \rho]$ generates the EIT signal.

For $\Omega = 0$ there is no excitation out of the ground state, and any initial state will ultimately relax to the ground state. The adiabatic solution in this limit is therefore simply the ground state, $\rho^{\alpha\beta} = \delta_{\alpha 0} \delta_{\beta 0}$; in particular $\rho^{10} = 0$. For small but finite $\Omega$, $\rho^{10}(\Omega)$ is linear in $\Omega$ and $\rho^{10}(\Omega)/\Omega$ is a constant, independent of $s$, proportional to the vapor linear polarizeability tensor whose imaginary (absorptive) part produces the beam attenuation. However, the result (\ref{2.37}) is actually quite general, including nonlinear atomic polarization effects present for larger $\Omega$.

In the linear regime, or for sufficiently thin cells for which $\Omega \simeq \Omega(0)$ may be taken as uniform, one obtains the simple result
\begin{equation}
R_P(L) = L\mathrm{Im}\left[\frac{\rho^{10}(\Omega(0))}{\Omega(0)} \right].
\label{2.40}
\end{equation}
One typically finds that this approximation is valid for $L \alt 1$ cm. The general case may be treated by developing a self-consistent system of ODEs for $\Omega(s)$, $\rho_{10}^\mathrm{th}(s)$, and $R_P(s)$ \cite{W2024}. However, for simplicity all of the physical estimates in this paper will be based on the uniform limit (\ref{2.40}).

\subsection{Static and dynamic linear response}
\label{sec:statdynlinresp}

A convenient measure for characterizing Rydberg sensor operation is via the EIT dynamic linear response sensitivity. Thus, for small harmonic perturbation to some parameter $X$
\begin{equation}
X(t) = X + \delta X e^{-i\omega t}.
\label{2.41}
\end{equation}
one expects time dependent transmission perturbation
\begin{eqnarray}
\delta {\cal P}^\mathrm{th}_\mathrm{EIT}(t)
&\equiv& {\cal P}^\mathrm{th}_\mathrm{EIT}(t)
- {\cal P}^\mathrm{th}_\mathrm{EIT}(X)
\label{2.42} \\
&=& \delta X e^{-i\omega t} S_\mathrm{EIT}(X,\omega)
+ O(|\delta X|^2),
\nonumber
\end{eqnarray}
which serves to define the dynamic linear response function $S_\mathrm{EIT}(X,\omega)$. This function is used to construct the response to the more general dynamic perturbation (\ref{2.6})
\begin{eqnarray}
\delta {\cal P}^\mathrm{th}_\mathrm{EIT}(t) &=& e^{-i\Delta \omega_\mathrm{RF} t}
\int \frac{d\omega}{2\pi} S_\mathrm{EIT}(X,\Delta \omega_\mathrm{RF} + \omega)
\nonumber \\
&&\hskip0.7in \times\ \hat X^B_\mathrm{RF}(\omega) e^{-i\omega t}
\label{2.43}
\end{eqnarray}
in which $\hat X^B_\mathrm{RF}(\omega)$ is the Fourier transform of the base-banded signal $X^B_\mathrm{RF}(t)$. For very narrow-banded signals, compared to all atomic response bandwidths, one obtains the simple proportionality
\begin{equation}
\delta {\cal P}^\mathrm{th}_\mathrm{EIT}(t)
\simeq X^B_\mathrm{RF}(t) e^{-i\Delta \omega_\mathrm{RF} t}
S_\mathrm{EIT}(X,\Delta \omega_\mathrm{RF}).
\label{2.44}
\end{equation}
For Rydberg sensors, validity of (\ref{2.44}) is limited, especially if one seeks ``resonant'' settings to maximize the response, and the more general result (\ref{2.43}) will in this case produce strong signal distortion.

In the adiabatic limit, in which one additionally tunes $\Delta \omega_\mathrm{RF} \to 0$, the steady state linear response
\begin{eqnarray}
S^\mathrm{th}_\mathrm{EIT}(X) &\equiv&  S_\mathrm{EIT}(X,0)
= \frac{\partial {\cal P}^\mathrm{th}_\mathrm{EIT}}{\partial X}
\label{2.45}
\end{eqnarray}
is obtained by simple differentiation. From the exponential form (\ref{2.37}) one obtains
\begin{equation}
S^\mathrm{th}_\mathrm{EIT} = -\alpha_P
\frac{\partial R_P(L_\mathrm{cell})}{\partial X}
{\cal P}^\mathrm{th}_\mathrm{EIT}.
\label{2.46}
\end{equation}
In the linear or thin cell regime where (\ref{2.40}) is valid, the result simplifies to the form
\begin{equation}
S^\mathrm{th}_\mathrm{EIT} = -\frac{\alpha_P L}{\Omega}
\mathrm{Im}\left[\frac{\partial \rho^{10}(\Omega)}
{\partial X }\right] e^{-\alpha_P R_P(L)}.
\label{2.47}
\end{equation}

In what follows, we will be interested in $X = \Delta$ or $X = \Delta_R$ (AC Stark coupling) for the single Rydberg level cases (Sec.\ \ref{sec:1ph1Rylevellim} or \ref{sec:2phmfsoln}, respectively) and $X = \Omega_R$ for the resonant Rydberg level pair (Sec.\ \ref{sec:2Ryresonant}). For optimal RF sensor operation one seeks setups that maximize the physical response ratio $|\delta X|/|{\bf E}_\mathrm{RF}|$. Although Rydberg states lead to strong enhancements for all cases, the last case [setup (c)] leads to far better performance \cite{NIST2019a,NIST2019b,Jing2020,Waterloo2021,MITRE2021,NIST2023,Romalis2024,Wu2024,Sandidge2024,Warsaw2024}.

\subsection{Explicit component equations of motion}
\label{sec:cmptmotioneqn}

In order to derive an explicit set of equations for the single atom density matrix components, we write out the latter in the form
\begin{equation}
\hat \rho_k = \sum_{\alpha = 1}^3 \rho_k^{\alpha\alpha} \hat n^\alpha
+ \sum_{\alpha \neq \beta} \rho_k^{\alpha\beta} \hat \sigma^{\alpha\beta},
\label{2.48}
\end{equation}
and the mean field Hamiltonian (\ref{2.25}) may similarly be written out in the form
\begin{eqnarray}
\hat H_k &=& -(\Delta_k + V_{k1}) \hat n^1 - (\Delta_k + \Delta_{R,k} + V_{k2}) \hat n^2
\nonumber \\
&&-\ \frac{1}{2} (\Omega_k \hat \sigma^{01} + \Omega_k^* \hat \sigma^{10})
- \frac{1}{2} (\Omega_{R,k} \hat \sigma^{12} + \Omega_{R,k}^* \hat \sigma^{21}).
\nonumber
\label{2.49}
\end{eqnarray}
Details of the somewhat tedious derivation based on these forms are relegated to App.\ \ref{app:dmmotioneq}. The most general componentwise result, following from (\ref{A10}), is
\begin{widetext}
\begin{eqnarray}
(\partial_t + \Gamma_1) \rho^{11}
&=& \frac{i}{2}(\Omega_R^* \rho^{12} - \Omega_R \rho^{21})
- \frac{i}{2}(\Omega^* \rho^{01} - \Omega \rho^{10}) + \Gamma_3 \rho^{22}
\nonumber \\
(\partial_t + \Gamma_2 + \Gamma_3) \rho^{22}
&=& -\frac{i}{2}(\Omega_R^* \rho^{12} - \Omega_R \rho^{21})
\nonumber \\
\left[\partial_t - i (\Delta + V_1)
+ \frac{\Gamma_1 + K_1}{2} \right] \rho^{01}
&=& \frac{i}{2}\left[\Omega(\rho^{00} - \rho^{11})  + \Omega_R^* \rho^{02} \right]
\nonumber \\
\left[\partial_t  - i (\Delta + \Delta_R + V_2)
+ \frac{\Gamma_2 + \Gamma_3 + K_2}{2} \right] \rho^{02}
&=& \frac{i}{2} (\Omega_R \rho^{01} - \Omega \rho^{12})
\nonumber \\
\left[\partial_t - i(\Delta_R + V_2 - V_1)
+ \frac{\Gamma_1 + \Gamma_2 + \Gamma_3 + K_1 + K_2}{2} \right] \rho^{12}
&=& \frac{i}{2}\left[\Omega_R(\rho^{11} - \rho^{22}) - \Omega^* \rho^{02} \right]
\label{2.50}
\end{eqnarray}
\end{widetext}
with $\rho^{00} = 1 - \rho^{11} - \rho^{22}$ and $\rho^{\beta\alpha} = \rho^{\alpha\beta *}$ determining the remaining four quantities. The atom index $k$ continues to be suppressed here. These are the basic equations, specialized to the appropriate setup, that will be solved in later sections.

\subsection{Motion-induced Doppler and thermal averages}
\label{sec:Dopplerthermave}

Recalling via (\ref{2.26}) that the interaction coefficients $V_1$ and $V_2$ depend self-consistently on the mean Rydberg state occupancies $\bar n_1$, $\bar n_2$, solutions to (\ref{2.50}) must be derived by developing (nonlinear) closure conditions for the former. This will be done first in the steady state limit, in which the time derivatives all vanish. Although the general time-dependent equations are therefore also nonlinear, an exact dynamic linear response theory will then be developed, valid for weak perturbations of the potential driving terms $\Delta(t)$, $\Delta_R(t)$, $\Omega(t)$, $\Omega_R(t)$.

The closure conditions are obtained by first solving (\ref{2.50}) for each atom $l$ for fixed values of the atom velocity $v_l$ and then averaging the results. This procedure is supported by the accuracy of the mean field approximation whose volume average suppresses thermal fluctuations in the effective interaction. Assuming statistical equilibrium of the atom position degrees of freedom, the usual ergodic hypothesis produces
\begin{eqnarray}
\hat \rho_\mathrm{th} &=& \frac{1}{N_a} \sum_{l=1}^{N_a} \hat \rho({\bf v}_l)
\nonumber \\
&=& \int d{\bf v} P_\mathrm{th}({\bf v}) \hat \rho({\bf v})
\label{2.51}
\end{eqnarray}
with Maxwell distribution
\begin{equation}
P_\mathrm{th}({\bf v}) = \frac{e^{-|{\bf v}|^2/2v_\mathrm{th}^2}}{(2\pi v_\mathrm{th})^{3/2}},\ \
v_\mathrm{th} = \sqrt{k_B T/m_a},
\label{2.52}
\end{equation}
in which $m_a$ is the atomic mass. Of course, the result $\hat \rho_\mathrm{th}(\bar n^1,\bar n^2)$ depends implicitly on the assumed values for the thermal averages $\bar n^\alpha$ appearing in the interaction parameters (\ref{2.26}). The self consistency conditions are therefore
\begin{eqnarray}
\bar n^\alpha &=& \mathrm{tr}[\hat \rho_\mathrm{th}(\bar n^1,\bar n^2) \hat n^\alpha]
= \langle \alpha| \hat \rho_\mathrm{th}(\bar n^1,\bar n^2) |\alpha \rangle
\nonumber \\
&=& \rho^{\alpha\alpha}_\mathrm{th}(\bar n^1,\bar n^2),\ \ \alpha = 1,2.
\label{2.53}
\end{eqnarray}
This includes full time dependent $\bar n^\alpha(t)$ implicitly appearing on both sides.

An important practical aspect of the theory to be developed is that, for the steady state and linear response results that are the focus of this paper, the thermal average (\ref{2.51}) can be performed analytically, vastly simplifying the subsequent numerical solution of the mean field equations (\ref{2.53}).

\section{Single photon, single Rydberg level limit}
\label{sec:1ph1Rylevellim}

We specialize now to setup (a) in Fig.\ \ref{fig:atomsetup} by setting $\Omega_R = 0$, which depopulates the second Rydberg level, $\rho^{02} = \rho^{12} = \rho^{22} = 0$ (at least for large times, in the general case of a nonzero initial condition). The model reverts to a $2\times 2$ matrix (spin-$\frac{1}{2}$) form, previously studied in an effective zero temperature limit \cite{CRWAW2013,MLDGL2014}, approximated by limiting attention to the small $\sim$\,1\% subset of sufficiently slow-moving atoms, $v \alt 1$ m/s. Here we will develop the full finite temperature theory.

In this limit, equations (\ref{2.50}) reduce to the pair of equations
\begin{eqnarray}
(\partial_t + \Gamma) n = - \frac{i}{2}(\Omega^* \rho^{01} - \Omega \rho^{01 *})
&=& \mathrm{Im}(\Omega^* \rho^{01})
\nonumber \\
\left[\partial_t - i (\Delta + V) + \frac{\Gamma + K}{2} \right] \rho^{01}
&=& \frac{i}{2} \Omega(n^0 - n)
\nonumber \\
\label{3.1}
\end{eqnarray}
where we simplify the notation using $\Gamma = \Gamma_1$, $K = K_1$, $n \equiv n^1 = \rho^{11}$, $n^0 = \rho^{00} = 1 - n$, and we have substituted $\rho^{10} = \rho^{01 *}$. We continue to suppress the atom index $k$. The interaction parameter reduces to
\begin{equation}
V \equiv V_1 = -V_\mathrm{eff} \bar n, \ \
V_\mathrm{eff} = V^{11}_\mathrm{eff}.
\label{3.2}
\end{equation}

Beginning with the steady state case, $\partial_t \rho^{11} = \partial_t \rho^{12} = 0$, it is convenient to adopt a spin-type notation for the real and imaginary parts of $\rho^{01}$ \cite{CRWAW2013,MLDGL2014}
\begin{equation}
\rho^{01} = \frac{S^x + i S^y}{2},\ \ \rho^{10} = \frac{S^x - i S^y}{2},\ \
\Omega = \Omega^x + i \Omega^y,
\label{3.3}
\end{equation}
which leads to the three real equations
\begin{eqnarray}
\Omega^x S^y - \Omega^y S^x &=& 2\Gamma n
\nonumber \\
\frac{\Gamma + K}{2} S^x + (\Delta + V)  S^y &=& \Omega^y (2n - 1)
\nonumber \\
\frac{\Gamma + K}{2} S^y - (\Delta + V)  S^x &=& -\Omega^x (2n - 1).
\label{3.4}
\end{eqnarray}
One may also define $S^z = 2 n - 1$, lying in the interval $[-1, 1]$, but it is convenient to continue to use $n$. The last two may be reorganized in the forms
\begin{eqnarray}
\frac{\Gamma + K}{2} (\Omega^x S^x + \Omega^y S^y) &=& -(\Delta + V) (\Omega^x S^y - \Omega^y S^x)
\nonumber \\
&=& -2(\Delta + V) \Gamma n
\nonumber \\
(\Delta + V) (\Omega^x S^x + \Omega^y S^y)
&=& \frac{\Gamma + K}{2} (\Omega^x S^y - \Omega^y S^x)
\nonumber \\
&&+\ |\Omega|^2(2n-1)
\label{3.5} \\
&=& (\Gamma + K)\Gamma n  + |\Omega|^2(2n-1),
\nonumber
\end{eqnarray}
and combining them leads to
\begin{eqnarray}
n &=& \frac{1}{2 + \frac{\Gamma(\Gamma+K)}{|\Omega|^2}
+ \frac{4\Gamma}{(\Gamma + K)} \frac{(\Delta + V)^2}{|\Omega|^2}}
\nonumber \\
\hat {\bm \Omega}^\perp \cdot {\bf S} &\equiv&
\frac{\Omega^x S^y - \Omega^y S^x}{|\Omega|} = \frac{2\Gamma}{|\Omega|} n
\nonumber \\
\hat {\bm \Omega} \cdot {\bf S} &\equiv& \frac{\Omega^x S^x + \Omega^y S^y}{|\Omega|}
= -\frac{4\Gamma (\Delta + V)}{|\Omega|(\Gamma + K)} n.
\label{3.6}
\end{eqnarray}
The last two represent the components of $(S^x,S^y)$ parallel to and orthogonal to the vector ${\bm \Omega} = (\Omega^x,\Omega^y)$.

\subsection{Exact thermally averaged mean field equation}
\label{sec:thermavemf}

At this point the atom label needs to be restored in order to derive a closure equation for $\bar n$. We idealize to a uniform vapor, with only the atom velocity varying through the detuning Doppler shift $\Delta(v) = \Delta + k_P v$, and only the velocity component $v = \hat {\bf k}_P \cdot {\bf v}$ along the probe beam direction enters. Defining the thermally scaled dimensionless quantities
\begin{equation}
u = \frac{v}{v_\mathrm{th}},\
\Delta_\mathrm{th} = \frac{\Delta}{k_P v_\mathrm{th}},\
V_\mathrm{th} = -\frac{V_\mathrm{eff}}{k_P v_\mathrm{th}},\
\Omega_\mathrm{th} = \frac{|\Omega|}{k_P v_\mathrm{th}},
\label{3.7}
\end{equation}
one may write (\ref{3.6}) in the conveniently scaled form
\begin{equation}
n(u) = \frac{1}{a} \frac{1}{1 + (u + u_{\bar n})^2/\nu^2}
\label{3.8}
\end{equation}
in which $\bar n$ enters through the combination
\begin{equation}
u_{\bar n} = \Delta_\mathrm{th} + V_\mathrm{th} \bar n
\label{3.9}
\end{equation}
and we define the dimensionless combinations
\begin{eqnarray}
a &=& 2 + \frac{\Gamma(\Gamma+K)}{|\Omega|^2}
\nonumber \\
\nu &=& \Omega_\mathrm{th} \sqrt{\frac{(\Gamma + K) a}{4\Gamma}}
\equiv \frac{v_0}{v_\mathrm{th}}
\nonumber \\
v_0 &=& \frac{\Gamma + K}{2k_P}
\sqrt{1 + \frac{2|\Omega|^2}{\Gamma(\Gamma + K)}}.
\label{3.10}
\end{eqnarray}
One may interpret $v_0$ here as an intrinsic atomic velocity measure derived from an underlying Lorentzian linewidth.

The result (\ref{3.8}) produces the Rydberg level occupancy for given atom speed if $\bar n$ is known. To obtain the self consistent equation for $\bar n$ itself one performs the (1D, in this case) Gaussian average over both sides of (\ref{3.8}) to obtain
\begin{equation}
\bar n = \frac{1}{a} \Phi_\nu(u_{\bar n})
\label{3.11}
\end{equation}
in which
\begin{eqnarray}
\Phi_\nu(w) &=& \int \frac{du}{\sqrt{2\pi}} \frac{e^{-u^2/2}}{1 + (w-u)^2/\nu^2}
\nonumber \\
&=& \nu \mathrm{Im} f(w + i\nu)
\nonumber \\
f(\zeta) &=& \int \frac{du}{\sqrt{2\pi}} \frac{e^{-u^2/2}}{u - \zeta}.
\label{3.12}
\end{eqnarray}
Note that since the right hand side of (\ref{3.8}) is positive and bounded above by $1/a$, an immediate consequence is that $0 < \bar n < 1/a < \frac{1}{2}$ indeed lies in the physical interval $[0, 1]$. The change of variable $u \to -u$ here compared to (\ref{3.8}) more conveniently exhibits the convolution form of $f(w + i\nu)$, which then yields the Fourier transform
\begin{eqnarray}
\hat f(q,\nu) &\equiv& \int dw e^{iq w} f(w + i\nu)
\nonumber \\
&=& -2\pi i e^{-q^2/2} e^{q \nu} \theta(-\nu q) \mathrm{sgn}(q),
\label{3.13}
\end{eqnarray}
in which $\theta(x) = 1$ for $x > 0$, vanishing for $x < 0$, is the step function. Inverse Fourier transforming, by completing the square in the exponent and taking appropriate care with the complex plane contours, yields the desired analytic (error function) form
\begin{eqnarray}
f(\zeta) &=& i\sqrt{\frac{\pi}{2}} e^{-\zeta^2/2}
\left[\mathrm{erf}\left(\frac{i\zeta}{\sqrt{2}} \right) \pm 1 \right],\ \
\mp \mathrm{Im}(\zeta) > 0
\nonumber \\
\mathrm{erf}(z) &=& \frac{2}{\sqrt{\pi}} \int_0^z e^{-u^2} du = 1 - \mathrm{erfc}(z)
\label{3.14}
\end{eqnarray}
The discontinuity across $\mathrm{Im}(\zeta) = 0$ corresponds to the principal value contribution to (\ref{3.12}) when $\zeta$ crosses the real axis. For the general case, with two Rydberg levels, we will need this identity for both signs of $\mathrm{Im}(\zeta)$, but here $\mathrm{Im}(\zeta) = \nu > 0$, and one obtains
\begin{eqnarray}
f(\zeta) &=& -i\sqrt{\frac{\pi}{2}} e^{-\zeta^2/2}
\mathrm{erfc}\left(\frac{i\zeta}{\sqrt{2}} \right)
\label{3.15} \\
\Phi_\nu(w) &=& -\nu \sqrt{\frac{\pi}{2}}
\mathrm{Re}\left[e^{(i w - \nu)^2/2}
\mathrm{erfc}\left(\frac{i w - \nu}{\sqrt{2}} \right) \right].
\nonumber
\end{eqnarray}
in which $\mathrm{erfc}(x) = 1- \mathrm{erf}(x)$ is the complimentary error function.

\begin{figure*}

\includegraphics[width=6.5in, viewport = 0 0 940 360, clip]{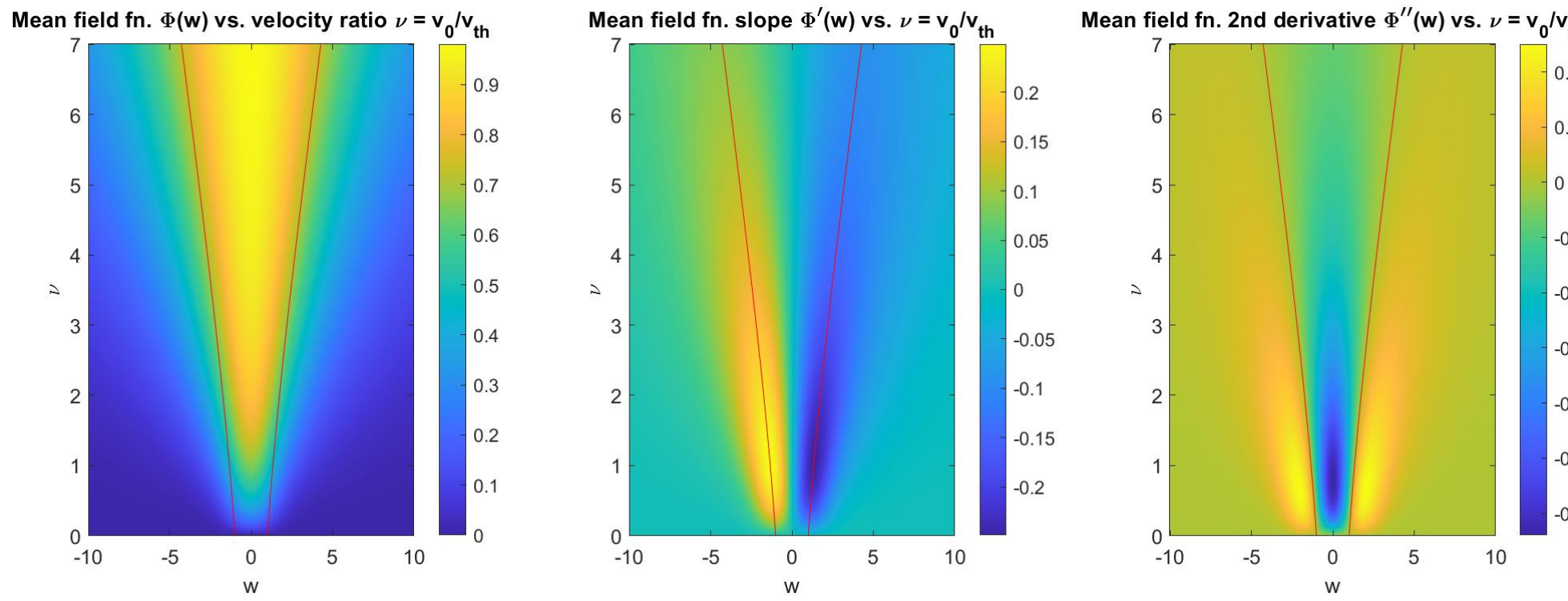}

\caption{2D color plots of $\Phi_\nu(w)$ (\textbf{left}), slope $\Phi_\nu'(w)$ (\textbf{center}), and second derivative $\Phi_\nu''(w)$ (\textbf{right}). The low temperature limit corresponds to large $\nu \gg 1$. The red lines represent the contours $\Phi_\nu''(w) = 0$, corresponding, as seen, to extrema of $\Phi_\nu'(w)$ that through tuning of the physical parameters determine the critical line (\ref{3.30}).}

\label{fig:PhiPhipPhipp}
\end{figure*}

\begin{figure*}

\includegraphics[width=6.0in, viewport = 0 0 950 470, clip]{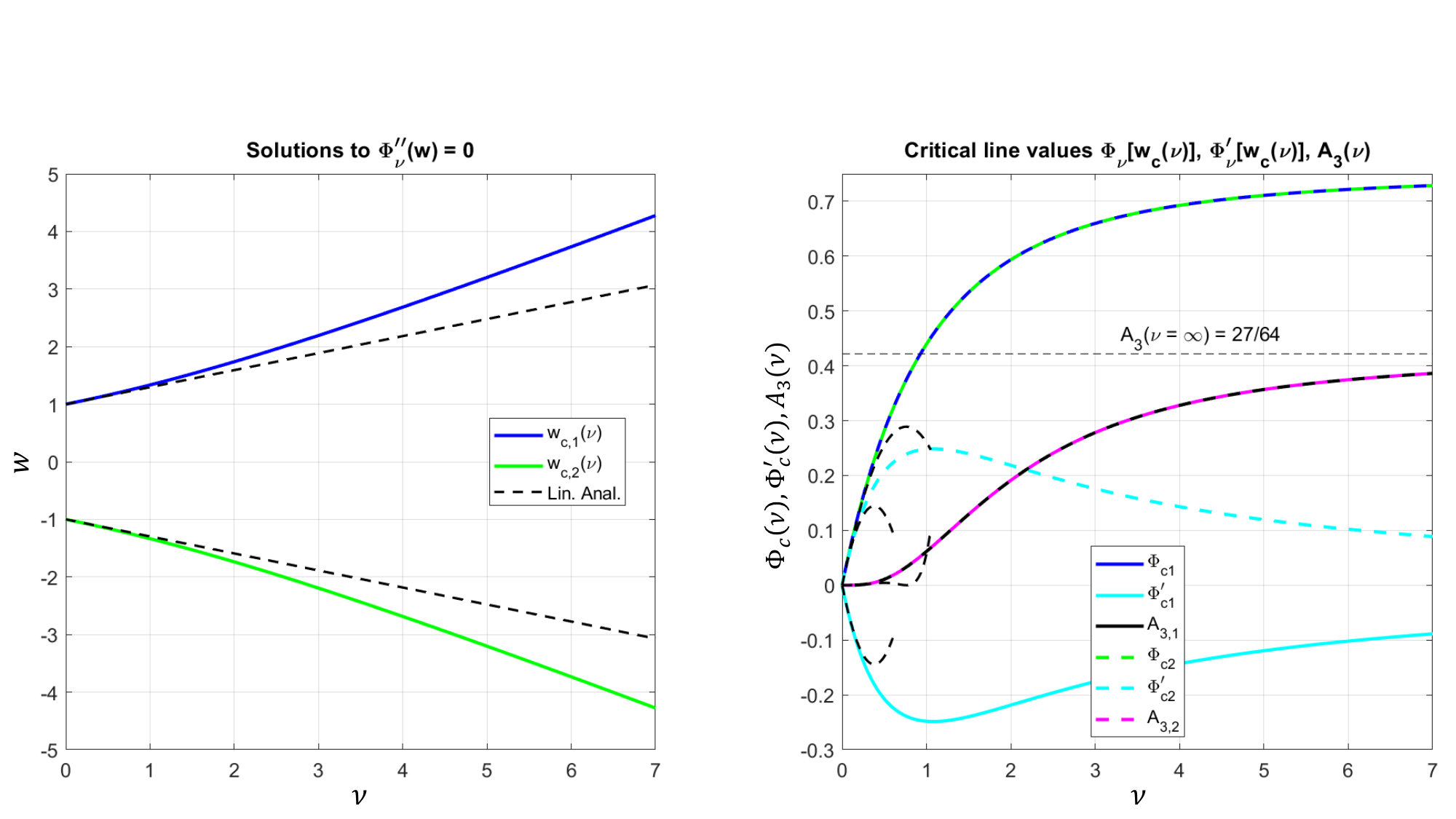}

\caption{\textbf{Left panel:} Critical curves $w_{c,1}(\nu) = -w_{c,2}(\nu)$ defined by the (symmetric) roots of the second derivative equation (\ref{3.29}), plotted also as red contours in Fig.\ \ref{fig:PhiPhipPhipp}. The dashed black lines show the analytic linear order result (\ref{B7}). \textbf{Right panel:} Critical line data derived from these curves. Plotted are quantities $\Phi_\nu[w_{c,1}(\nu)] = \Phi_\nu[w_{c,2}(\nu)]$, $\Phi_\nu'[w_{c,1}(\nu)] = -\Phi_\nu'[w_{c,2}(\nu)]$, and the combination $A_{3,1}(\nu) = A_{3,2}(\nu)$ defined by (\ref{3.32}). As seen in (\ref{3.31})--(\ref{3.35}) these quantities fully describe the relation between the critical values of the dimensionless physical parameters $a,b,c$, defined in (\ref{3.10}) and (\ref{3.17}), as thermal effects vary through the characteristic velocity ratio $\nu = v_0/v_\mathrm{th}$. The black dashed lines show the small $\nu$ asymptotic results derived in App.\ \ref{app:1ph1RyhighTasymp}. In particular, $A_3 \sim \nu^4$ while the others vanish linearly.}

\label{fig:critlinedat}
\end{figure*}

\subsection{Zero temperature limit}
\label{sec:zeroTlim}

In zero temperature limit the Maxwell distribution suppresses all nonzero $v$. Equivalently, for large $\nu, u_n$ the $u$-dependence in the denominator of (\ref{3.8}) is negligible for $u = O(1)$, the Gaussian integral becomes trivial, and (\ref{3.11}) reduces to
\begin{equation}
\bar n = \frac{1}{a} \frac{1}{1 + u_{\bar n}^2/\nu^2} = \frac{1}{a + b(c + \bar n)^2}
\label{3.16}
\end{equation}
with
\begin{equation}
b = \frac{4\Gamma}{\Gamma + K} \frac{V_\mathrm{eff}^2}{|\Omega|^2},\ \
c = -\frac{\Delta}{V_\mathrm{eff}},
\label{3.17}
\end{equation}
equivalent to (\ref{3.6}) with $v = 0$. This yields $\bar n$ as the roots of the cubic polynomial
\begin{equation}
p(\bar n) = \bar n[a + b(c + \bar n)^2] - 1,
\label{3.18}
\end{equation}
and equivalently corresponds to minimizing the effective free energy
\begin{equation}
F(\bar n) = \frac{1}{4} b \bar n^4 + \frac{2}{3} b c \bar n^3
+ \frac{1}{2} (a + bc^2) \bar n^2 - \bar n,
\label{3.19}
\end{equation}
which has a very familiar quartic mean field form \cite{MLDGL2014}.

A critical point corresponds to free energy of the form
\begin{equation}
F(\bar n) = \frac{1}{4} b(\bar n - n_c)^4 + F_c
\label{3.20}
\end{equation}
with unique quartic minimum at $\bar n = n_c$. Matching to (\ref{3.17}) one obtains
\begin{equation}
b = -\frac{27}{8c^3},\ \ a = -\frac{9}{8c},\ \ n_c = -\frac{2c}{3}.
\label{3.21}
\end{equation}
Since $b > 0$ is required for stability, this constrains $c < 0$ (hence $V_\mathrm{eff}/\Delta > 0$). Since $a > 2$ one also obtains $c > -\frac{9}{16}$ and $0 < n_c < \frac{3}{8}$. From (\ref{3.17}), one sees that varying $c$ is equivalent to varying the detuning, and obtaining the critical values of $a$ and $b$ is accomplished, respectively, by tuning $|\Omega|$ in the second line of (\ref{3.10}) and $|V_\mathrm{eff}/\Omega|$ in the first of (\ref{3.17}) (e.g., by varying the vapor density).

A first order line corresponds to the form
\begin{equation}
F(\bar n) = \frac{1}{4} b (\bar n - n_0)^2 - \frac{1}{2} d(\bar n - n_0)^2 + F_0
\label{3.22}
\end{equation}
which produces a pair of degenerate minima
\begin{equation}
\bar n_\pm = n_0 \pm \sqrt{\frac{d}{b}}.
\label{3.23}
\end{equation}
Matching to (\ref{3.21}) one obtains
\begin{equation}
b = -\frac{27}{8c^3} + \frac{9d}{4c^2},\ \ a = -\frac{9}{8c} - \frac{d}{4},\ \ n_0 = n_c = -\frac{2c}{3}.
\label{3.24}
\end{equation}
in which one may view $d$ as an additional free parameter moving one along the first order line (still constrained here by $a > 2$), and vanishing at the critical point.

Spinodal boundaries, the endpoints of hysteresis loops, correspond (within the mean field description) to the disappearance of one the free energy local minima as one increasingly lifts the degeneracy moving away from the first order line. At such a point the cubic polynomial (\ref{3.18}) takes the form
\begin{equation}
p(\bar n) = b(\bar n - r)^2 (\bar n - s),
\label{3.25}
\end{equation}
producing a pair of degenerate roots $\bar n = r$, hence cubic inflection point in $F(\bar n)$, and a single stable minimum at $\bar n = s$. Matching to (\ref{3.19}) produces two independent solutions
\begin{eqnarray}
r_+ &=& n_c + Q,\ \ s_+ = n_c - 2 Q
\nonumber \\
r_- &=& n_c - Q,\ \ s_- =  n_c + 2 Q
\label{3.26}
\end{eqnarray}
in which $n_c = -2c/3$ as before and
\begin{equation}
Q = \frac{1}{3} \sqrt{c^2 - 3\alpha},\ \ \alpha \equiv \frac{a}{b}.
\label{3.27}
\end{equation}
These correspond, respectively, to disappearance of the right and left hand local (and higher) free energy minimum. A convenient choice here is to view, for given fixed $c$, the ratio $\alpha = a/b \leq c^2/3$ as the tunable parameter moving one along the spinodal lines. The parameter values $b(c,\alpha)$ and $a(c,\alpha) = \alpha b(c,\alpha)$ along the two lines follow in the form
\begin{equation}
b_\pm(c,\alpha) = \frac{3}{(n_c \pm Q)[\alpha + c(c \pm 4Q)] \pm (\alpha + c^2)Q}.
\label{3.28}
\end{equation}
They join at the critical point $\alpha = c^2/3$, $Q = 0$, reproducing (\ref{3.24}),

Previous comparisons to experiment have focused on the above model, under the assumption that it is effectively obeyed by the small subset of slow moving atoms, $v \alt 1$ m/s (Doppler shifts $\alt 2$ MHz) \cite{CRWAW2013}. This approximation will be placed in proper context in the full finite $T$ analysis to follow. In particular, fits of the interaction strength $V_\mathrm{eff}$ to data based on comparisons to solutions of (\ref{3.16}) yield $O(10^2)$ times smaller values than implied by the full theory. Thus, interactions in the true thermal vapor are far stronger than the naive estimate, but stabilization of the bistatic phase is correspondingly diluted by the atom motion.

\begin{figure*}
\includegraphics[width=6.5in,viewport = 0 0 920 430,clip]{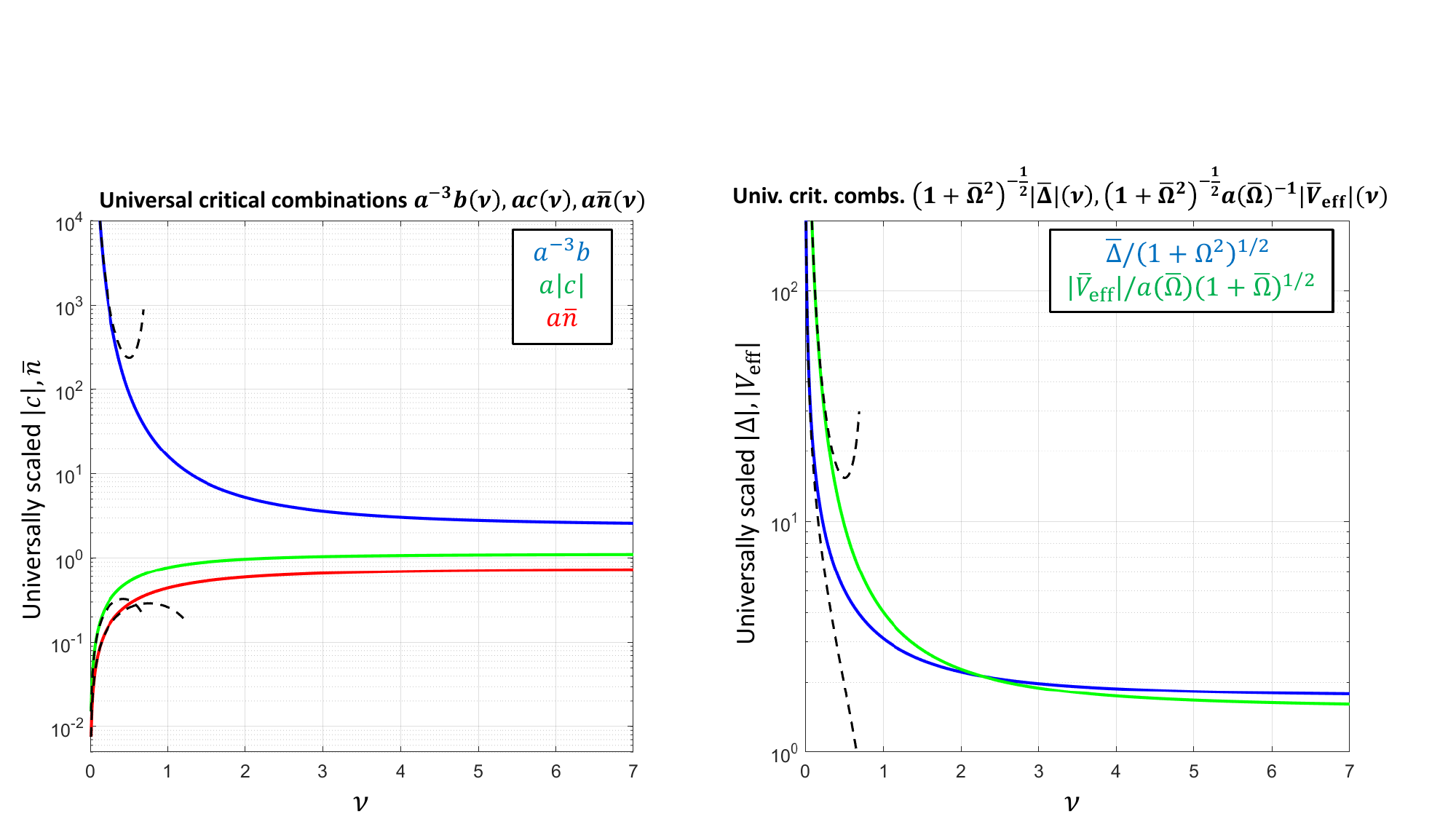}

\caption{Results for universally scaled critical line parameters, plotted on semi-log scale to better expose their behavior and relative magnitudes. \textbf{Left:} Critical values $b/a^3 = 1/A_3(\nu)$, $ac = C(\nu)$, and $a\bar n = \Phi_c(\nu)$ obtained from (\ref{4.1}).  \textbf{Right:} Magnitudes of the critical line universally scaled laser detuning $\Delta_U = \bar \Delta/\sqrt{1 + \bar\Omega^2} = -\mathrm{sgn}(V_\mathrm{eff}) C(\nu)/\sqrt{A_3(\nu)}$ and interaction parameter $V_U = \bar V_\mathrm{eff}/a(\bar \Omega)\sqrt{1 + \bar\Omega^2} = \mathrm{sgn}(V_\mathrm{eff})/\sqrt{A_3(\nu)}$ obtained from (\ref{4.8}) and defined as well in (\ref{B10}) ($\bar \Omega$ is written as $\Omega_\mathrm{sc}$ in the plot labels). In both panels, the black dashed lines show the small $\nu$ asymptotic results derived in (\ref{B8})--(\ref{B10}).}

\label{fig:clinephys}
\end{figure*}

\subsection{Finite temperature critical line}
\label{sec:finiteTcritline}

We next present exact solutions of the full thermal equation (\ref{3.11}), deriving first the critical line. Critical points are defined by simultaneous enforcement, along with (\ref{3.11}), of
\begin{eqnarray}
1 &=& \frac{1}{a} \partial_{\bar n} \Phi_\nu(u_{\bar n})
= \frac{V_\mathrm{th}}{a} \Phi_\nu'(u_{\bar n})
\nonumber \\
0 &=& \Phi_\nu''(u_{\bar n}).
\label{3.29}
\end{eqnarray}
The first condition ensures existence of at least one root while the second guarantees a single root, with local cubic structure of $\Phi_\nu(w)$ in the neighborhood of the root, analogous to the condition $d=0$ in the $T=0$ solution (\ref{3.23}) and (\ref{3.24}). In Fig.\ \ref{fig:PhiPhipPhipp}, 2D color plots of $\Phi_\nu(w)$ and its first two derivatives are shown. The red lines correspond to contours $w_c(\nu)$ defined by
\begin{eqnarray}
&&\Phi_\nu''[w_c(\nu)] = 0
\label{3.30} \\
&&\Leftrightarrow\ w_c = \pm \sqrt{1 + \nu^2 - \nu
\frac{1 + 2 w_c \mathrm{Im}[f(w_c + i\nu)]}{\mathrm{Re}[f(w_c + i\nu)]}}.
\nonumber
\end{eqnarray}
Since $\Phi_\nu(w) > 0$ is an even function there are two symmetrically related solutions $\pm |w_c(\nu)|$. Here the second equation is obtained by reorganizing the terms in the second line of the derivative identities (\ref{B2}), and forms a convenient basis for a numerical iterative solution: at each step of the iteration (starting, e.g., with the $\nu = 0$ solutions $w_c = \pm 1$), the approximation $w_{c,n}$ at step $n$ is substituted into the right hand side to obtain the update $w_{c,n+1}$. These are the solutions shown in the left panel of Fig.\ \ref{fig:critlinedat}.

Simultaneously enforcing the first of conditions (\ref{3.30}) leads to the requirement
\begin{eqnarray}
\Phi'_c(\nu) &\equiv& \Phi_\nu'[w_c(\nu)]
\nonumber \\
&=& \frac{a}{V_\mathrm{th}} = \mathrm{sgn}(V_\mathrm{th})
\frac{a^{3/2}}{\nu \sqrt{b}}.
\label{3.31}
\end{eqnarray}
in which the definition (\ref{3.10}) of $\nu$ has been used. In particular, since $\Phi_\nu(w) > 0$ is even and decreasing with increasing $|w|$, one has $\mathrm{sgn}[\Phi_\nu'(w)] = -\mathrm{sgn}(w)$ and it follows that $\mathrm{sgn}[w_c(\nu)] = -\mathrm{sgn}(V_\mathrm{th}) = \mathrm{sgn}(V_\mathrm{eff})$, and the selected solution therefore obeys $V_\mathrm{th} w_c(\nu) < 0$. Squaring both sides, one obtains the condition
\begin{equation}
\frac{a^3}{b} = \nu^2 \Phi'_c(\nu)^2 \equiv A_3(\nu)
\label{3.32}
\end{equation}
Any choice of $a, b$ constrained by this ratio corresponds to a valid critical point so long as one also enforces the final mean field condition (\ref{3.11}), now written in the form
\begin{eqnarray}
\Phi_c(\nu) &\equiv& \Phi_\nu[w_c(\nu)]
\nonumber \\
&=& a \left(\frac{w_c(\nu)}{V_\mathrm{th}} - c \right)
\nonumber \\
&=& -\frac{\sqrt{A_3(\nu)} |w_c(\nu)|}{\nu} - a c,
\label{3.33}
\end{eqnarray}
where we note again that $\mathrm{sgn}(V_\mathrm{th}) w_c(\nu) = -|w_c(\nu)|$. Since $c$ remains fully unconstrained, one may trivially enforce this equation by choosing
\begin{eqnarray}
c(a,\nu) &=& \frac{C(\nu)}{a}
\nonumber \\
C(\nu) &=& -\frac{\sqrt{A_3(\nu)} |w_c(\nu)|}{\nu} - \Phi_c(\nu)
\nonumber \\
&=& -[|\Phi_c'(\nu)||w_c(\nu)| + \Phi_c(\nu)].
\label{3.34}
\end{eqnarray}
One sees that the combination $ac$ on the critical line is uniquely determined by $\nu$. The right panel of Fig.\ \ref{fig:critlinedat} shows numerical results for $\Phi_c(\nu)$, $\Phi_c'(\nu)$ and $A_3(\nu)$. The small $\nu$ asymptotic forms for these quantities, as well as for $w_c(\nu)$, are derived in App.\ \ref{app:1ph1RyhighTasymp} and shown as well.

The parameters $a,c,\nu$ are all directly controlled by the laser and atomic vapor parameters, which may then be adjusted to find the critical line. Thus, one sees from (\ref{3.17}) that $c$ is directly controllable through the laser detuning $\Delta$. Similarly, the value of $a$ defined in (\ref{3.10}) is directly specified by known (or at least independently measurable) atomic decay parameters $\Gamma,K$ and the also known laser Rabi frequency $\Omega$ (governed by the laser intensity and the transition dipole moment). One may similarly express
\begin{equation}
\nu = |V_\mathrm{th}| \sqrt{\frac{a}{b}}
= \sqrt{\frac{\Gamma + K}{\Gamma}}
\frac{|\Omega|\sqrt{a(\Omega)}}{2 k_P v_\mathrm{th}},
\label{3.35}
\end{equation}
which happily does not actually depend on the very poorly known interaction strength $V_\mathrm{eff}$, but does depend on temperature through $v_\mathrm{th}$.

\begin{figure*}

\includegraphics[width=3.3in,viewport = 0 0 550 540,clip]{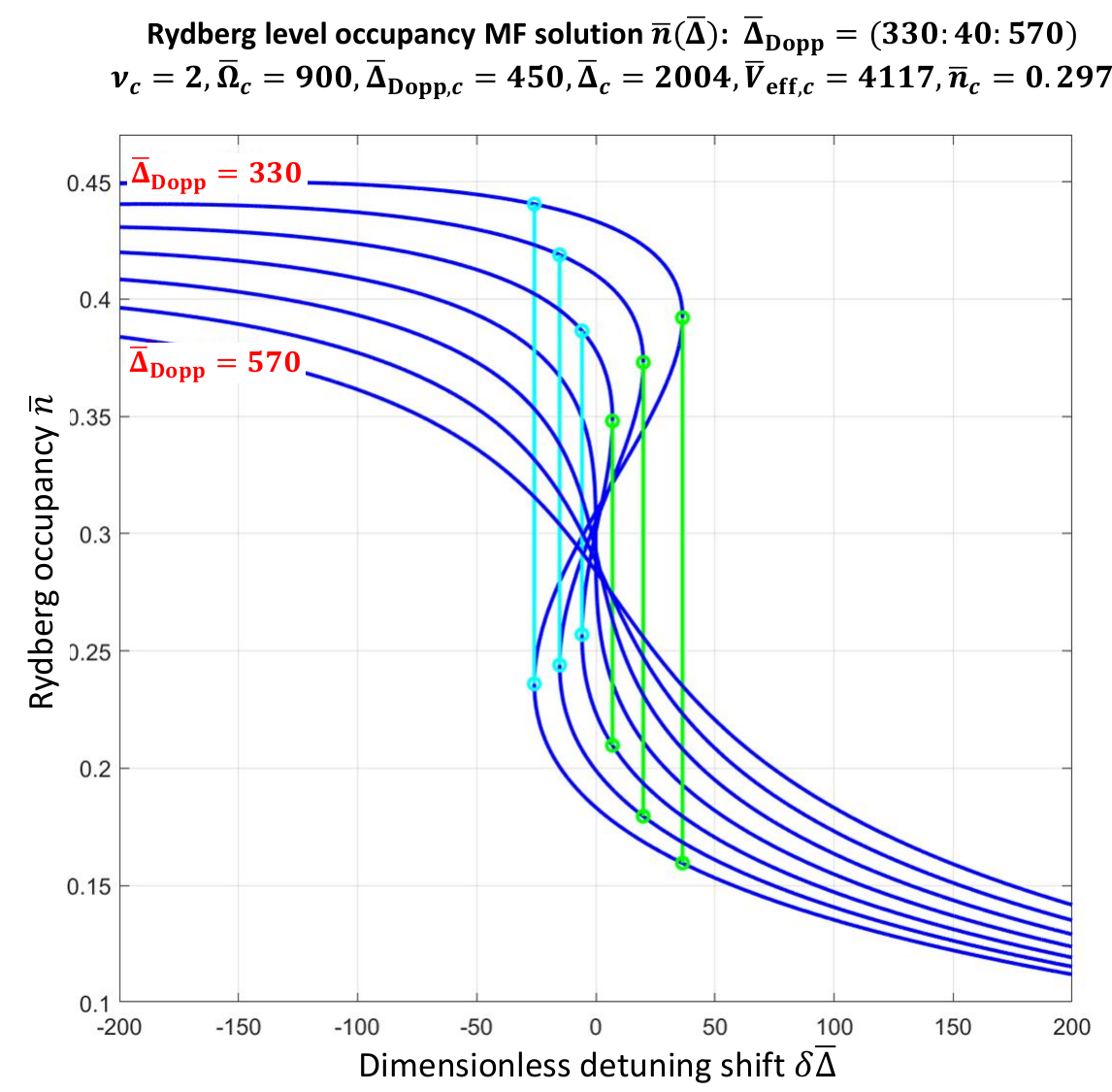}
\quad
\includegraphics[width=3.3in,viewport = 0 0 550 540,clip]{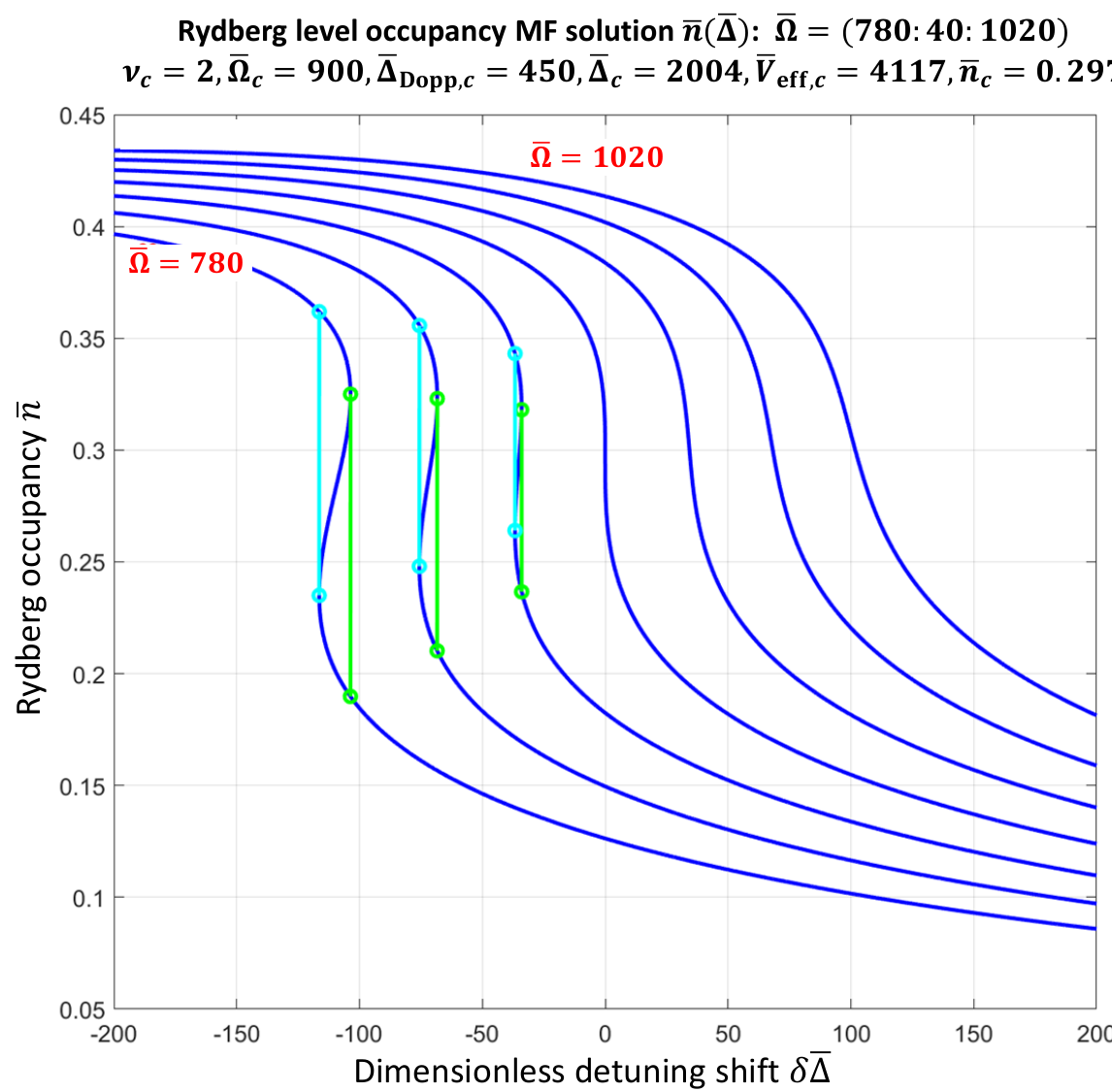}

\caption{Example mean field equation solutions in the vicinity of the critical point defined by scaled parameter values $\nu_c = 2$, $\bar \Omega_c = 900$, and $\bar \Delta_{\mathrm{Dopp},c} = 450$ which serve, as described in Sec.\ \ref{sec:critscale}, to define the scaled critical values $\bar \Delta_c \simeq 2004, \bar V_{\mathrm{eff},c} \simeq 4117$ (scaled parameters are indicated by `sc' instead of overbars in the plot labels). \textbf{Left:} Solutions for different scaled Doppler detuning values of $-120 \leq \bar \Delta_\mathrm{Dopp} - \bar \Delta_{\mathrm{Dopp},c} \leq 120$ in increments of 40 (from top curve to bottom curve on the left, as indicated by the labels). The parameters $\bar V_\mathrm{eff} = \bar V_{\mathrm{eff},c}$, $\bar \Omega = \bar \Omega_c$ are fixed at the previously defined critical values, while $\nu(\bar \Omega_c,\bar \Delta_\mathrm{Dopp})$ is updated according to the second line of (\ref{3.35}). It is seen that both the real and imaginary parts of the complex argument $\zeta = u_{\bar n} + i\nu$ scale as $1/\bar \Delta_\mathrm{Dopp}$. As described in Sec.\ \ref{subsec:scandetune}, the hysteretic behavior in the bistable phase, $\bar \Delta_\mathrm{Dopp} < \bar \Delta_{\mathrm{Dopp},c}$, is indicated by the vertical lines. Thus, scanning upwards from small $\delta \bar \Delta$, the system will discontinuously jump (green line) from the higher to the lower $\bar n$ solution at the higher spinodal point $\delta \bar\Delta_1^\mathrm{sp}(\bar \Delta_\mathrm{Dopp})$ where the upper solution disappears. Conversely, the system follows the cyan line when scanning downwards from higher values of $\delta \bar \Delta$, discontinuously jumping at the lower spinodal point $\delta \bar\Delta_2^\mathrm{sp}(\bar \Delta_\mathrm{Dopp})$ where the lower solution disappears. \textbf{Right:} Each curve now corresponds to a different dimensionless Rabi frequency value $-120 \leq \bar \Omega - \bar \Omega_c \leq 120$ in increments of 40 (from bottom curve to top curve on the left, as indicated by the labels), at fixed $\bar \Delta_\mathrm{Dopp} = \bar \Delta^c_\mathrm{Dopp}$. The bistable phase appears for $\bar \Omega < \bar \Omega_c$. Both $a(\bar \Omega)$ and $\bar \nu(\Omega,\bar \Delta^c_\mathrm{Dopp})$ are updated according to the first two lines of (\ref{4.8}).}

\label{fig:nroots}
\end{figure*}

\section{Critical scaling, bistable phase, and contact with experiment}
\label{sec:critscale}

The various dimensionless scaling forms derived in the previous section provide a convenient formulation for exploration of the bistable phase. They also allow proper contact with experiment through optimal fitting of theoretical free parameters to data, at the same time quantifying any limitations of the mean field approximation.

\subsection{Mean field predictions for critical line scaling relations}
\label{sec:MFcritlinescaling}

We assume that the intrinsic atomic dissipation parameters $\Gamma$ and $K$ are known from prior experiments, independent of bistable phase physics, as are the transition dipole moments that allow one, via (\ref{2.4}), to derive the Rabi frequency $\Omega$ from the laser intensity. The value of $\Omega$ then fixes $a = a(\Omega)$ via the first of (\ref{3.10}). The combination of $\Omega$ and cell temperature $T$ determine the ratio $\Omega_\mathrm{th}$ defined in (\ref{3.7}), and the value $\nu(\Omega,v_\mathrm{th})$ then follows from the second line of (\ref{3.10}). This value of $\nu$ is substituted into the mean field equation solution to determine all of the critical line parameters in (\ref{3.30})--(\ref{3.34}). The critical values
\begin{eqnarray}
b &=& \frac{4\Gamma}{\Gamma + K} \frac{V_\mathrm{eff}^2}{\Omega^2}
= \frac{a(\Omega)^3}{A_3(\nu)}
\nonumber \\
c &=& -\frac{\Delta}{V_\mathrm{eff}}
= \frac{C(\nu)}{a(\Omega)}
\nonumber \\
n_c &=& \frac{\Phi_c(\nu)}{a(\Omega)}
\label{4.1}
\end{eqnarray}
then follow. Although both $b$ and $c$ depend on the typically unknown fitting parameter $V_\mathrm{eff}$, the combination
\begin{eqnarray}
c \sqrt{b} &=& \mathrm{sgn}(c)
\sqrt{\frac{2\Gamma}{\Gamma + K}} \frac{|\Delta|}{|\Omega|}
\nonumber \\
&=& \frac{C(\nu)}{\sqrt{A_3(\nu)}} \sqrt{a(\Omega)}
\label{4.2}
\end{eqnarray}
is accurately known and hence fixes the critical point ratio $|\Delta|/|\Omega|$. Consistency of this scaling relation between experimentally derived values of $\Omega, \Delta, \nu$ at the critical point provides a first test of the accuracy of the mean field approximation.

Substituting the resulting critical value $|\Delta(\Omega)|$ (for the given value $\Omega$ chosen above) into the second of (\ref{4.1}) one obtains the desired fitted value
\begin{eqnarray}
V_\mathrm{eff} &=& -\frac{\Delta(\Omega)}{c(\Omega,\nu)}
\nonumber \\
&=& -\mathrm{sgn}(V_\mathrm{eff}) |\Omega|
\sqrt{\frac{\Gamma + K}{2\Gamma}} \sqrt{\frac{a(\Omega)^3}{A_3(\nu)}}.
\label{4.3}
\end{eqnarray}
From the definition (\ref{3.17}) one obtains
\begin{equation}
\mathrm{sgn}(\Delta) = -\mathrm{sgn}(V_\mathrm{eff}) \mathrm{sgn}(c).
\label{4.4}
\end{equation}
in which $\mathrm{sgn}(V_\mathrm{eff})$ is determined by the microscopic interaction physics. From (\ref{3.34}), one sees that the critical value $\mathrm{sgn}(c) = \mathrm{sgn}(C) = -1$ is universal, in particular independent of the sign choice for $w_c(\nu)$, and it follows that the critical laser detuning has sign opposite to that of the effective atomic interaction:
\begin{equation}
\mathrm{sgn}(\Delta) = -\mathrm{sgn}(V_\mathrm{eff}).
\label{4.5}
\end{equation}

Note that these results are fully consistent with the zero temperature theory, $\nu \to \infty$, hence $w(\nu) \to \infty$, where the critical line solution (\ref{3.21}) yields the critical ratio finite limit
\begin{equation}
A_3 = \frac{a^3}{b} = \frac{27}{64}.
\label{4.6}
\end{equation}
If $a$ is considered the control parameter, through $\Omega$ in (\ref{3.10}), the critical values of $b$ and $c$ follow and again produce consistent fits for $V_\mathrm{eff}$ for known values of $\Omega,\Delta$.

\subsection{Universal dissipation scaling combinations}
\label{sec:univdissipscaling}

The results here depend on the choice of dissipation parameters, which may depend strongly on the chosen Rydberg level and on the experimental setup (e.g., $K$ generally depends on laser beam width via transit time broadening effect). A further test of the mean field approximation, and consistency of the estimates for these parameters, is obtained by defining the alternative dimensionless combinations
\begin{eqnarray}
&&\bar \Omega = \frac{\Omega}{\sqrt{\Gamma(\Gamma+K)/2}},\ \
\bar \Delta = \frac{2\Delta}{\Gamma + K}
\nonumber \\
&&\bar V_\mathrm{eff} = \frac{2V_\mathrm{eff}}{\Gamma + K},\ \
\bar \Delta_\mathrm{Dopp} =  \frac{2 k_P v_\mathrm{th}}{\Gamma + K}
\label{4.7}
\end{eqnarray}
From (\ref{3.10}) and (\ref{3.17}) one obtains the fully universal critical line scaling relations
\begin{eqnarray}
a(\bar \Omega) &=& 2 \left(1 + \frac{1}{|\bar \Omega|^2}\right)
\nonumber \\
\nu &=& \sqrt{\frac{|\bar \Omega|^2 a(\bar \Omega)}{2\bar \Delta_\mathrm{Dopp}^2}}
= \frac{\sqrt{1 + |\bar \Omega|^2}}{\bar \Delta_\mathrm{Dopp}}
\nonumber \\
\bar \Delta &=& -\frac{1}{\sqrt{2}} \mathrm{sgn}(V_\mathrm{eff}) c\sqrt{b} |\bar \Omega|
\nonumber \\
&=& -\mathrm{sgn}(V_\mathrm{eff}) \frac{C(\nu)}{\sqrt{A_3(\nu)}}
\sqrt{1 + |\bar \Omega|^2}
\nonumber \\
\bar V_\mathrm{eff} &=& -\frac{\bar \Delta}{c}
= \mathrm{sgn}(V_\mathrm{eff}) \frac{a(\bar \Omega) \sqrt{1 + |\bar \Omega|^2}}{\sqrt{A_3(\nu)}}.
\label{4.8}
\end{eqnarray}
which in principle contain no free parameters. The degree to which these relations agree with experiment (e.g., over some range of temperatures) provides a further test of the theory and/or an opportunity for optimal adjustment of the estimates for $\Gamma, K$.

The critical line properties derived from these quantities are shown in Fig.\ \ref{fig:clinephys}. The left panel shows the scaled quantities $b/a^3$ and $ac$ obtained from (\ref{4.1}), and which depend only on $\nu$. The universally scaled versions $\Delta_U = \bar \Delta/\sqrt{1+\bar\Omega^2}$, $V_U = \bar V_\mathrm{eff}/a\sqrt{1+\bar\Omega^2}$, obtained from last two lines of (\ref{3.34}), exhibited as well in (\ref{B10}), and depending only on $\nu$, are plotted in the right panel of Fig.\ \ref{fig:clinephys}. The sign choice $\mathrm{sgn}(\bar \Delta) = \mathrm{sgn}(\Delta)$ again follows from (\ref{4.5}). The small $\nu$ asymptotic forms for these are also derived in App.\ \ref{app:1ph1RyhighTasymp}.

\subsection{Bistable phase, hysteresis behavior, and sensor operation}
\label{sec:biphasehystsensor}

We next move beyond the critical point to the broader bistable phase diagram, formulated in terms of the universal combinations (\ref{4.7}), and use the results to characterize the sensitivity of experimental electric field measurements. We begin with investigation of the variation of the Rydberg level population $\bar n$ along various parameter trajectories, including through the critical point, seeking to maximize the variation rate. However, direct measurement of $\bar n$ is difficult (e.g., through an ionization measurement, applying a microwave field tuned to the ionization threshold of the Rydberg state \cite{CRWAW2013}). Indirect EIT measurement (see Fig.\ \ref{fig:atomsetup} and Sec.\ \ref{sec:probeEIT}) is more convenient. It is a quantity that can be measured extremely accurately, using standard photon counter technology, and has zero effect on the Rydberg state preparation process since the measurement is based on ``unused'' light that has exited the vapor. As described in Sec.\ \ref{sec:probeEITsigsense}, the probe laser absorption measurement is predicted to show similar sharp features as one scans through the phase diagram. We consider here examples in which the laser detuning $\Delta$ is scanned and explore possible settings of the remaining parameters in order to optimize sensor performance.

\subsubsection{Scaled mean field equation}
\label{subsec:scaledmfeq}

Using the scaled variables (\ref{4.7}) the mean field equation (\ref{3.11}) may be written in the form
\begin{equation}
\Phi_\nu(u_{\bar n}) = \beta u_{\bar n} + \gamma
\label{4.9}
\end{equation}
with coefficients
\begin{eqnarray}
\beta &=& \frac{a(\bar \Omega)}{V_\mathrm{th}}
= -\frac{a(\bar \Omega) \bar \Delta_\mathrm{Dopp}}{\bar V_\mathrm{eff}}
\nonumber \\
\gamma &=& -\frac{a(\bar \Omega) \Delta_\mathrm{th}}{V_\mathrm{th}}
= \frac{a(\bar \Omega) \bar \Delta}{\bar V_\mathrm{eff}}.
\label{4.10}
\end{eqnarray}
Solving (\ref{4.9}) produces a three-parameter family of solutions $u_{\bar n} = U(\beta,\gamma,\nu)$, with physical Rydberg state occupation number
\begin{equation}
\bar n = \frac{\beta U(\beta,\gamma,\nu) + \gamma}{a(\bar \Omega)}
\label{4.11}
\end{equation}
obtained by inverting (\ref{3.9}). From (\ref{3.30}), (\ref{3.31}) and the second line of (\ref{4.1}), for given $\nu$ the critical point values are, respectively,
\begin{equation}
u_{\bar n, c} = U_c = w_c(\nu),\ \ \beta_c = \Phi_c'(\nu),\ \ \gamma_c = -C(\nu).
\label{4.12}
\end{equation}
These quantities are included in Figs.\ \ref{fig:critlinedat} and \ref{fig:clinephys}.

\subsubsection{Detuning scans through the critical region}
\label{subsec:scandetune}

In Fig.\ \ref{fig:nroots} are shown example results for detuning scans in the neighborhood of the critical point defined by $\nu_c = 2$, $\bar \Omega_c = 900$, and $\bar \Delta_{\mathrm{Dopp},c} = 450$ which leads via the above relations to $\bar \Delta_c = 2004$, $\bar V_{\mathrm{eff},c} = 4117$, and $\bar n_c = 0.297$.

The left panel of the figure shows mean field solutions $\bar n(\bar \Delta_c + \delta \bar \Delta)$, derived from (\ref{4.10}) and (\ref{4.15}), for a sequence of $\bar \Delta_\mathrm{Dopp}$ values at fixed $\bar \Omega = \bar \Omega_c$. Note that, via the second line of (\ref{4.8}), $\nu$ also takes a different value on each curve. The $\bar \Delta_\mathrm{Dopp} = \bar \Delta_{\mathrm{Dopp},c}$ curve passes through the critical point at $\delta \bar \Delta = 0$, but otherwise lies outside the bistatic phase. The contour shows a clear $\delta \bar n = n - n_c \sim \delta \bar \Delta^{1/3}$ ``critical isotherm,'' and the extremely large sensitivity of $\bar n$ to small changes in $\bar \Delta$ is a potential basis for sensor operation. For this to work, the signal perturbation must be coupled to a change in $\bar \Delta$. For example, as described in Sec.\ \ref{sec:probeEIT}, an external RF electric field will, via the Stark shift, perturb the energy level difference between the ground and Rydberg states. This shift will appear as a detuning change even as the laser frequency remains fixed.

For $\bar \Delta_\mathrm{Dopp} > \bar \Delta^c_\mathrm{Dopp}$ the solution remains unique and there is no singularity, hence it lies entirely outside the bistatic phase. This is consistent with one's intuition that larger $\bar \Delta_\mathrm{Dopp}$, hence higher temperature, corresponds to larger Doppler spread, hence reduced atomic coherence needed to support the bistatic phase. On the other hand, for $\bar \Delta_\mathrm{Dopp} < \bar \Delta^c_\mathrm{Dopp}$ there is an interval $\delta \bar\Delta_1^\mathrm{sp}(\bar \Delta_\mathrm{Dopp}) \leq \delta \bar \Delta \leq \delta \bar\Delta_2^\mathrm{sp}(\bar \Delta_\mathrm{Dopp})$ on which there are multiple (three) solutions for $\bar n$. The middle solution is unstable, while the outer pair constitute the two locally stable bistatic phase solutions.

The boundaries $\delta \bar \Delta^\mathrm{sp}_{1,2}(\bar \Delta_\mathrm{Dopp})$ of this interval are the spinodal points beyond which the solution again becomes unique. If one scans upwards from smaller $\delta \bar \Delta$ one will follow the upper part of the curve and eventually encounter $\delta \bar \Delta_2^\mathrm{sp}(\bar \Delta_\mathrm{Dopp})$, beyond which the system state jumps discontinuously to the lower part of the curve (green vertical lines in the plot). If one scans in the opposite direction, one will follow the lower part of the curve until one encounters the discontinuous jump at $\delta \bar\Delta_1^\mathrm{sp}(\bar \Delta_\mathrm{Dopp})$ to the upper part of the curve (cyan lines). This is classic hysteretic behavior. Since the behavior is not reversible, such a setting may require regular resets, hence not be optimal for continuous sensor operation. It may be preferable to exploit the critical or near critical curve $\bar \Delta_\mathrm{Dopp} \geq \bar \Delta^c_\mathrm{Dopp}$.

The right panel of the figure similarly shows solutions $\bar n(\bar \Delta_c + \delta \bar \Delta)$ for varying Rabi frequency $\bar \Omega$ (hence laser amplitude) at fixed $\bar \Delta_\mathrm{Dopp} = \bar \Delta^c_\mathrm{Dopp}$. The critical contour $\bar \Omega = \bar \Omega_c$, $\bar \Delta_\mathrm{Dopp} = \bar \Delta_{\mathrm{Dopp},c}$ is the same in both panels. The value of $\bar \Omega$ enters the mean field calculation through $a(\bar \Omega)$ and $\nu(\bar \Omega,\bar \Delta_\mathrm{Dopp})$. As seen, the bistable phase appears for $\bar \Omega < \bar \Omega_c$.

\subsection{Probe beam EIT response and signal detection sensitivity}
\label{sec:probeEITsigsense}

We return now to the EIT effect described in Sec.\ \ref{sec:probeEIT}, exhibiting explicitly the form of the matrix element $\rho^{01}$ governing the probe beam measured intensity.

\subsubsection{Off-diagonal density matrix elements and EIT response}
\label{subsec:DMoffdiagEITresp}

Using the spin representation (\ref{3.3}) and the steady state solution (\ref{3.5}) one obtains the thermal averages
\begin{eqnarray}
\bar S^y &=& \frac{2\Gamma}{|\Omega|} \bar n
\nonumber \\
\bar S^x &=& -\frac{4\Gamma}{(\Gamma + K) |\Omega|} \langle (\Delta + V) n \rangle_\mathrm{th}
\nonumber \\
&=& \sqrt{\frac{4\Gamma}{(\Gamma + K) a(\Omega)}} \Psi_\nu(u_{\bar n})
\label{4.13}
\end{eqnarray}
in which
\begin{eqnarray}
\Psi_\nu(w) &=& \int \frac{du}{\sqrt{2\pi}} e^{-u^2/2}
\frac{(u - w)/\nu}{1 + (u - w)^2/\nu^2}
\nonumber \\
&=& \nu \mathrm{Re} [f(w + i\nu)]
\label{4.14}
\end{eqnarray}
with additional parameters defined by as usual by (\ref{3.7})--(\ref{3.10}) and $f(\zeta)$ by the second line of (\ref{3.12}). Combining these, one obtains
\begin{eqnarray}
\frac{\bar \rho^{01}}{\Omega} &=& \frac{\bar S^x - i \bar S^y}{2|\Omega|}
\label{4.15} \\
&=& -i \frac{\Gamma \bar n}{|\Omega|^2}
+ \sqrt{\frac{\Gamma}{(\Gamma + K) |\Omega|^2 a(\Omega)}} \Psi_\nu(u_{\bar n}).
\nonumber
\end{eqnarray}

From the above result one obtains immediately from (\ref{2.36})--(\ref{2.40})
\begin{equation}
R_P(L) = \Gamma \int_0^L ds \frac{\bar n(s)}{|\Omega(s)|^2}
\approx \frac{\Gamma \bar n L}{|\Omega|^2}.
\label{4.16}
\end{equation}
It follows that the decay rate, given by the imaginary part of (\ref{4.15}), is directly proportional to the Rydberg level occupancy, and hence, as desired, will directly reflect the bistable phase properties. The approximate form here treats the vapor as homogeneous, relevant to the linear response regime (small $\Omega$, where $\bar n \propto |\Omega|^2$) or more generally to thin cells. The latter can be seen directly in the limit $|\bar \Omega| \ll 1$ where, from the second line of (\ref{4.8}), $\nu \to 1/\bar \Delta_\mathrm{Dopp}$ is independent of $\bar \Omega$ and $a \approx 2/|\bar \Omega|^2$. The mean field equation (\ref{3.11}) then leads to
\begin{equation}
\frac{2\bar n}{|\bar \Omega|^2}
\approx \Phi_\nu\left(\frac{\bar \Delta
+ \bar n \bar V_\mathrm{eff}}{\bar \Delta_\mathrm{Dopp}} \right)
\to \Phi_\nu\left(\frac{\bar \Delta}{\bar \Delta_\mathrm{Dopp}} \right)
\label{4.17}
\end{equation}
in the limit where $|\bar \Omega|^2 \bar V_\mathrm{eff} \ll 1$ as well. The ratio on the left is therefore independent of $\bar \Omega$ as is then the EIT response exponent $\alpha_P R_P(L)$.

Of course, the bistable phase is absent in this limit since it requires finite critical line value of the combination $\sqrt{b/a^3} \approx \frac{1}{2} |\bar \Omega|^2 \bar V_\mathrm{eff}$ (see the blue curve in the left panel of Fig.\ \ref{fig:clinephys}), hence requiring rapidly growing, $\bar V_\mathrm{eff} \sim 1/|\bar\Omega|^2$, rather than having a physically sensible fixed value. More generally, in the bistable phase, $\Omega$ is playing an additional role, controlling the location in the phase diagram---see the right panel of Fig.\ \ref{fig:nroots}. This behavior clearly lies outside the linear response regime. In a nontrivially thick cell, the position in the phase diagram will also vary significantly with $\bar \Omega(s)$, possibly crossing the phase boundary as the latter decreases along the beam path.

Although the new function $\Psi_\nu$ does not enter the attenuation portion of the optical transmission, it does strongly affect the index of refraction. It therefore plays a substantial role in cavity-enhanced setups \cite{Wang2023} where multiple reflections between mirrors leads to interference effects depending on the phase \begin{equation}
\phi(L) \approx \frac{1}{2}
\mathrm{Re}\left(\frac{\bar\rho^{10}}{\Omega} \right) \alpha_P L
\label{4.18}
\end{equation}
accumulated through a cell of length of $L$.

\begin{figure*}

\includegraphics[width=5.5in,viewport = 0 0 960 460,clip]{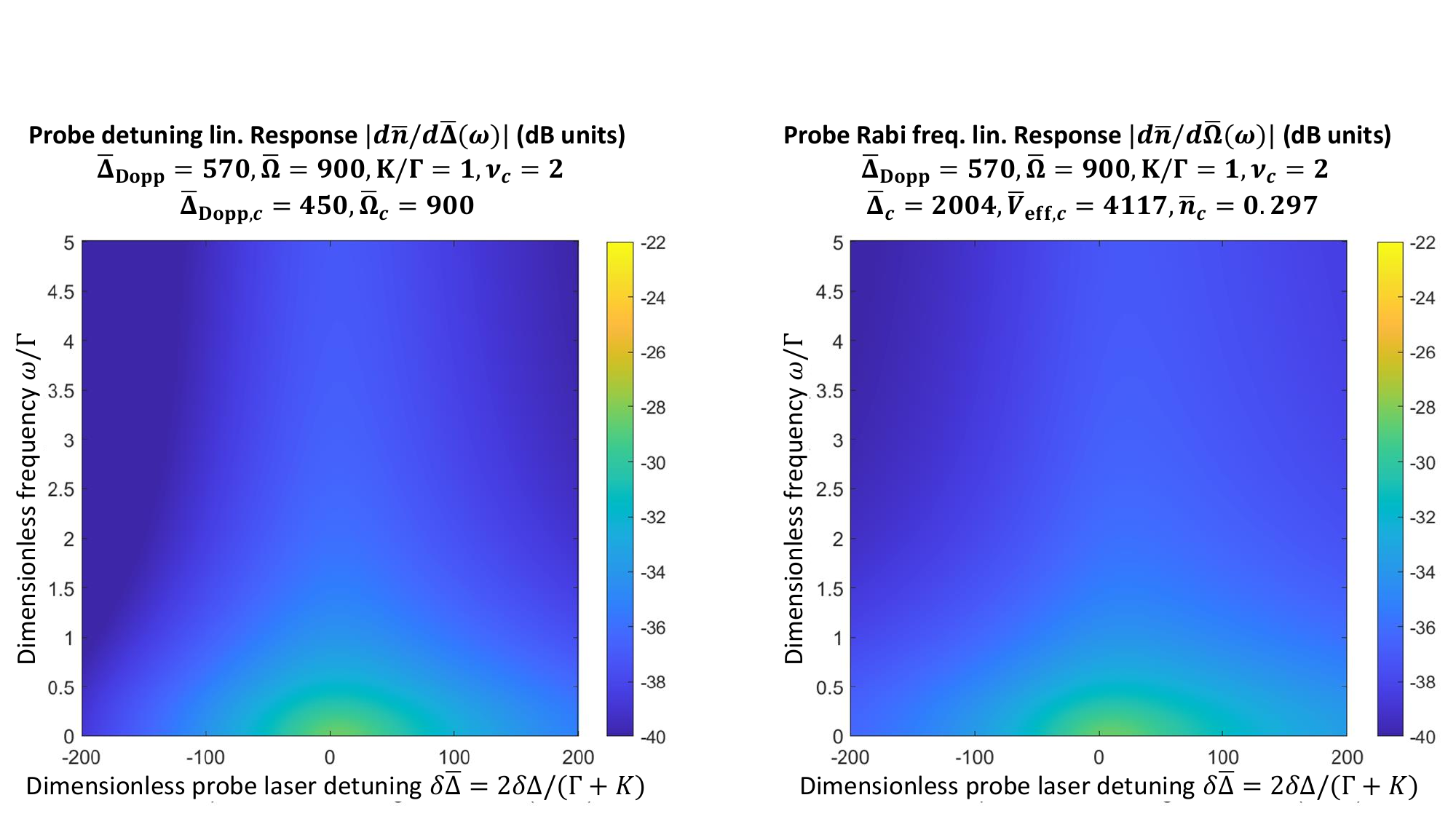}
\includegraphics[width=5.5in,viewport = 0 0 960 460,clip]{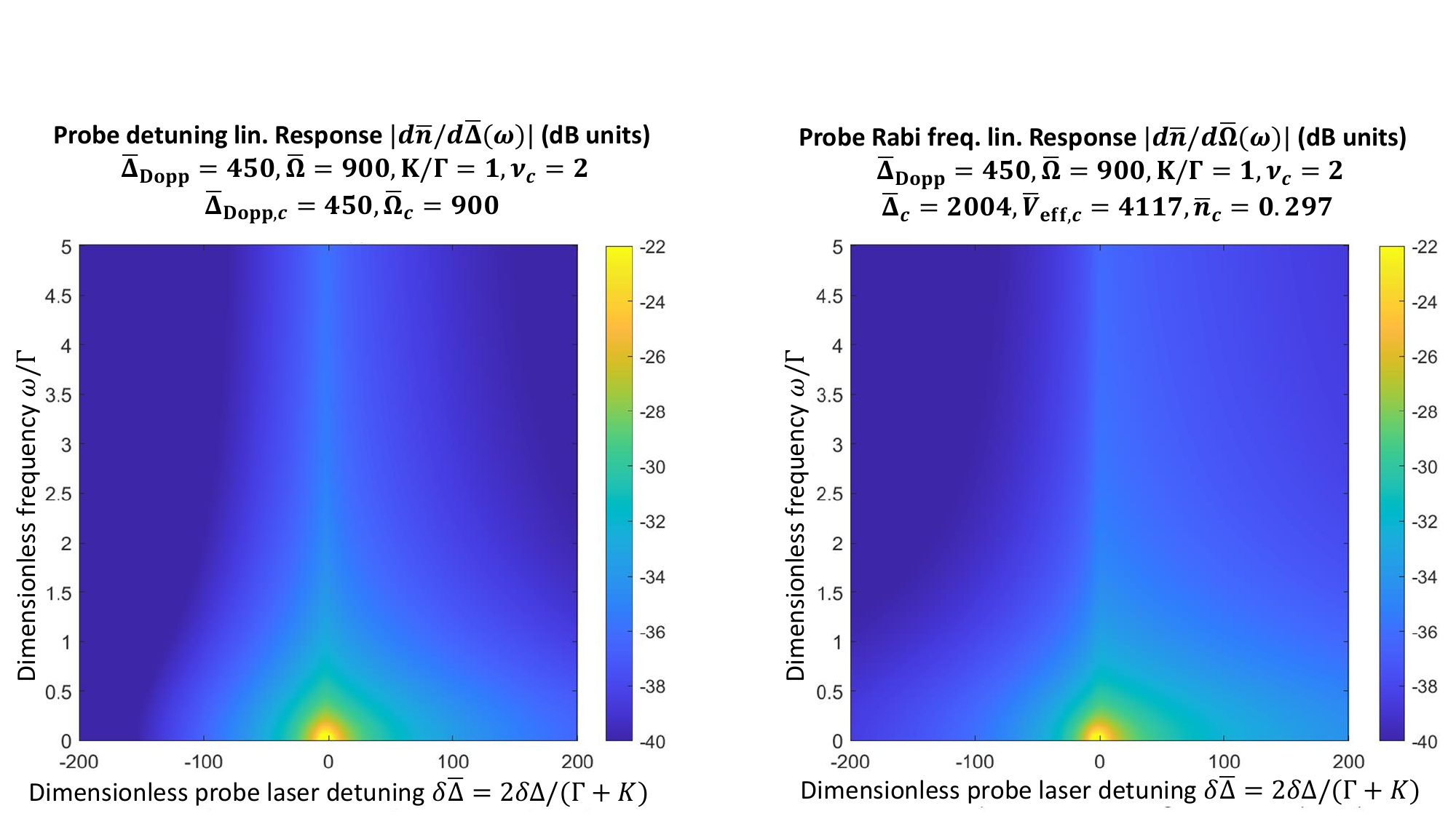}
\includegraphics[width=5.5in,viewport = 0 0 960 460,clip]{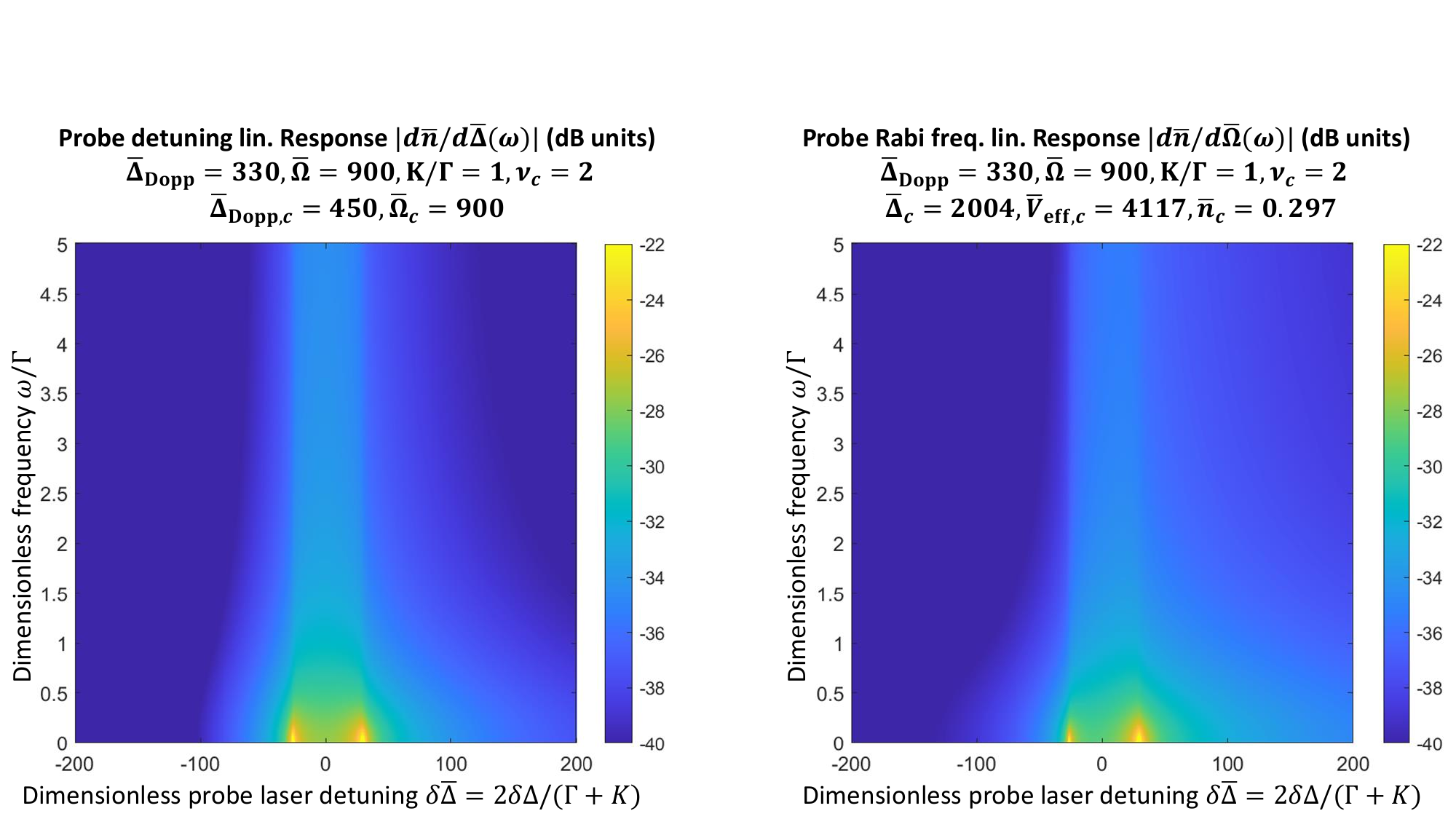}

\caption{Rydberg level occupancy detuning (left column) and Rabi frequency (right column) linear response spectra (\ref{5.1}), in the detuning--frequency plane, along the lines $\bar \Delta_\mathrm{Dopp} = 330, 450, 570$, bottom to top, shown in the left panel of Fig.\ \ref{fig:nroots}. Remaining parameters are as described in that figure. For improved clarity, the response function color scales are the magnitudes in logarithmic dB units $10\log_{10}|\partial \bar n/\partial \bar \Omega|$, $10\log_{10}|\partial \bar n/\partial \bar \Delta|$. The center row, $\bar \Delta_\mathrm{Dopp} = 450$, includes the critical point $\delta \bar \Delta = 0$ where both response functions diverge at zero frequency, consistent with the vertical slope of the corresponding Fig.\ \ref{fig:nroots} curve at that point. The bottom row, $\bar \Delta_\mathrm{Dopp} = 330$, passes through the bistable phase, with divergent zero frequency response at the two spinodal points, also consistent with the vertical slope of the corresponding Fig.\ \ref{fig:nroots} curve.}

\label{fig:linresponse_Dopp}
\end{figure*}

\begin{figure*}

\includegraphics[width=5.75in,viewport = 0 0 960 465,clip]{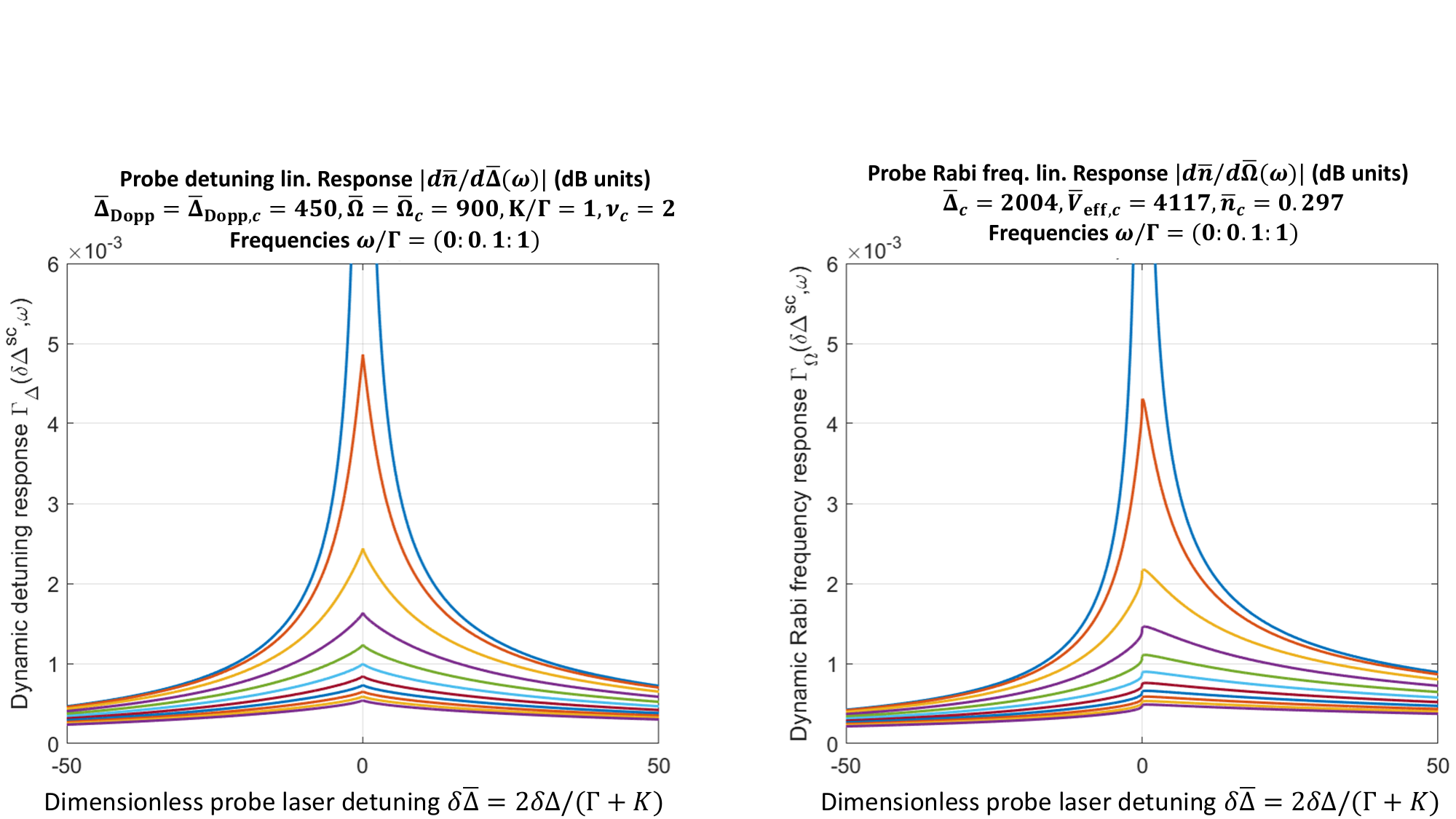}

\caption{Dynamic linear response curves for selected fixed frequency 1D transects, $0 \leq \omega/\Gamma \leq 1$ in steps of 0.1 from top curve to bottom curve, obtained from the middle row of Fig.\ \ref{fig:linresponse_Dopp} (critical line $\bar \Delta = 450, \bar \Omega = 900$ shown in Fig.\ \ref{fig:nroots}). The detuning (left) and Rabi frequency (right) response magnitudes are plotted now on a linear scale. The topmost, zero frequency curve actually diverges at the critical point $\delta \bar \Delta = 0$.}

\label{fig:linresponse_Dopp1D}
\end{figure*}

\subsubsection{Some physical estimates}
\label{subsec:physest}

Example parameter values $\lambda_P = 400$ nm, $|{\bm \mu}_{01}| = 10^{-29}$ C-m, $\rho_a = 10^{11}$ cm$^{-3}$ lead to $\alpha_P = 1.7 \times 10^9$ cm$^{-1}$s$^{-1}$. Taking $|\Omega|/\Gamma = 600$, $\Omega = 2\pi(100\ \mathrm{MHz})$, one obtains
\begin{equation}
\frac{\alpha_P \Gamma}{|\Omega|^2} = 0.9\ \mathrm{cm}^{-1}.
\label{4.19}
\end{equation}
It follows that for perturbations around $\bar n \simeq 0.3$, as seen in Fig.\ \ref{fig:nroots}, cell lengths $L_\mathrm{cell} \sim 3$ cm will produce a strong EIT response. Obviously the true inputs here may vary widely, but this value is within the rather broad 2 mm--10 cm range seen in the literature \cite{CRWAW2013,Ding2022,Wang2023}.

Experiments will typically lie in the high temperature regime, defined here by $\nu \simeq |\bar \Omega|/\bar \Delta_\mathrm{Dopp} \sim |\Omega|/k_P v_\mathrm{th} < 1$. For common setups, $|\Omega|/2\pi = 100$ MHz is considered a large value, with $|\Omega|/2\pi \sim 10$ MHz more typical. Even for the former it follows that for typical experiments $\nu \alt 0.2$. Referring to the green curve in the right panel of Fig.\ \ref{fig:clinephys}, accessing the bistable phase therefore requires rather large values $\bar V_\mathrm{eff}/2\pi \sim 10^2|\Omega|/2\pi \sim 10$ GHz.

The bistable phase has been experimentally observed for various ranges of laser parameters in this temperature regime \cite{CRWAW2013,Ding2022,Wang2023}. The model fits, on the other hand, were made to a zero temperature theory, hence effectively modeling only perhaps the $O(1\%)$ fraction of sufficiently slow moving atoms. The fitted values of $\Gamma/2\pi$, $K/2\pi$ are in the 1 MHz range, but much smaller values of $V_\mathrm{eff}/2\pi < 100$ MHz are found. However, since $V_\mathrm{eff} = \rho_a V_\mathrm{tot}$ scales with total number density [see (\ref{2.28})], one certainly expects a $\sim$\,2 order of magnitude decrease in the critical value of $V_\mathrm{eff}$ if one uses the zero temperature theory with the corresponding smaller number density. On the other hand, the full finite temperature theory presented here requires one to substitute the full number density, consistently leading to the 1--2 order of magnitude larger value of $V_\mathrm{eff}$, while consistently also maintaining a common value of $V_\mathrm{tot}$ derived from the microscopic scattering physics of the Stark shift-induced dipole potential. Note that $\rho_a(T)$ has strong, exponentially activated, temperature dependence as well---much stronger than that contained in $v_\mathrm{th} \propto \sqrt{T}$---so that $V_\mathrm{eff}(T)$ is exponentially activated as well, providing an additional route to increasing the effective interaction strength.

\subsubsection{Signal detection metrics}
\label{subsec:sigdetectmetric}

We now revisit signal detection metrics, defined generally in Sec.\ \ref{sec:statdynlinresp}, using $X = \Delta$ and based on the explicit absorption rate form (\ref{4.16}). For weak RF fields with frequencies much lower than any atomic resonance, the leading frequency shift is given by the DC Stark form \cite{G2005,ARL2020}
\begin{equation}
\Delta_\mathrm{RF}(t) = -\frac{1}{2} \alpha_N |{\bf E}_\mathrm{RF}(t)|^2
\label{4.20}
\end{equation}
with coefficient
\begin{equation}
\alpha_N \approx \frac{2 \mu_N^2}{\Delta \varepsilon_N}
= \frac{(a_B e)^2 N^7}{\hbar \varepsilon_\infty},
\label{4.21}
\end{equation}
where $\mu_N = e a_B N^2$ is the Rydberg level dipole moment (with Bohr radius $a_B$) and $\Delta \varepsilon_N = \varepsilon_\infty/N^2 - \varepsilon_\infty/(N+1)^2 \simeq 2 \varepsilon_\infty/N^3$ is the neighboring energy level difference (with hydrogen atom Rydberg energy $\varepsilon_\infty = e^2/8\pi\epsilon_0 a_B$). This result is based on second order perturbation theory since the first order term $\propto {\bf E}_\mathrm{RF} \cdot \langle 1| e {\bf r} |1 \rangle$ vanishes for states such as these with definite parity. The estimate for $\alpha$ is a crude approximation to the usual second order sum over intermediate states, hence only valid to within a factor of two or so \cite{ARL2020}.

The quadratic dependence (\ref{4.20}) leads to rather poor performance. A substantial improvement is obtained by applying an additional strong DC bias field ${\bf E}_0$ to obtain the linear form
\begin{eqnarray}
\Delta_\mathrm{RF}(t) &=& -\frac{1}{2} \alpha_N |{\bf E}_0 + {\bf E}_\mathrm{RF}(t)|^2
\nonumber \\
&\simeq& -\frac{1}{2} \alpha_N |{\bf E}_0|^2 - \alpha_N {\bf E}_0 \cdot {\bf E}_\mathrm{RF}(t).
\label{4.22}
\end{eqnarray}
Adjusting ${\bf E}_0$ to be parallel to the signal polarization axis is clearly desirable. The frequency-domain EIT linear response (\ref{2.42}) takes the form
\begin{equation}
\delta {\cal P}_\mathrm{EIT}^\mathrm{th}(t)
= - \alpha_N {\bf E}_0 \cdot {\bf E}_\mathrm{RF} e^{-i\omega t}
S_\mathrm{EIT}(\Delta - \textstyle{\frac{1}{2}} \alpha_N |{\bf E}_0|^2,\omega).
\label{4.23}
\end{equation}
The $N^7$ scaling highlights the advantage of working with higher Rydberg levels. Using (\ref{4.16}), the divergence of the slope
\begin{equation}
\frac{\partial R_P}{\partial \Delta}
= \frac{\Gamma L}{|\Omega|^2} \frac{\partial \bar n}{\partial \Delta}
\label{4.24}
\end{equation}
entering the sensitivity result (\ref{2.47}), evident at the critical and hysteresis points in Fig.\ \ref{fig:nroots}, serves to define the bistable phase enhancement.

By contrast, the resonant Rydberg level pair case treated in Sec.\ \ref{sec:2Ryresonant}, and using $X = \Omega_R$, is tuned to operate in the opposite limit, where the RF frequency nearly matches the level spacing. As discussed in the Introduction, this resonant coupling leads to the current records for Rydberg sensor RF signal detection. The proposed additional order of magnitude or more bistability enhancement will be discussed in Sec.\ \ref{sec:1ph2Rydlr}.

\begin{figure*}

\includegraphics[width=5.5in,viewport = 0 0 960 440,clip]{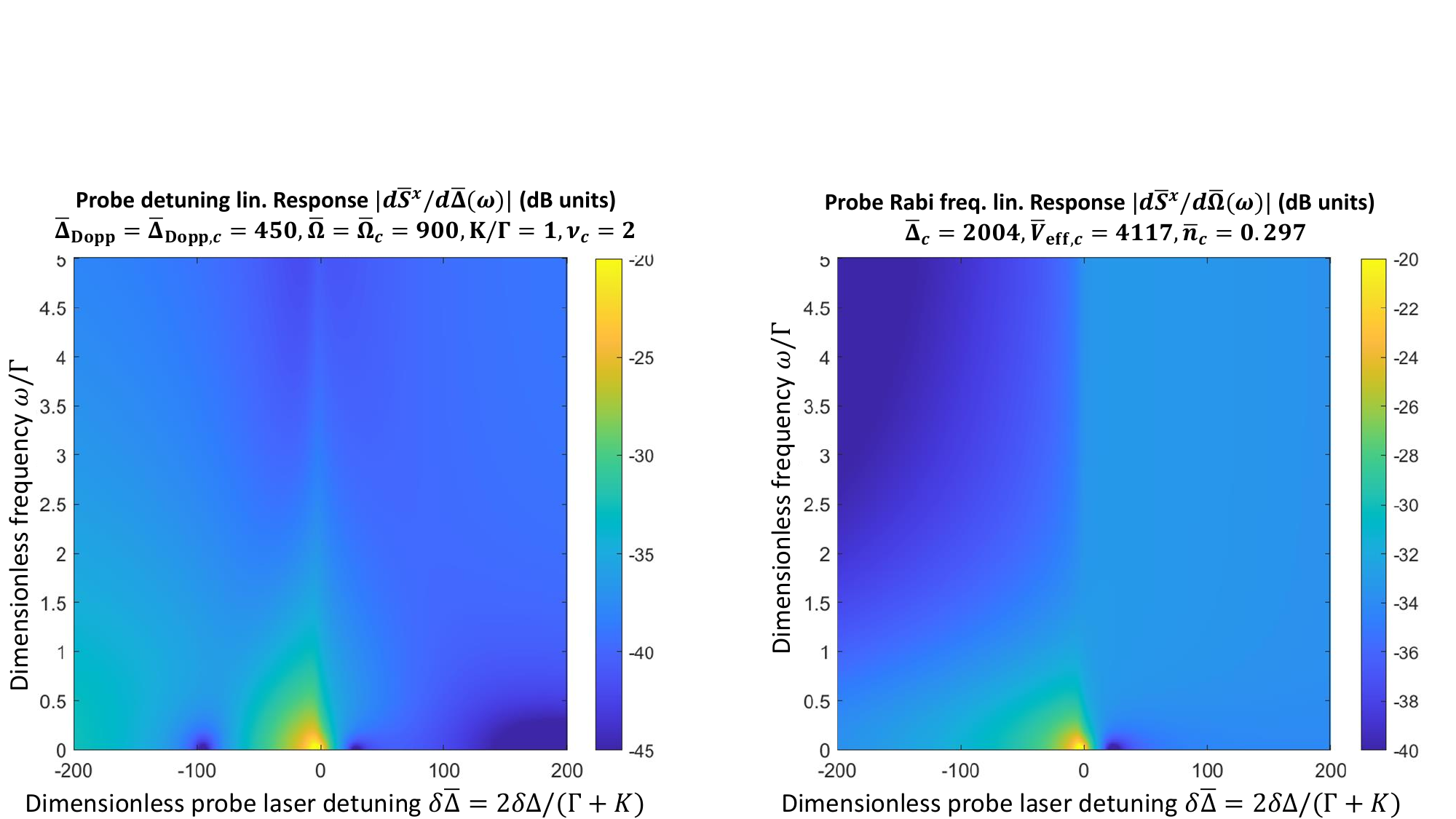}
\includegraphics[width=5.5in,viewport = 0 0 960 440,clip]{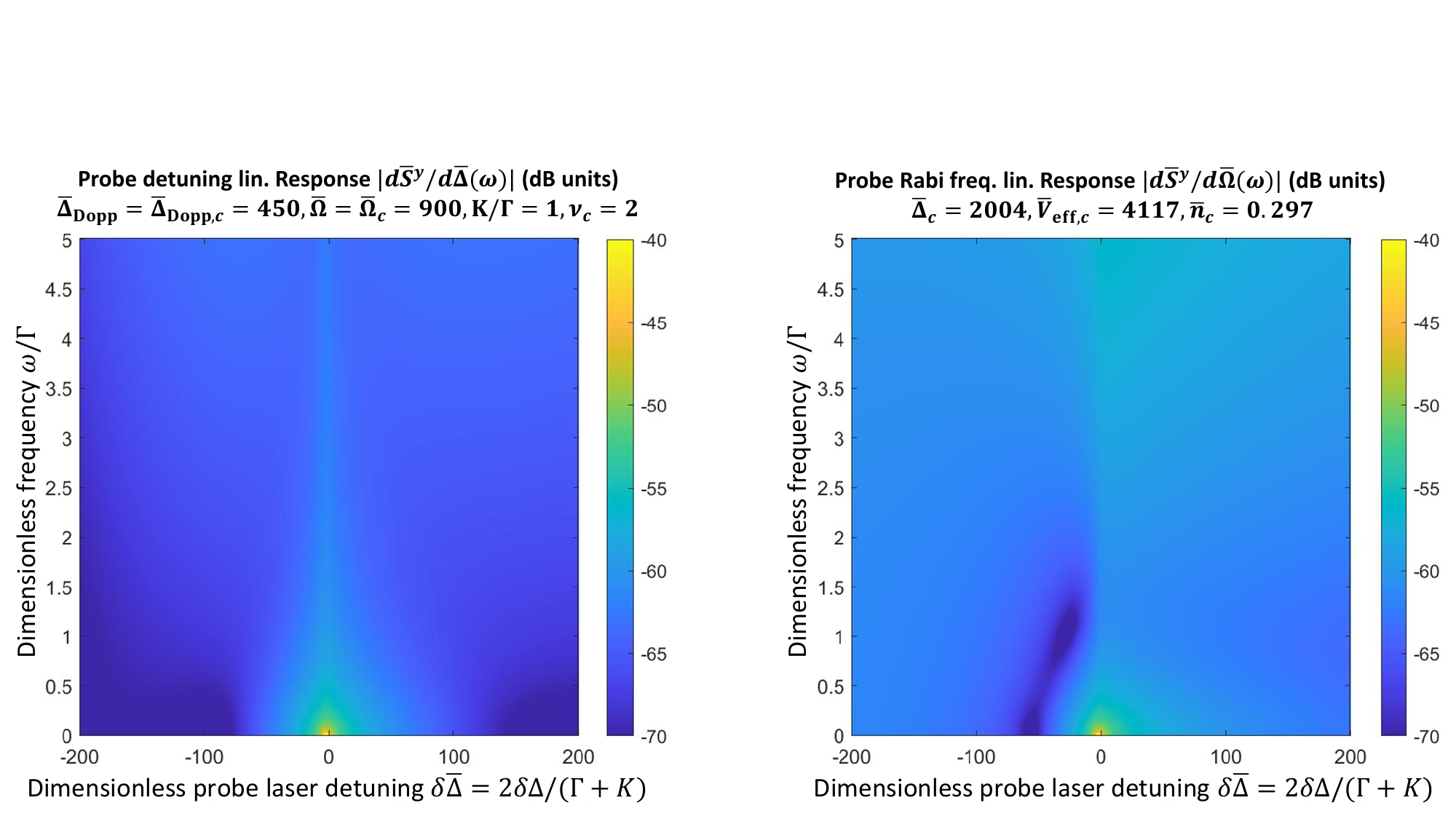}

\caption{Spin components $\bar S^x$ (top row) and $\bar S^y$ (bottom row) detuning (left column) and Rabi frequency (right column) linear response spectra (\ref{5.3}), in the detuning--frequency plane, along the critical line $\bar \Delta = 450, \bar \Omega = 900$ shown in Fig.\ \ref{fig:nroots}. Remaining parameters are as described in that figure. For improved clarity, the response function color scales are the magnitudes in logarithmic dB units $10\log_{10}|\partial \bar S^{x,y}/\partial \bar \Omega|$, $10\log_{10}|\partial \bar S^{x,y}/\partial \bar \Delta|$.}

\label{fig:linresponse_Sxy}
\end{figure*}

\section{Single photon, single Rydberg level dynamic linear response}
\label{sec:dynlinresp1}

We consider now extension of the sensor response to finite frequency, generalizing (\ref{4.24}) by going beyond the stationary limit treated so far, and confirming strong amplification near the critical and hysteresis points. To this end, consider the equations of motion (\ref{3.1}), including the mean field closure approximation (\ref{3.2}) for the interaction parameter, which makes the equations nonlinear. The time dependence induced by an incident external field is assumed to enter through time dependence of one or more of the parameters appearing on the right hand side of (\ref{3.1}):
\begin{eqnarray}
\Delta_k(t) &=& \Delta - k_P v_k + \delta \Delta_k(t)
\nonumber \\
\Omega_k(t) &=& \Omega_k + \delta \Omega_k(t).
\label{5.1}
\end{eqnarray}
Here, $\delta \Delta_k(t)$ reflects the same Stark shift perturbation considered in Sec.\ \ref{subsec:sigdetectmetric} while $\delta \Omega_k(t)$ represents a laser amplitude modulation. Although a full theory would account for nonuniform spatial illumination through the volume of the cell, in the examples below we will specialize to the case of uniform $\delta \Delta_k(t) = \delta \Delta(t)$, $\Omega_k(t) = \Omega + \delta \Omega(t)$ which simplifies the thermal average.

The general dynamic linear response formalism, along with its specialization to the mean field description, is developed in App.\ \ref{app:dynlinresp}. The explicit closed (thermally averaged) forms for the single Rydberg level case presented here are derived in App.\ \ref{app:1ph1Rydynexact}. Specifically, equation (\ref{D16}) defines the frequency domain linear response in terms of convenient scaled variables (\ref{D5}) and (\ref{D13}) based on the spin notation (\ref{3.3}) and (\ref{D4}). The full analytic forms follow by reduction of the frequency-dependent thermal averages (\ref{D14}) to generalizations (\ref{D18})--(\ref{D20}) of the steady state error function form.

As an example, we consider the linear response of the Rydberg level occupancy
\begin{eqnarray}
\frac{\delta \bar n}{\delta \bar \Delta(\omega)}
&=& \frac{1}{2}\frac{\delta \bar S^z}{\delta \Delta(\omega)}
= \frac{1}{2(1 - \frac{2i\omega}{\Gamma + K})} \frac{\delta \tilde S^z}{\delta \tilde \Delta(\omega)}
\nonumber \\
&=& \frac{1}{2(1 - \frac{2i\omega}{\Gamma + K})}
\frac{\tilde f^z_1}{1 - \frac{1}{2} \tilde V_\mathrm{eff} \tilde \Sigma^z}
\nonumber \\
\frac{\delta \bar n}{\delta \bar \Omega(\omega)}
&=& \frac{1}{2 \sqrt{(1 - \frac{2i\omega}{\Gamma + K})(1 - \frac{i\omega}{\Gamma})}}
\frac{\delta \tilde S^z}{\delta \tilde \Omega(\omega)}
\label{5.2} \\
&=& \frac{1}{2 \sqrt{(1 - \frac{2i\omega}{\Gamma + K})(1 - \frac{i\omega}{\Gamma})}}
\frac{\tilde f^z_2}{1 - \frac{1}{2} \tilde V_\mathrm{eff} \tilde \Sigma^z}, \ \ \ \ \ \
\nonumber
\end{eqnarray}
and of the other two spin components
\begin{eqnarray}
\frac{\delta \bar S^{x,y}}{\delta \bar \Delta(\omega)}
&=& \frac{1}{\tilde \gamma(\omega)(1 - \frac{2i\omega}{\Gamma + K})}
\frac{\delta \tilde S^{x,y}}{\delta \tilde \Delta(\omega)}
\nonumber \\
&=& \frac{1}{\tilde \gamma(\omega)(1 - \frac{2i\omega}{\Gamma + K})}
\nonumber \\
&&\times \left(f_1^{x,y}
+ \frac{\frac{1}{2} \tilde V_\mathrm{eff} \tilde \Sigma^{x,y} \tilde f^z_1}
{1 - \frac{1}{2} \tilde V_\mathrm{eff} \tilde \Sigma^z} \right)
\nonumber \\
\frac{\delta \bar S^{x,y}}{\delta \bar \Omega(\omega)}
&=& \frac{1}{\tilde \gamma(\omega)
\sqrt{(1 - \frac{2i\omega}{\Gamma + K})(1 - \frac{i\omega}{\Gamma})}}
\frac{\delta \tilde S^{x,y}}{\delta \tilde \Omega(\omega)}
\nonumber \\
&=& \frac{1}{\tilde \gamma(\omega)
\sqrt{(1 - \frac{2i\omega}{\Gamma + K})(1 - \frac{i\omega}{\Gamma})}}
\nonumber \\
&&\times \left(f^{x,y}_2
+ \frac{\frac{1}{2} \tilde V_\mathrm{eff} \tilde \Sigma^{x,y} \tilde f^z_2}
{1 - \frac{1}{2} \tilde V_\mathrm{eff} \tilde \Sigma^z} \right),
\label{5.3}
\end{eqnarray}
to detuning and Rabi frequency perturbations. These expressions follow from (\ref{D16}), the steady state scaling (\ref{4.7}), and the dynamic scaling of $\delta \Delta, \delta \Omega$ in the first two lines of (\ref{D5}).

Results for these quantities, evaluated along selected contours from the left panel of Fig.\ \ref{fig:nroots}, are shown in Figs.\ \ref{fig:linresponse_Dopp}--\ref{fig:linresponse_Sxy}. Results using the right panel contours (not shown) are similar. The key highlights are the divergences of the responses at the critical point and at the spinodal points in the static limit $\omega \to 0$, consistent with the vertical slopes of the corresponding Fig.\ \ref{fig:nroots} curves at these points. These divergences are rendered finite at finite frequency, but continue to display strong amplification out to $\omega/\Gamma \sim 0.5$.

\begin{figure}

\includegraphics[width = 3.3in,viewport = 0 0 550 520 ,clip]{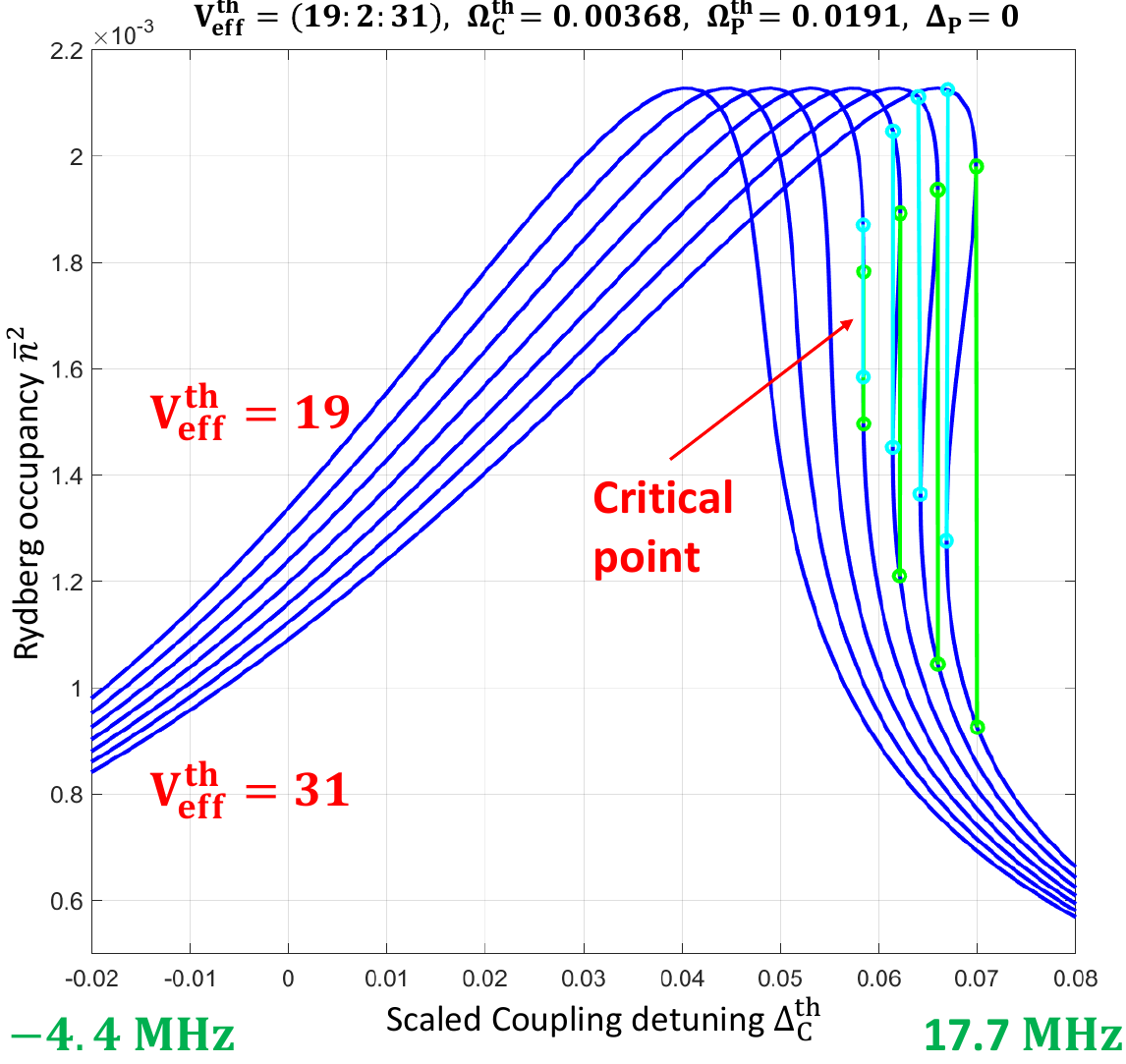}

\caption{Physically motivated (as described in Sec.\ \ref{sec:2phmfsoln} and Table \ref{tab:2ph1RyPhysparms}) example two photon, single Rydberg level mean field solutions. Plotted is the mean Rydberg level occupancy $\bar n^2$ vs.\ Doppler-scaled [see (\ref{6.2})] coupling laser detuning $\Delta_C^\mathrm{th}$ for a sequence of similarly scaled effective interaction values (\ref{6.1}): $V^\mathrm{th}_\mathrm{eff} = 19, 21, 23, \ldots, 31$. Larger magnitude values correspond to increasingly hysteretic behavior (indicated, as in Fig.\ \ref{fig:nroots}, by the vertical lines), with critical value $V^\mathrm{th}_{\mathrm{eff},c} \simeq 24$ intermediate between the third and fourth curves, and whose line passes through the critical point $\Delta_{C,c}^\mathrm{th} \simeq 0.056$, $n_{\mathrm{Ry},c} \simeq 1.7 \times 10^{-3}$. As indicated in the plot title, the probe detuning is taken to vanish, $\Delta_P = 0$, and the remaining Doppler-scaled parameter values (derived from the physical numbers in Table \ref{tab:2ph1RyPhysparms}) are $(\Omega_P^\mathrm{th}, \Omega_C^\mathrm{th}) = (0.0191,0.00368)$ along with decay rates $(\Gamma_1^\mathrm{th}, \Gamma_2^\mathrm{th}, \Gamma_3^\mathrm{th}) = (0.03, 3.4\times 10^{-4}, 9.2\times 10^{-6})$. As indicated, the detuning covers the physical range $-4.4 \leq \Delta_C/2\pi \leq 17.7$ MHz.}

\label{fig:2ph1RyPhysEg}
\end{figure}

\begin{table}
\begin{tabular}{ll}
Probe power $P_P$ & 34.7 $\mu$W \\
Coupling power $P_C$ & 110 mW \\
Beam areas $A_P = A_C$ & 1 mm$^2$ \\
Probe trans.\ dip. moment $\mu_P/ea_B$ & 2.0421 \\
Coupling trans.\ dip. moment $\mu_C/ea_B$ & 0.007 \\
Probe Rabi freq.\ $\Omega_P/2\pi$ & 4.23 MHz \\
Coupling Rabi freq.\ $\Omega_C/2\pi$ & 0.816 MHz \\
Decay rate $(\Gamma_1 = \gamma_{1 \to 0})/2\pi$ & 6.6 MHz \\
Decay rate $(\Gamma_2 = \gamma_{2 \to 0})/2\pi$ & 75 kHz \\
Decay rate $(\Gamma_3 = \gamma_{2 \to 1})/2\pi$ & 2 kHz \\
Temperature $T$ & 39.9 C \\
Atom therm.\ velocity $v_\mathrm{th} = \sqrt{k_B T/m_a}$ & 173 m/s \\
Probe wavelength $\lambda_P$ & 780 nm \\
Probe therm.\ Dopp.\ $k_Pv_\mathrm{th}/2\pi = v_\mathrm{th}/\lambda_P$ & 222 MHz \\
Eqm.\ vapor number density $\rho_a(T)$ & $5.98 \times 10^{10}$ cm$^{-3}$
\end{tabular}

\caption{Physical parameters corresponding to the results shown in Figs.\ \ref{fig:2ph1RyPhysEg}, and motivating those in Fig.\ \ref{fig:2ph1Ry} as well, using a $^{87}$Rb vapor cell. Using (\ref{2.4}), laser power, beam area, and Rabi frequency are related through the electric field by $E = h\Omega/\mu = \sqrt{\eta_0 P/A}$ where $\eta_0 = \sqrt{\mu_0/\epsilon_0} = 377\ \Omega$ is the vacuum impedance. The Rydberg level decay to ground is taken to be dominated by transit time broadening, $\Gamma_2 = 2.71\, v_\mathrm{th}/\sqrt{A_P}$; the other two are intrinsic to the atom. The transition dipole moments are scaled by the hydrogen Bohr radius value $e a_B = 8.4784 \times 10^{-30}$ C-m. The value $\mu_C$ is very small due to the very small physical overlap between core and Rydberg states, hence the much larger required coupling laser power.}

\label{tab:2ph1RyPhysparms}
\end{table}

\section{Mean field solutions for two photon, single Rydberg level setup}
\label{sec:2phmfsoln}

The general formulation of the mean field equations for three-level setups is developed in App.\ \ref{app:3statemfform}, based on the dynamical matrix (\ref{E4}), as defined in App.\ \ref{app:genmultiph}. The explicit evaluation of the thermal averages is described by (\ref{C33})--(\ref{C39}). The specialization to the single Rydberg level setup (b) in Fig.\ \ref{fig:atomsetup} is obtained by setting (see Sec.\ \ref{sec:mfapprox})
\begin{equation}
V_1 = 0,\ V_2 = -V_\mathrm{eff} \bar n^2,\
V_\mathrm{eff} \equiv V_\mathrm{eff}^{22},
\label{6.1}
\end{equation}
in the equations of motion (\ref{2.50}), and recalling the sign convention (\ref{2.26}). The more conventional notation $(\Delta, \Omega) \to (\Delta_P,\Omega_P)$, $(\Delta_R, \Omega_R) \to (\Delta_C,\Omega_C)$ adopted in this section now refers to the probe and coupling lasers, respectively. The laser beams are assumed collinear, thereby maximizing active vapor volume, with Doppler shifts (\ref{E6}) depending only on the single along-beam velocity component. Since it now follows that only $\bar n^2$ appears on the right hand side of the dynamical matrix (\ref{E4}), the mean field equation is obtained by solving the self-consistent equation obtained from the thermal average of $n^2$.

\subsection{Example two-photon bistable phase results}
\label{2phbistable}

An example numerical solution to this equation is shown in Fig.\ \ref{fig:2ph1RyPhysEg}, motivated by some recent preliminary experimental results \cite{foot:CPcomm}. The physical parameters are listed in Table \ref{tab:2ph1RyPhysparms}, and produce the thermal Doppler-scaled parameters defined, similar to (\ref{3.7}), by
\begin{equation}
\Delta_\alpha^\mathrm{th} = \frac{\Delta_\alpha}{k_P v_\mathrm{th}},\
\Omega_\alpha^\mathrm{th} = \frac{\Omega_\alpha}{k_P v_\mathrm{th}},\
V^\mathrm{th}_\mathrm{eff} = \frac{V_\mathrm{eff}}{k_P v_\mathrm{th}},
\label{6.2}
\end{equation}
and similarly for the scaled dissipation parameters $\Gamma^\mathrm{th}_\alpha, K^\mathrm{th}_\alpha$. The equilibrium vapor number density $\rho_a$  is strongly temperature-dependent, exponentially activated from the liquid or solid phase deposited on the cell wall (e.g., increasing by a factor of 2.4 between $30^\circ$ and $40^\circ$ C), and provides the main experimental control over the effective interaction via the relation (\ref{2.35}) with the physical pair potential. The effective interaction parameter is not directly measured in the experiments, hence can only be estimated by matching with a predicted $\bar n^2(\Delta_C,V_\mathrm{eff})$ curve. As a very rough estimate, using scaled the critical value $V^\mathrm{th}_\mathrm{eff} = 24$ one obtains using the values in the table,
\begin{equation}
\frac{V_\mathrm{eff}}{2\pi}
= \frac{V^\mathrm{th}_\mathrm{eff} v_\mathrm{th}}{\lambda_P}
= 5.3\ \mathrm{GHz}.
\label{6.3}
\end{equation}
From (\ref{6.1}), and using the critical value $\bar n_c^2 = 1.7 \times 10^{-3}$ from Fig.\ \ref{fig:2ph1RyPhysEg}, this leads to the interaction-induced detuning
\begin{equation}
\frac{V_2}{2\pi} = -\frac{V_\mathrm{eff}}{2\pi} \bar n_c^2 = -9.1\ \mathrm{MHz},
\label{6.4}
\end{equation}
which is comparable in magnitude to the critical coupling detuning value
\begin{equation}
\frac{\Delta_C}{2\pi} = 0.056 \frac{v_\mathrm{th}}{\lambda_P} = 12.4\ \mathrm{MHz}.
\label{6.5}
\end{equation}

\begin{figure}

\includegraphics[width = 3.3in,viewport = 0 0 565 520, clip]{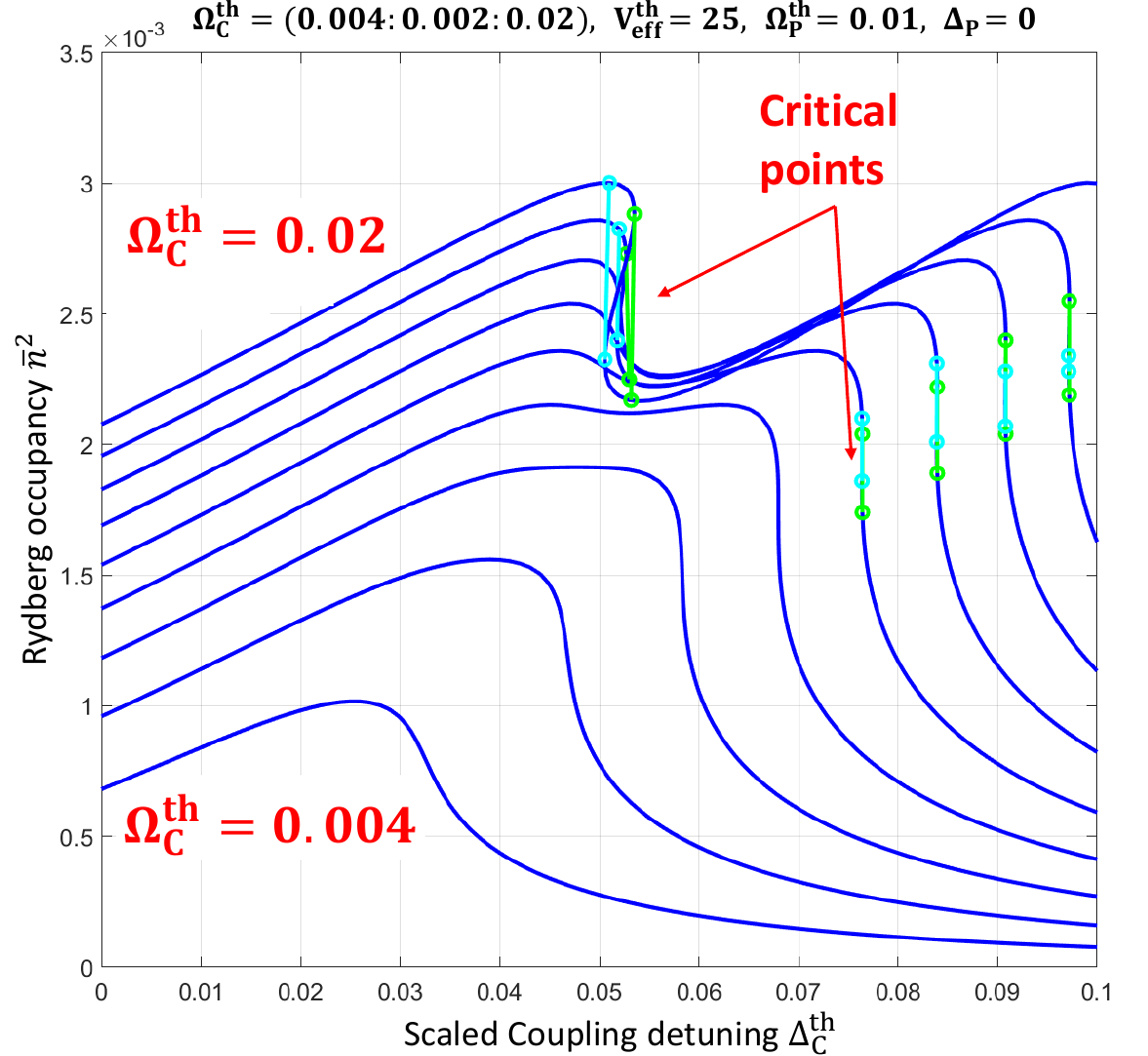}

\caption{Example two photon, single Rydberg level mean field solutions showing multiple bistable regions, with (Doppler-scaled) critical points at $(\Omega_C^\mathrm{th}, \Delta_C^\mathrm{th}, \bar n^2)_c \simeq (0.017, 0.052, 0.0026)$ and $(0.011, 0.073, 0.0019)$. In contrast to Fig.\ \ref{fig:2ph1RyPhysEg}, the interaction $V_\mathrm{eff}^\mathrm{th} = 25$ is fixed and instead the coupling Rabi frequency is scanned over the range $\Omega_C^\mathrm{th} = 0.004, 0.006, 0.008, \ldots, 0.02$. The value $\Omega_P^\mathrm{th} = 0.01$ is somewhat smaller than in Fig.\ \ref{fig:2ph1RyPhysEg}, explaining the absence of a hysteresis loop on the $\Omega_C^\mathrm{th} = 0.004$ curve. All other parameters are the same as in Fig.\ \ref{fig:2ph1RyPhysEg}.}

\label{fig:2ph1Ry}
\end{figure}

A second example is shown in Fig.\ \ref{fig:2ph1Ry}, exhibiting the potentially increased complexity of the phase diagram, with multiple critical points and bistable regions, for the three level vs.\ two level case analyzed in Sec.\ \ref{sec:1ph1Rylevellim}. The more complex behavior, compared to that emerging there from only two states and one laser, is due to the more complex resonant line shapes, involving varying coherent superposition of three different states [based, e.g., on eigenstates of Hamiltonians of the form (\ref{2.25})], with variation of two different sets of laser parameters. The result is the occurrence of multiple steep responses $n_\mathrm{Ry}$ vs.\  $\Delta_C$ that may then be perturbed into hysteresis loops by way of the additional interaction effect. Of course, these curves represent just a 2D slice through the higher dimensional phase diagram in which these bistable regions could well connect up.

\begin{figure*}

\includegraphics[width = 3.5in,viewport = 0 0 645 535, clip]{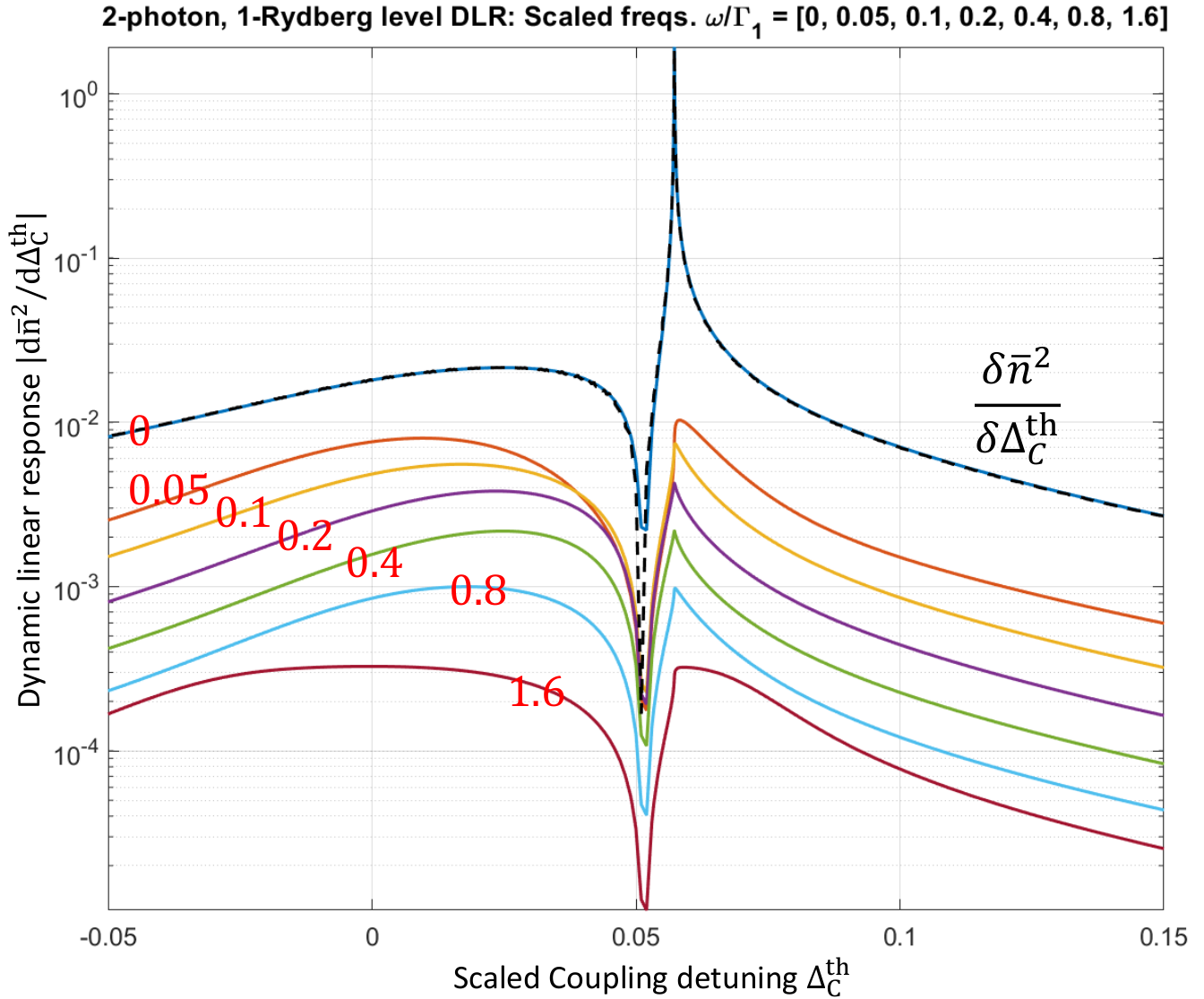}
\includegraphics[width = 5.0in,viewport = 0 0 940 475, clip]{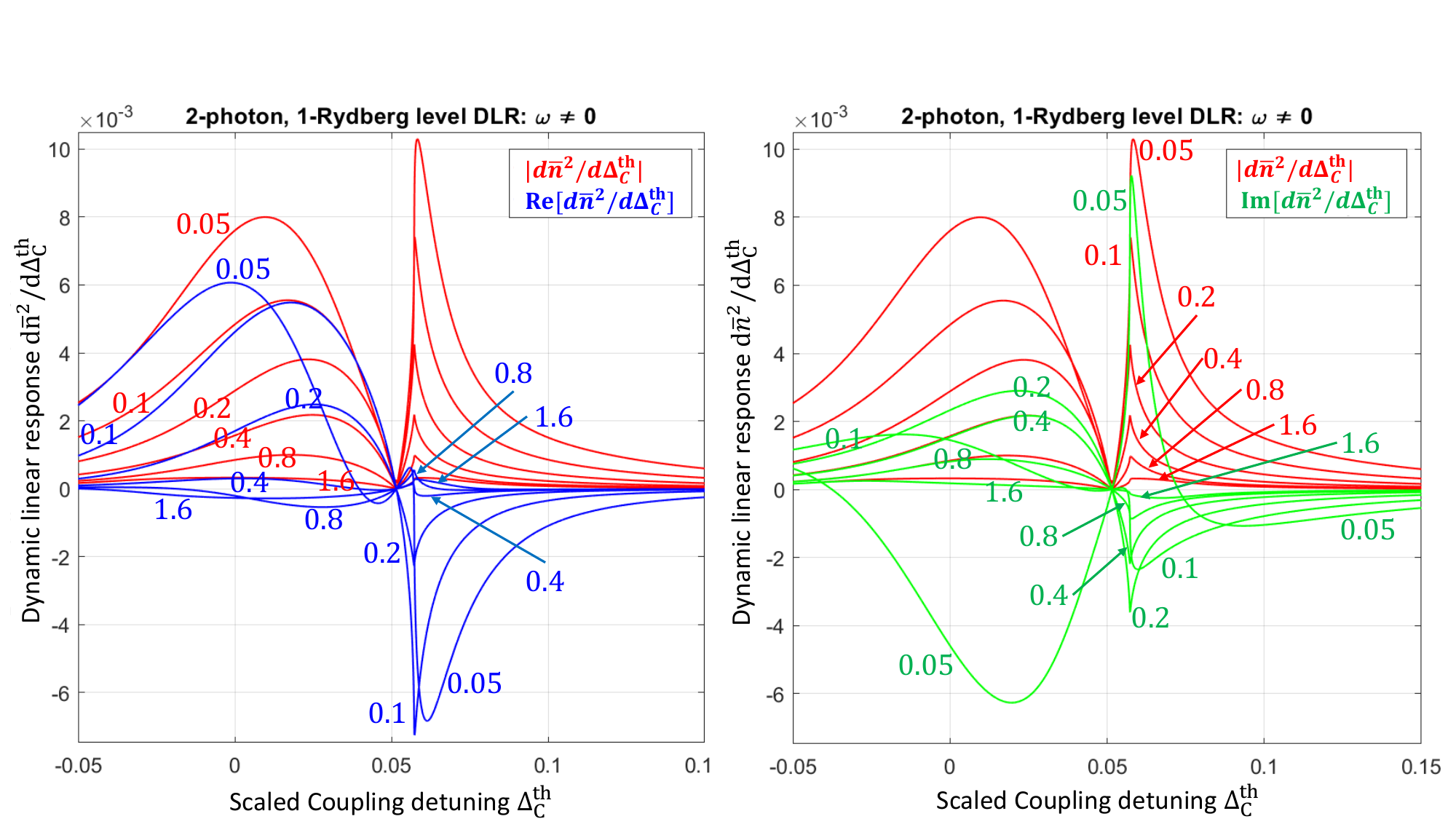}

\caption{Example two photon, single Rydberg level dynamic linear response for a sequence of scaled frequencies $0 \leq \omega/\Gamma_1 \leq 1.6$, as indicated by the plot labels. \textbf{Top:} Semi-log plot of the magnitude of the (scaled) response function $\partial \bar n^2/\partial \Delta^\mathrm{th}_C(\omega)$ along the mean field solution curve $V^\mathrm{th}_\mathrm{eff} = 23$ in Fig.\ \ref{fig:2ph1RyPhysEg} (just outside the bistable phase), whose caption then specifies the values of all remaining parameters. The $\omega = 0$ curve is the derivative of the curve in Fig.\ \ref{fig:2ph1RyPhysEg}, as confirmed by the numerical derivative shown as the overlapping black dashed line. The real parts all change sign near the mean field solution maximum at $\Delta_{C,c}^\mathrm{th} \simeq 0.0573$. \textbf{Bottom:} Linear scale plots of the nonzero frequency curves. The magnitude (red) is plotted along with the real part (blue, left) and imaginary part (green, right).}

\label{fig:2ph1Rydlr}
\end{figure*}

\subsection{Comparison with macroscopic polarization model}
\label{sec:macropolcompare}

The interaction parameter value (\ref{6.4}) may be compared to that derived from the polarization model developed in Sec.\ \ref{sec:macroRypol}.  The Rydberg level mean occupancy $\bar n_\mathrm{Ry}^c \approx 1.7 \times 10^{-3}$ seen in Fig.\ \ref{fig:2ph1Ry} is quite small, consistent with the $\sim$\,1\% estimate for the fraction of sufficiently slow-moving resonant atoms, but now properly accounting for the full thermal velocity distribution. One obtains
\begin{equation}
\bar \rho_\mathrm{Ry} = \rho_a \bar n_\mathrm{Ry}^c = 1.02 \times 10^8\ \mathrm{cm}^{-3},\ \
R_\mathrm{Ry} = 21.4\ \mu\mathrm{m}.
\label{6.6}
\end{equation}
Using again Rydberg quantum number $N = 50$, one obtains from (\ref{2.34}) the estimate
\begin{equation}
\frac{V_2}{2\pi} = -2.59\ \mathrm{MHz}.
\label{6.7}
\end{equation}
Though a factor of $\sim$\,3.5 smaller in magnitude than (\ref{6.4}), it is at least of similar order and there may be experimental cell design parameters that would help account for the difference.

\subsection{Single Rydberg level two-photon dynamic linear response}
\label{sec:2ph1Rydlr}

We next consider the frequency dependent response $\delta n^2(\omega)$ to detuning perturbations $\delta \Delta_C(\omega)$. The general result is described in App.\ \ref{app:dynlinresp}, with dynamical array inputs (\ref{E4}), with the same notational adjustments $(\Delta, \Omega, \Delta_R, \Omega_R) \to (\Delta_P, \Omega_P, \Delta_C, \Omega_C)$. The perturbation matrix therefore corresponds in this case to ${[\partial_{\Delta_R} {\bf A}]}_0$ in (\ref{E9}).

\begin{figure}
\includegraphics[width = 3.3in,viewport = 0 0 850 540, clip]{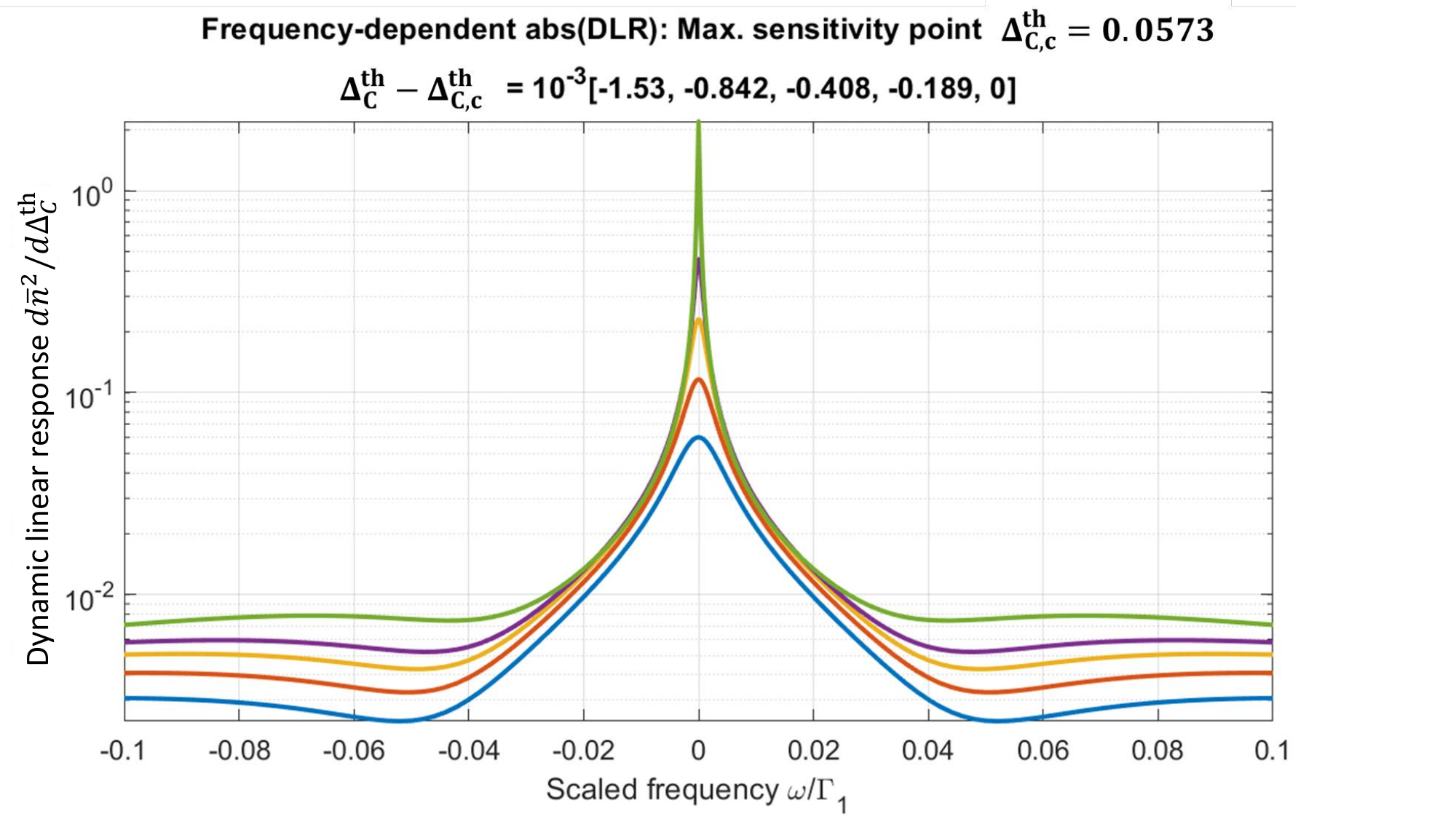}

\caption{Example two photon, single Rydberg level dynamic linear response magnitude vs.\ frequency for a sequence of Doppler-scaled coupling detuning $\Delta^\mathrm{th}_C$ values close to the $\Delta_{C,c}^\mathrm{th} \simeq 0.0573$ peak in the top panel of Fig.\ \ref{fig:2ph1Rydlr} (equivalently, the steepest slope point of the $V^\mathrm{th}_\mathrm{eff} = 23$ curve Fig.\ \ref{fig:2ph1RyPhysEg}). The sharpest peak corresponds to $\Delta^\mathrm{th}_{C,c}$, with decreasingly sharp response as the plot values $\Delta^\mathrm{th}_C - \Delta_{C,c}^\mathrm{th} = [0, -0.189, -0.408, -0.842, -1.53] \times 10^{-3}$ decrease. Similar behavior occurs for increasing positive values (not shown). The peak would actually diverge at the critical point, or at any hysteresis boundary point as seen in Figs.\ \ref{fig:2ph1RyPhysEg} and \ref{fig:2ph1Ry}, where the slope of $\bar n^2(\Delta_C)$ diverges.}

\label{fig:2ph1Rydlrfreq}
\end{figure}

Results for the (Doppler-scaled) response function $\partial \bar n^2/\partial \Delta^\mathrm{th}_C(\omega)$, based on the $V^\mathrm{th}_\mathrm{eff} = -23$ curve in Fig.\ \ref{fig:2ph1RyPhysEg}, are shown in Figs.\ \ref{fig:2ph1Rydlr} and \ref{fig:2ph1Rydlrfreq}. The former displays $\Delta^\mathrm{th}_C$ scans at a sequence of frequencies (described in the caption), showing the strongest response at $\omega = 0$, and rapidly weakening for increasing $|\omega| > 0$ (on the scale of the probe transition decay rate $\Gamma_1$). The latter shows frequency scans at a sequence of $\Delta^\mathrm{th}_C$ values (described in the caption) close to the maximum slope point of the $\bar n^2(\Delta^\mathrm{th}_C)$ curve in Fig.\ \ref{fig:2ph1RyPhysEg}. The peak around zero frequency, which would actually diverge at a critical or hysteresis point, is observed to shrink and broaden, as expected, as one moves away from this point.

The curves in these two figures illustrate the very strong constraints on signal sensitivity enhancement obtained by operating close to a bistable transition point. The strong DC enhancement has a very limited dynamic range (narrow linear response regime), and the finite frequency response has a correspondingly narrow instantaneous bandwidth (beyond which strong signal distortion will occur).

The EIT response density matrix element $\rho^{01}$ [see (\ref{2.36})--(\ref{2.40})] is more complicated than the two-level result (\ref{4.16}), but is straightforward to compute from the mean field solution---following from the general form derived in App.\ \ref{app:steadymfeq}, with input arrays defined explicitly by (\ref{E1})--(\ref{E4}). Supporting computations will be left for future comparisons with experiment.

\begin{figure*}
\includegraphics[width=3.5in,viewport = 0 0 715 500,clip]{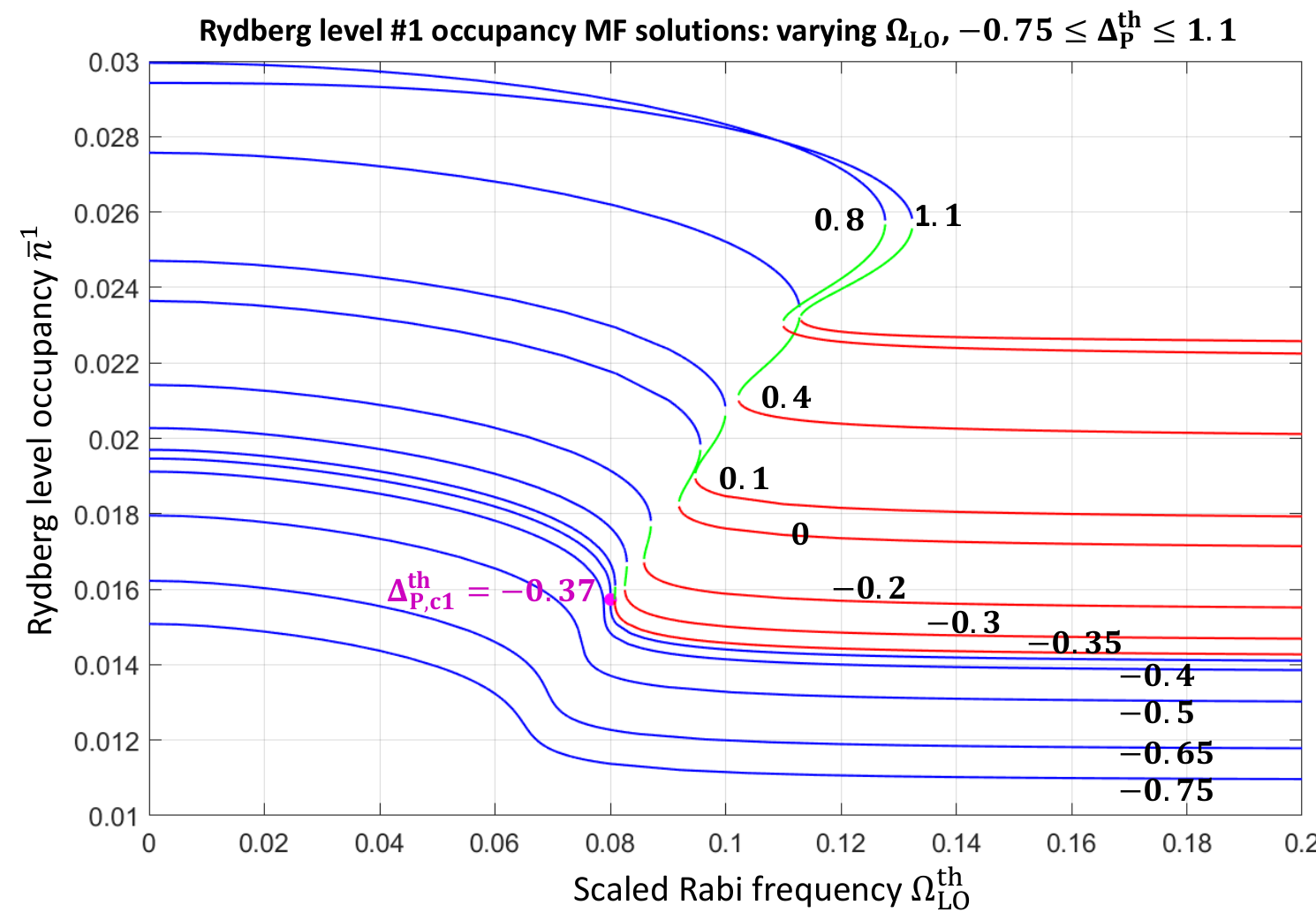}
\includegraphics[width=3.5in,viewport = 0 0 715 500,clip]{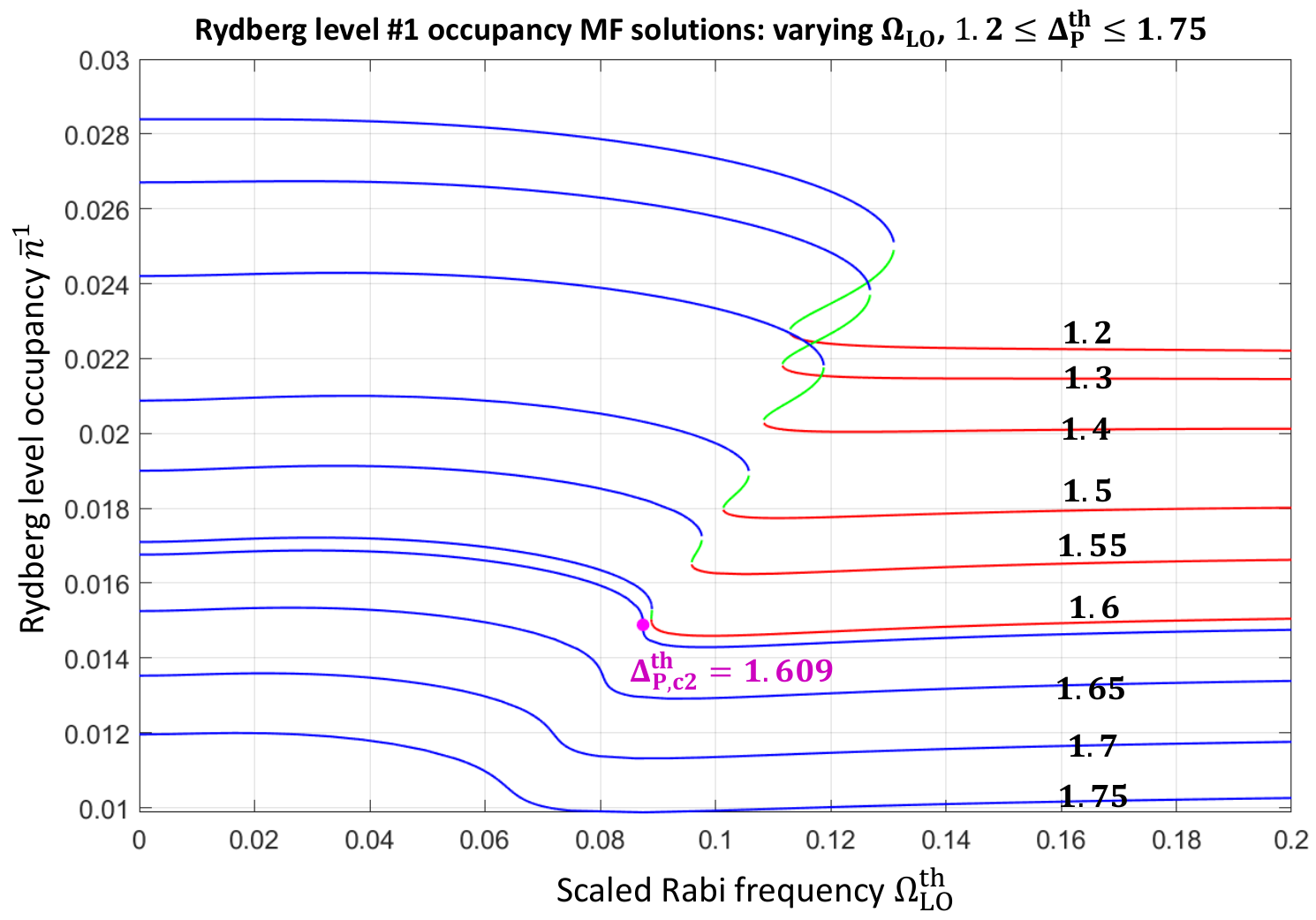}

\includegraphics[width=3.5in,viewport = 0 0 715 500,clip]{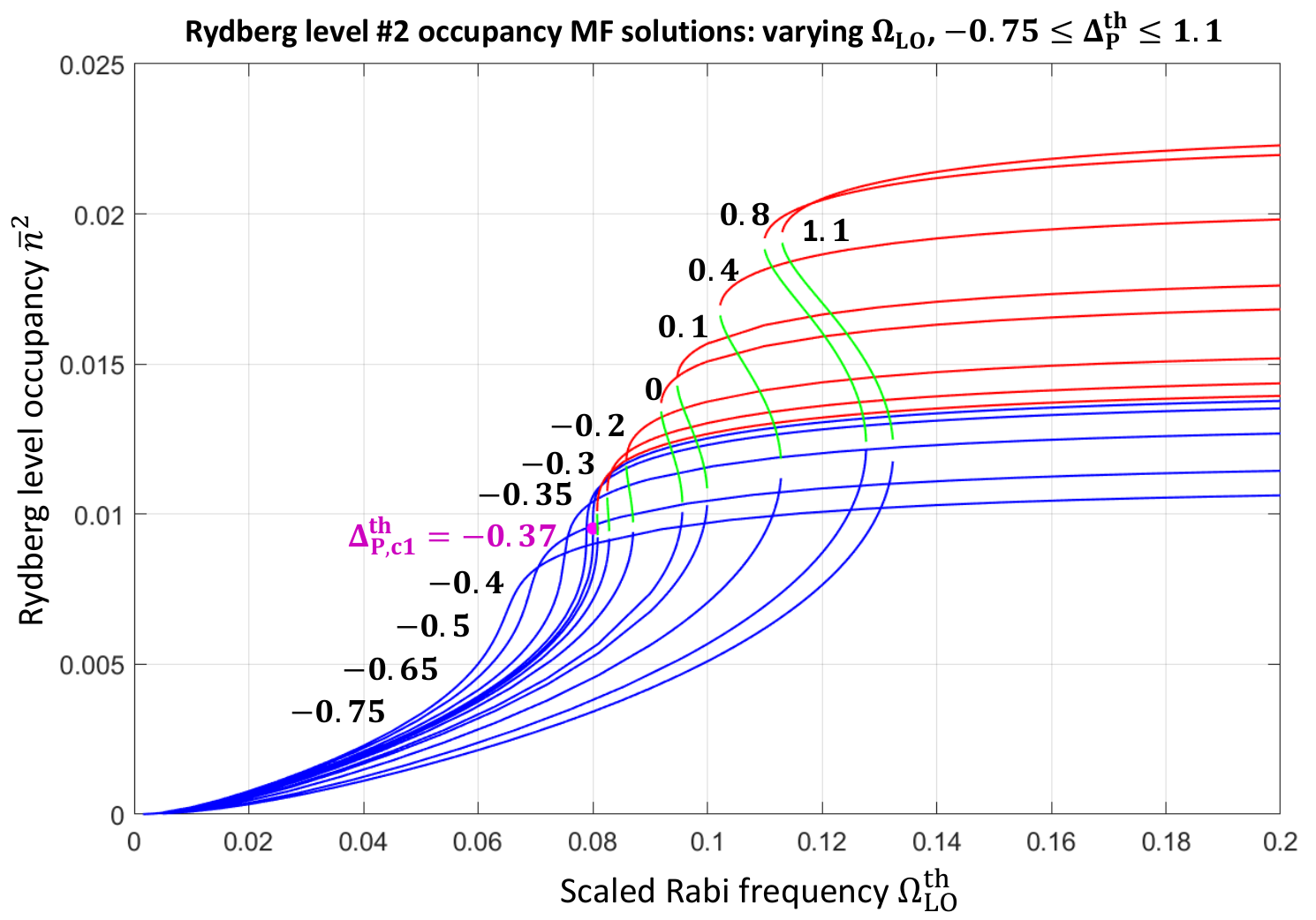}
\includegraphics[width=3.5in,viewport = 0 0 715 500,clip]{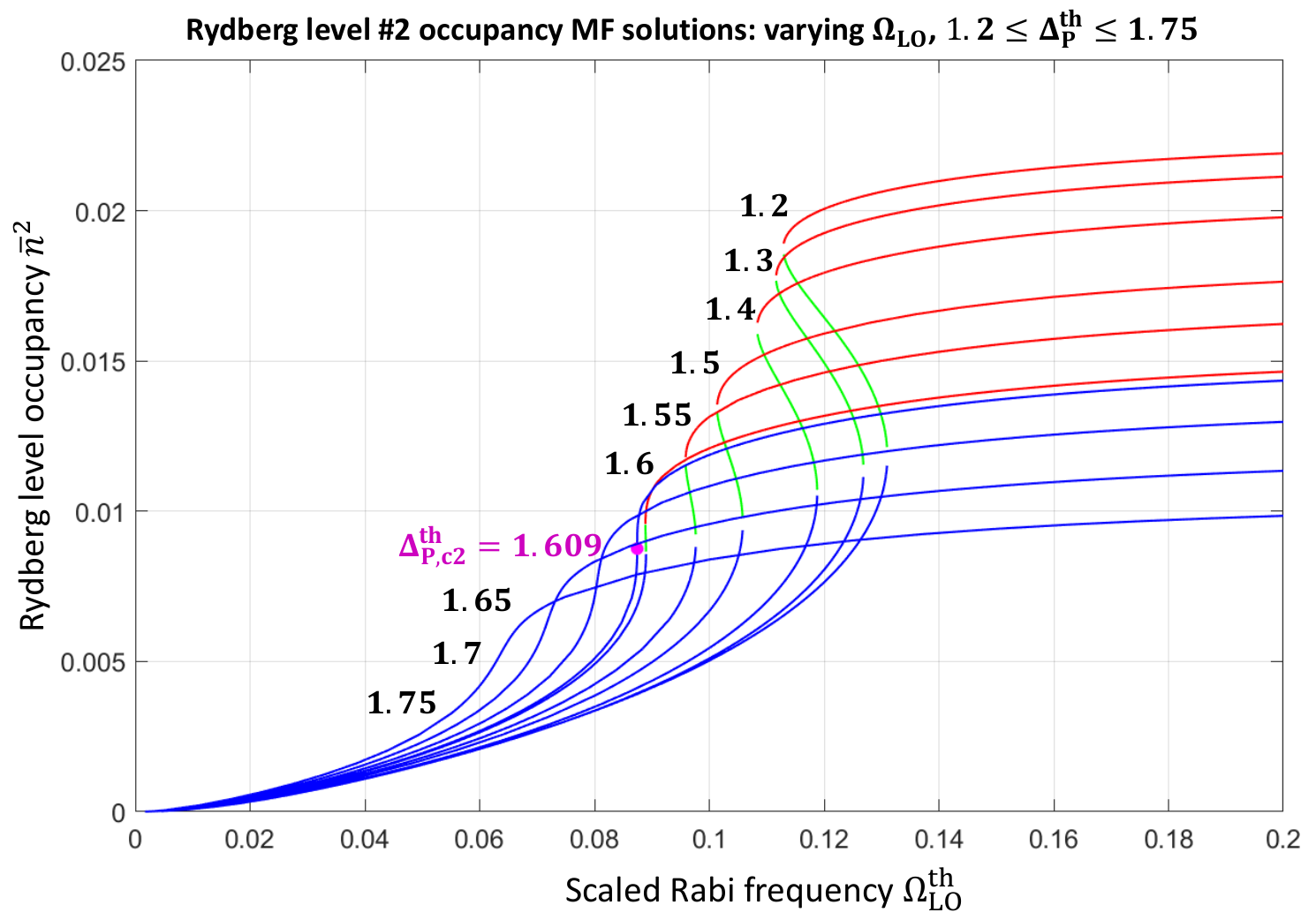}

\caption{Example mean field solutions for the case of a resonant Rydberg level pair. Similar in magnitude to the values used in Figs.\ \ref{fig:2ph1RyPhysEg} and \ref{fig:2ph1Ry}, scaled parameters are $\Omega^\mathrm{th}_P = 0.05$, $\Delta^\mathrm{th}_\mathrm{LO} = 0$, dissipation parameters $\Gamma^\mathrm{th}_1 = \Gamma^\mathrm{th}_2 = K^\mathrm{th}_1 = K^\mathrm{th}_2 = 0.002$, $\Gamma^\mathrm{th}_3 = 0$, and interaction parameters $V_\mathrm{eff}^{\mathrm{th},11} = V_\mathrm{eff}^{\mathrm{th},22} = -30$, $V_\mathrm{eff}^{\mathrm{th},12} = V_\mathrm{eff}^{\mathrm{th},21} = -15$. Plotted are the Rydberg level occupancies $n_1^\mathrm{Ry}(\Omega^\mathrm{th}_R)$ (top row) and $n_2^\mathrm{Ry}(\Omega^\mathrm{th}_R)$ (bottom row) vs. scaled local oscillator Rabi frequency $\Omega^\mathrm{th}_\mathrm{LO}$ for a sequence of scaled probe laser detunings $-0.75 \leq \Delta_P^\mathrm{th} \leq 1.1$ (left column) and $1.2 \leq \Delta_P^\mathrm{th} \leq 1.75$ (right column), labeled on the plots, with the two ranges plotted separately for clarity in order to avoid overlap. The bistable region exists in the range $\Delta^\mathrm{th}_{P,c1} \leq \Delta_P^\mathrm{th} \leq \Delta^\mathrm{th}_{P,c2}$, bounded by critical points $(\Delta^\mathrm{th}_{P,c1}, \Omega^\mathrm{th}_{\mathrm{LO},c1}, \bar n_c^1, \bar n_c^2) = (-0.37, 8.005 \times 10^{-2}, 1.57 \times 10^{-2}, 9.50 \times 10^{-3})$, $(\Delta^\mathrm{th}_{P,c2}, \Omega^\mathrm{th}_{\mathrm{LO},c2}, \bar n^1_c, \bar n^2_c) = (1.609, 0.087, 1.49 \times 10^{-2}, 8.75 \times 10^{-3})$ (magenta dots). Within this range, the three line colors highlight the stable (blue, red) and unstable (green) solutions of the mean field equations. As usual, hysteretic behavior corresponds to discontinuous jumps between the red and blue curves as one scans in the corresponding direction beyond their endpoints.}

\label{fig:MF2Ry}

\end{figure*}

\section{Resonant Rydberg level pair}
\label{sec:2Ryresonant}

We turn finally to solutions of the mean field equations (\ref{E5}), including the full Rydberg level pair interaction matrix appearing in (\ref{2.26}). We adopt in this section, along with the probe beam notation $(\Delta_P, \Omega_P)$, the more conventional local oscillator (LO) notation $(\Delta_R,\Omega_R) \to (\Delta_\mathrm{LO},\Omega_\mathrm{LO})$. Scaled parameters, similar to (\ref{6.2}), continue to be defined by dividing by the probe thermal Doppler shift $k_P v_\mathrm{th}$.

\subsection{Example single photon, Rydberg level pair bistable phase}
\label{sec:eg1ph2Rybistable}

Since the mean field interaction parameters $V_\alpha$ now depend on both Rydberg level occupancies $\bar n^1, \bar n^2$, the two equations must be solved simultaneously. Numerically, this is accomplished by solving the first equation for $\bar n^1(\bar n^2)$ for given $\bar n^2$ and then inserting this into the second equation and solve $\bar n^2 = \bar n^2(\bar n^1(\bar n^2), \bar n^2)$. Performing this in the opposite order is also permitted, and sometimes is found to improve numerical stability.

An example phase diagram is shown in Fig.\ \ref{fig:MF2Ry}, verifying existence the bistable phase under the reasonable assumption that both levels experience similar interaction strengths, in this case choosing scaled effective parameters (see Sec.\ \ref{sec:mfapprox}) $V^{\mathrm{th},11}_\mathrm{eff} = V^{\mathrm{th},22}_\mathrm{eff}  = 2V^{\mathrm{th},12}_\mathrm{eff}  = 2V^{\mathrm{th},21}_\mathrm{eff}  = 30$, similar in magnitude to the value for $V^\mathrm{th}_\mathrm{eff}$, defined by (\ref{6.1}) and (\ref{6.2}) and used in Figs.\ \ref{fig:2ph1RyPhysEg} and \ref{fig:2ph1Ry}. Unlike the two-photon case which includes a much larger probe transition linewidth (see Table \ref{tab:2ph1RyPhysparms}), all dissipation parameters here are characteristic of much slower Rydberg level decay and dephasing processes. For simplicity we simply set all $\Gamma^\mathrm{th}_\alpha, K^\mathrm{th}_\alpha = 1/500$. In both cases, more realistic choices are left for future comparisons with experiment.

In this example we assume vanishing LO detuning, $\Delta_\mathrm{LO} = 0$, and fixed probe beam amplitude $\Omega^\mathrm{th}_P = 0.05$, corresponding to a reasonable physical value $\Omega_P/2\pi \sim 10$ MHz. The different curves in Fig.\ \ref{fig:MF2Ry} then correspond to scanning the local oscillator strength $0 \leq \Omega^\mathrm{th}_\mathrm{LO} \leq 0.2$ for a sequence of the probe beam detuning values in the range $-0.75 \leq \Delta^\mathrm{th}_P \leq 1.75$---broken up into two subintervals (left and right columns) to avoid overlapping curves. The bistable phase in this example is seen to be reentrant, spanning $\Delta_{P,c1}^\mathrm{th} \leq \Delta_P^\mathrm{th} \leq \Delta_{P,c2}^\mathrm{th}$ with bounding critical points $(\Delta_{P,c1}^\mathrm{th}, \Omega_{\mathrm{LO},c1}^\mathrm{th}) \simeq (-0.37, 0.080)$, $(\Delta_{P,c2}^\mathrm{th}, \Omega_{\mathrm{LO},c2}^\mathrm{th}) \simeq (1.609, 0.087)$.

\begin{figure*}
\includegraphics[width=6.0in,viewport = 0 0 730 540,clip]{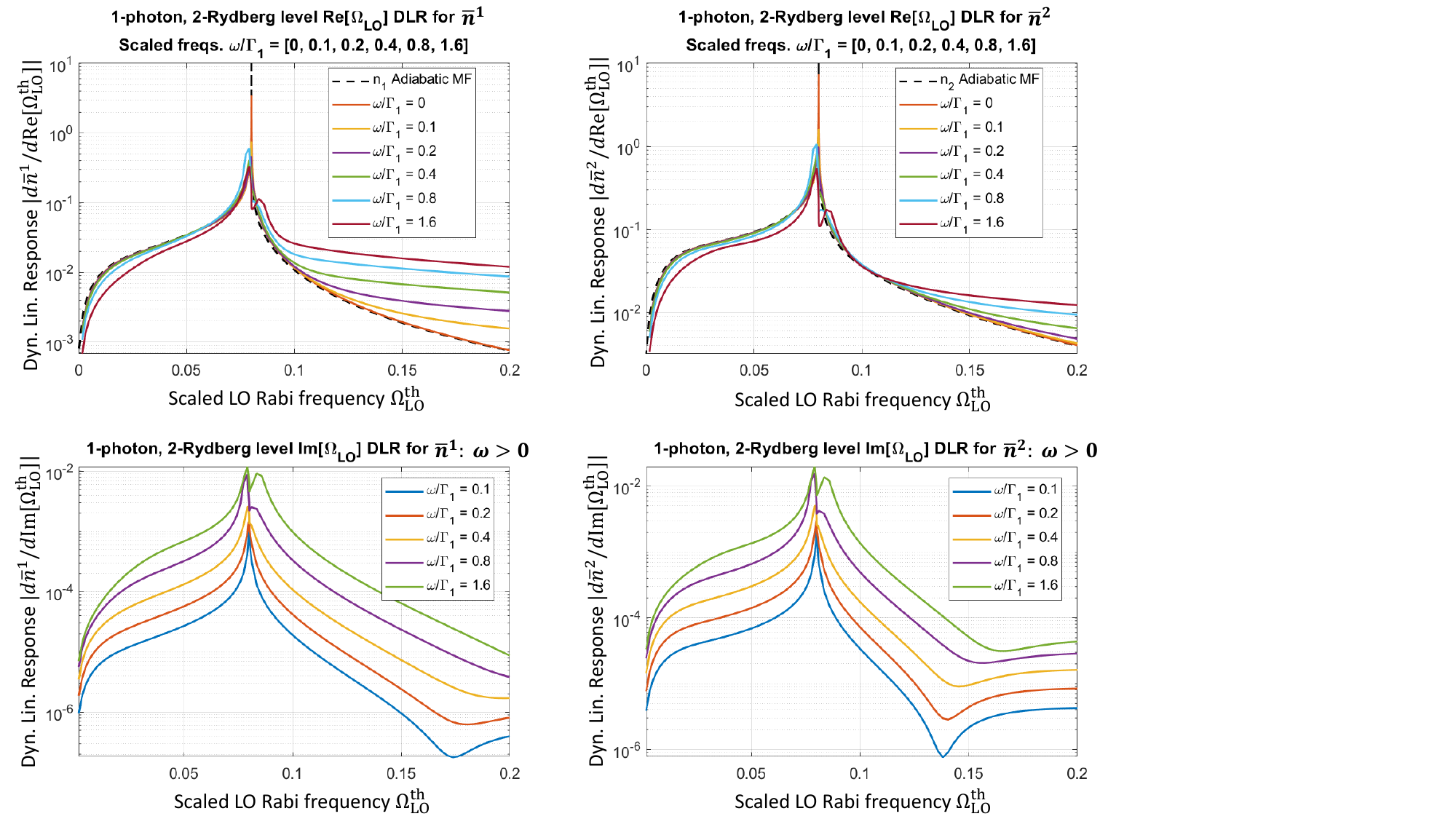}

\caption{Similar to Fig.\ \ref{fig:2ph1Rydlr}, example dynamical linear response solutions vs.\ scaled local oscillator Rabi frequency $\Omega^\mathrm{th}_\mathrm{LO}$ for the case of a resonant Rydberg level pair, along the $\Delta_P^\mathrm{th} = \Delta_{P,c1}^\mathrm{th} \simeq -0.37$ line in the left column of Fig.\ \ref{fig:MF2Ry}, passing through the critical point $(\Omega^\mathrm{th}_{\mathrm{LO},c1}, \bar n_{1,c}, \bar n_{2,c}) = (8.005 \times 10^{-2}, 1.57 \times 10^{-2}, 9.50 \times 10^{-3})$. All other parameters are as described in the caption of that figure. The four panels correspond to the (log-scale magnitudes) of linear response of the two Rydberg level population fractions $\bar n^1$ (left column) and $\bar n^2$ (right column) to perturbations $\delta \Omega^{\mathrm{th}\,\prime}_\mathrm{LO}(\omega)$ (upper row) and $\delta \Omega^{\mathrm{th}\,\prime\prime}_\mathrm{LO}(\omega)$ (lower row) of the real and imaginary parts of $\Omega^\mathrm{th}_\mathrm{LO}$, respectively. Each line corresponds to a different scaled frequency $\omega/\Gamma_1 = (0, 0.1, 0.2, 0.4, 0.8, 1.6)$. The $\omega = 0$ curves in the top row are the derivatives of the curves in the left column of Fig.\ \ref{fig:MF2Ry}, as confirmed by the numerical derivative shown as the overlapping black dashed line. The $\omega = 0$ curves in the bottom row are missing because the imaginary response vanishes identically in that limit.}

\label{fig:DLR2Ry_OmegaR}

\end{figure*}

\subsection{Rydberg level pair dynamic linear response and near-critical sensor enhancement}
\label{sec:1ph2Rydlr}

\begin{figure*}
\includegraphics[width=6.0in,viewport = 0 0 800 540,clip]{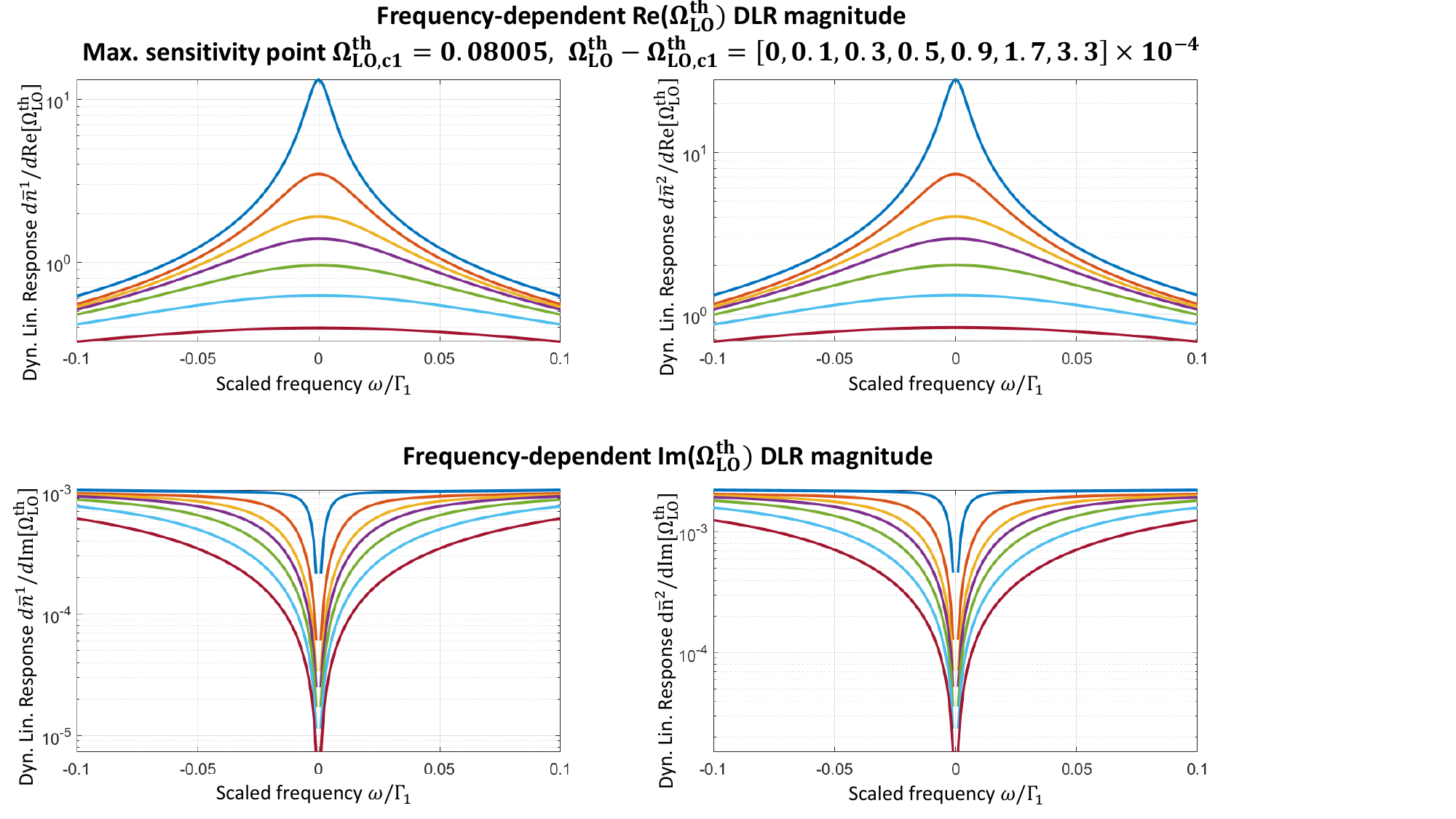}

\caption{Similar to Fig.\ \ref{fig:2ph1Rydlrfreq}, example dynamical linear response solutions vs.\ scaled frequency $\omega/\Gamma_1$ for the case of a resonant Rydberg level pair, at a sequence of near-critical points $\Omega^\mathrm{th}_\mathrm{LO} - \Omega_\mathrm{LO,c1}^\mathrm{th} = [0, 0.1, 0.3, 0.5, 0.9, 1.7, 3.3] \times 10^{-4}$ (in order of broadening response) along the $\Delta^\mathrm{th}_P = \Delta_{P,c1}^\mathrm{th} \simeq -0.37$ line (left column of Fig.\ \ref{fig:MF2Ry}; all other parameters are as described in the caption). The four panels correspond to the (log-scale magnitudes) of the four different linear response combinations described in the caption to Fig.\ \ref{fig:DLR2Ry_OmegaR}. The $\omega = 0$ peak would actually diverge exactly at the critical point, and also at any hysteresis point, where the slopes of $\bar n^1(\Omega_\mathrm{LO}), \bar n^2(\Omega_\mathrm{LO})$ diverge. The bottom row confirms that the imaginary response vanishes identically at $\omega = 0$.}

\label{fig:DLR2Ry_freq}

\end{figure*}

Potential bistability-enhanced sensing is obtained by fixing parameter values close to a critical or hysteresis point. Quantifying this, dynamical linear response results, analogous to those in Figs.\ \ref{fig:2ph1Rydlr} and \ref{fig:2ph1Rydlrfreq}, are shown in Figs.\ \ref{fig:DLR2Ry_OmegaR} and \ref{fig:DLR2Ry_freq} along the one of the curves $\Delta_P^\mathrm{th} = \Delta_{P,c1}^\mathrm{th}$ passing through a critical point. The four panels correspond to the four combinations of responses of $\bar n^1, \bar n^2$ to real and imaginary perturbation amplitudes $\delta \Omega^\mathrm{th}_\mathrm{LO}$ relative to the real, in this example, background value $\Omega^\mathrm{th}_\mathrm{LO} \to \Omega^\mathrm{th}_\mathrm{LO} + \delta \Omega^\mathrm{th}_\mathrm{LO} e^{-i\omega t}$.

Similar to those previous figures, a divergent response is seen at the critical point in the limit $\omega \to 0$. The finite frequency linewidth is governed by the dissipation scale $\Gamma_1$---which could clearly become more complex if they were not all chosen equal. Similar to Fig.\ \ref{fig:2ph1Rydlr}, the multipeak structure in Fig.\ \ref{fig:DLR2Ry_OmegaR} is due to evolving interference between the various three-level atomic resonant interactions with variation in the probe laser and local oscillator parameters.

Recall that the objective here is to take advantage of the already resonantly enhanced sensitivity to perturbations of the local oscillator $\Omega_\mathrm{LO}$ by additionally coupling it to a bistable phase transition. The geometry and parameter ranges of the critical and hysteresis curves is not tremendously different between Figs.\ \ref{fig:2ph1RyPhysEg} and \ref{fig:MF2Ry}. In both cases there is a 1--2 order of magnitude multiplier on the ``background'' sensitivity far from the critical point, depending on the desired measurement bandwidth (i.e., deliberate setting deviation from the critical point in order to broaden the linear response regime). In terms of absolute physical measurement sensitivity, the difference between the two is therefore not significantly in the multiplier, but in the background value. Thus, as is now well established \cite{NIST2019a,NIST2019b,Jing2020, Waterloo2021,MITRE2021,NIST2023,Romalis2024,Wu2024,Sandidge2024,Warsaw2024}, the resonant local oscillator setup has led to record sensitivity increases due to the greatly enhanced coupling of the incident field directly through the LO perturbation (\ref{2.6}) and the accompanying $O(N^2)$ Rydberg-enhanced dipole coupling (\ref{2.4}). Although, as summarized in Sec.\ \ref{subsec:sigdetectmetric}, the original approach based on the AC Stark effect produces an even stronger $O(N^7)$ enhancement (\ref{4.21}), the multiplying coefficient going into the final result (\ref{4.23}) is extremely small, and physically sensible values $N \sim 50$ are unable to compete \cite{ARL2020}.

We note again that a rigorous measurement analysis should be based on the probe beam EIT response density matrix element $\rho^{01}$, along with an evaluation of various noise sources to compute a proper SNR \cite{W2024}, rather than the Rydberg state populations $\bar n^{1,2}$ focused on here. However, supported by the two-level system discussion in Sec.\ \ref{sec:probeEITsigsense}, one does not expect this to change the basic conclusions. A complete analysis will be left for future comparisons with experiment.

\section{Conclusions}
\label{sec:conclude}

In this work the theory of bistatic behavior in Rydberg vapors \cite{CRWAW2013,MLDGL2014,Ding2022,Wang2023} has been extended to include (a) exact treatment of atom motion-induced Doppler; (b) a larger number of participating resonantly-interacting atomic levels, especially including local oscillator coupling of a Rydberg level pair; and (c) full treatment of dynamic effects within linear response theory.

The first effect leads to a $\sim$\,99\% effective reduction in participating atoms, hence at minimum is critical to quantitative modeling of the vapor thermodynamics and RF sensor operation---beyond an effective zero temperature treatment based on a simple $\sim$\,1\% scaling of the atomic density $\rho_a$. The contrast between the zero temperature and finite temperature results presented in Secs.\ \ref{sec:1ph1Rylevellim} and Ref.\ \ref{sec:critscale} serves to highlight this.

The three-level system model treated in Secs.\ \ref{sec:2phmfsoln} and \ref{sec:2Ryresonant} allows for the first time quantitative analysis of the proposed merging of ``conventional'' RF sensing enhancement, through local oscillator tuning of a resonant Rydberg level pair \cite{NIST2019a,NIST2019b,Jing2020,Waterloo2021,MITRE2021,NIST2023, Romalis2024,Wu2024,Sandidge2024,Warsaw2024}, with the near-critical (or near-hysteretic) bistatic susceptibility enhancement \cite{Ding2022,Wang2023}. The results described in Sec.\ \ref{sec:2Ryresonant} confirm this in principle, though experimental confirmation may place new constraints on the space of available model parameters that would then need to be thoroughly explored.

The dynamic linear response theory results are a critical requirement for sensor operation in the presence of more realistic, finite bandwidth signals and/or measurements based on finite dwell time. The results in Secs.\ \ref{sec:2ph1Rydlr} and Sec.\ \ref{sec:1ph2Rydlr} quantify the tradeoff between optimal (in principle unbounded) sensitivity, for near-critical setups and local oscillator tuned to a monotone signal, and rapid sensitivity fall-off on a combination of scales set by the distance from the critical (or hysteresis) point and the Rydberg state inverse-lifetime $1/\Gamma_1$ which determines the underlying atomic state linewidth. These results will certainly be impacted by numerous practical experimental issues, such vapor cell, laser beam, and local oscillator inhomogeneities and other imperfections that will need to be properly explored in future work.

As alluded to earlier, quantitative comparisons with experimental measurements will also rely on full treatment of the EIT density matrix element $\mathrm{Im}[\rho^{01}]$, which follows straightforwardly from the general formulation in App.\ \ref{app:3statemfform} [specifically applied to the fourth element of the dynamical variable vector (\ref{E2})]. The hard work solving the mean field equations (\ref{E5}) has already been done.

Less straightforward would be extending the theory to two- and three-photon versions of setup (c) in Fig.\ \ref{fig:atomsetup}. Here additional (dressing, coupling) lasers are used, for various practical and physical reasons \cite{RBTEY2011,Michigan2019,W2024}, to access the Rydberg levels through a chain of intermediate non-Rydberg states. An $N_L$-level system description leads to $N_L^2 - 1$ independent density matrix parameters, greatly increasing the numerical complexity. In addition, there will be complex interplay between the more intricate underlying multi-level resonant structure and the bistability interactions---influenced as well by the growing set of dissipation parameters (e.g., multi-scale Lorentzian line shapes). The basic physics of the bistable phase does not change, but the phase diagram will be strongly affected by any control feature that induces strong changes in the Rydberg state populations. On the other hand, a more complex phase diagram might also provide interesting opportunities for sensor optimization, e.g., seeking more forgiving tradeoffs between sensitivity and bandwidth adapted to the signal of interest.

Finally, we return to some questions related to the fundamentals of the mean field approximation. It was argued in Sec.\ \ref{sec:macroRypol} that the electric dipole nature of the Rydberg state should make this approximation exact, with effective interactions (\ref{2.26}) dominated by macroscopic averages over the excited vapor volume. These are the only major unknowns in the model---the single atom properties are well tabulated \cite{ARC}. It would be extremely interesting to verify this experimentally, especially testing the macroscopic Maxwell equation approach (\ref{2.31})--(\ref{2.34}) to their estimation and observing the predicted dependence on excited vapor volume geometry---also another likely source of inhomogeneity. It would also be interesting to test the limits of the mean field approximation at finite frequency, e.g., on the scale of the $\sim$\,10 $\mu$s mean transit time of atoms across the beam.

\appendix

\section{Density matrix equation of motion}
\label{app:dmmotioneq}

We provide here details of the derivation of the equation of motion (\ref{2.50}) for the density matrix components. The commutator entering the fundamental equation of motion (\ref{2.24}) is evaluated with the aid of the identities
\begin{eqnarray}
[\hat n^\alpha,\hat n^\beta] &=& 0
\nonumber \\
\left[\hat n^\alpha,\hat \sigma^{\mu\nu}\right] &=& \hat \sigma^{\mu\nu}
(\delta_{\alpha\mu} - \delta_{\alpha\nu})
\nonumber \\
\left[\hat \sigma^{\alpha\beta},\hat \sigma^{\mu\nu}\right]
&=& \delta_{\beta\mu} |\alpha\rangle \langle \nu|
- \delta_{\alpha\nu} |\mu\rangle \langle \beta|
\label{A1} \\
&=& \left\{\begin{array}{ll}
\hat \sigma^{\alpha\nu}, & \beta = \mu,\ \alpha \neq \nu \\
-\hat \sigma^{\mu\beta}, & \beta \neq \mu,\ \alpha = \nu \\
\hat n_\alpha - \hat n_\beta, & \beta = \mu,\ \alpha = \nu \\
0, & \beta \neq \mu,\ \alpha \neq \nu.
\end{array} \right.
\nonumber
\end{eqnarray}
Using these, one obtains
\begin{eqnarray}
\big[\hat n^0, \hat H \big] &=& -\frac{1}{2}(\Omega \hat \sigma^{01} - \Omega^* \hat \sigma^{10})
\nonumber \\
\big[\hat n^1, \hat H \big] &=& \frac{1}{2}(\Omega \hat \sigma^{01} - \Omega^* \hat \sigma^{10})
- \frac{1}{2}(\Omega_R \hat \sigma^{12} - \Omega_R^* \hat \sigma^{21})
\nonumber \\
\big[\hat n^2, {\hat H} \big] &=& \frac{1}{2}(\Omega_R \hat \sigma^{12} - \Omega_R^* \hat \sigma^{21})
\label{A2}
\end{eqnarray}
for the projection operators and
\begin{eqnarray}
\big[\hat \sigma^{01}, \hat H \big] &=& -(\big[\hat \sigma^{10}, \hat H \big])^\dagger
\nonumber \\
&=& -(\Delta + V_1) \hat \sigma^{01}
- \frac{1}{2} \Omega_R \hat \sigma^{02} - \frac{1}{2} \Omega^* (\hat n^0 - \hat n^1)
\nonumber \\
\big[\hat \sigma^{12}, \hat H \big] &=& -(\big[\hat \sigma^{21}, \hat H \big])^\dagger
\nonumber \\
&=& -(\Delta_R + V_2 - V_1) \hat \sigma^{12}
+ \frac{1}{2} \Omega \hat \sigma^{02}
\nonumber \\
&&-\ \frac{1}{2} \Omega_R^* (\hat n^1 - \hat n^2)
\label{A3} \\
\big[\hat \sigma^{02}, \hat H\big] &=& -(\big[\hat \sigma^{20}, \hat H\big])^\dagger
\nonumber \\
&=& -(\Delta + \Delta_R + V_2) \hat \sigma^{02}
+ \frac{1}{2} \Omega^* \hat \sigma^{12} - \frac{1}{2} \Omega_R^* \hat \sigma^{01}.
\nonumber
\end{eqnarray}
for the transition operators, in which the atom label $k$ is temporarily dropped to simplify the notation. One obtains
\begin{eqnarray}
[\hat \rho,H] &=& C_D[\hat \rho] + C_\Omega[\hat \rho] + C_\Delta[\hat \rho]
\nonumber \\
C_D[\hat \rho] &=& \sum_{\alpha=1}^3 \rho^{\alpha\alpha} \big[\hat n^\alpha, \hat H \big]
\nonumber \\
{\cal C}_\Omega[\hat \rho] + {\cal C}_\Delta[\hat \rho]
&=& \sum_{\alpha \neq \beta} \rho^{\alpha\beta}
\big[\hat \sigma^{\alpha\beta}, \hat H \big]
\label{A4}
\end{eqnarray}
with, respectively, diagonal, Rabi, and detuning (anti-Hermitean matrix) contributions
\begin{widetext}
\begin{eqnarray}
C_D[\hat \rho] &=& \frac{1}{2} \left[\begin{array}{ccc}
0 & -\Omega (\rho^{00} - \rho^{11}) & 0 \\
\Omega^* (\rho^{00} - \rho^{11}) & 0 & -\Omega_R (\rho^{11} - \rho^{22}) \\
0 & \Omega_R^* (\rho^{11} - \rho^{22}) & 0
\end{array}\right]
\nonumber \\
{\cal C}_\Omega[\hat \rho] &=& \frac{1}{2} \left[\begin{array}{ccc}
\Omega^* \rho^{01} - \Omega \rho^{10} & \Omega_R^* \rho^{02}
& \Omega_R \rho^{01} - \Omega \rho^{12} \\
-\Omega_R \rho^{20} & \Omega_R^* \rho^{12} - \Omega_R \rho^{21}
- (\Omega^* \rho^{01} - \Omega \rho^{10}) & -\Omega^* \rho^{02} \\
- (\Omega_R^* \rho^{10} - \Omega^* \rho^{21})
& \Omega \rho^{20} & -(\Omega_R^* \rho^{12} - \Omega_R \rho^{21})
\end{array}\right]
\nonumber \\
{\cal C}_\Delta[\hat \rho] &=& \left[\begin{array}{ccc}
0 & -(\Delta + V_1) \rho^{01} & -(\Delta + \Delta_R + V_2) \rho^{02} \\
(\Delta + V_1)\rho^{10} & 0 & -(\Delta_R + V_2 - V_1) \rho^{12} \\
(\Delta + \Delta_R + V_2) \rho^{20} & (\Delta_R + V_2 - V_1) \rho^{21} & 0
\end{array}\right].
\label{A5}
\end{eqnarray}
\end{widetext}
Next, from the Lindblad operators (\ref{2.17}) one obtains the dissipation term inputs
\begin{eqnarray}
(L_1^\alpha)^\dagger L_1^\alpha &=& \Gamma_\alpha \hat n^\alpha,\ \
(L_2^\alpha)^\dagger L_2^\alpha = K_\alpha \hat n^\alpha,\ \
L_3^\dagger L_3 = \Gamma_3 \hat n^2.
\nonumber \\
L_1^\alpha \hat \rho L_1^{\alpha \dagger}
&=& \Gamma_\alpha \hat \sigma^{0\alpha} \hat \rho \hat \sigma^{\alpha 0}
= \Gamma_\alpha \rho^{\alpha \alpha} \hat n^0
\nonumber \\
L_2^\alpha \hat \rho L_2^{\alpha \dagger}
&=& K_\alpha \hat n^\alpha \hat \rho \hat n^\alpha
= K_\alpha \rho^{\alpha \alpha} \hat n^\alpha
\nonumber \\
L_3 \hat \rho L_3^\dagger &=& \Gamma_3 \hat \sigma^{12} \hat \rho \sigma^{21}
= \Gamma_3 \rho^{22} \hat n^1.
\label{A6}
\end{eqnarray}
The second line accounts for the change in the ground state population due to decay from the two higher levels, the last two for the change in excited state populations, each at rates proportional to the relevant excited state probabilities $\rho^{11}, \rho^{22}$. One obtains as well the anticommutator identities
\begin{eqnarray}
\frac{1}{2} \{\hat n^\alpha,\hat \rho\} &=& \rho^{\alpha \alpha} \hat n^\alpha
+ \frac{1}{2} \hat R_\alpha[\hat \rho],\ \ \alpha = 1,2
\nonumber \\
\hat R_\alpha[\hat \rho] &\equiv& \sum_{\gamma (\neq \alpha)}
(\rho^{\gamma\alpha} \hat \sigma^{\gamma\alpha}
+ \rho^{\alpha\gamma} \hat \sigma^{\alpha\gamma}),
\label{A7}
\end{eqnarray}
with explicit forms
\begin{equation}
\hat R_1[\hat \rho] = \left(\begin{array}{ccc}
0 & \rho^{01} & 0 \\
\rho^{10} & 0 & \rho^{12} \\
0 & \rho^{21} & 0
\end{array} \right),\ \
\hat R_2[\hat \rho] = \left(\begin{array}{ccc}
0 & 0 & \rho^{02} \\
0 & 0 & \rho^{12} \\
\rho^{20} & \rho^{21} & 0
\end{array} \right),
\label{A8}
\end{equation}
which are just projected subsets of elements of $\hat \rho$. Using these results the full Lindblad operator (\ref{2.20}) may be assembled in the form
\begin{widetext}
\begin{eqnarray}
{\cal L}[\hat \rho] &=& \sum_{\alpha = 1}^2
\left[\Gamma_\alpha \rho^{\alpha\alpha} (\hat n^0 - \hat n^\alpha)
- \frac{1}{2} (\Gamma_\alpha + K_\alpha) \hat R_\alpha[\hat \rho] \right]
+ \Gamma_3 \left\{\rho^{22} (\hat n^1 - \hat n^2) - \frac{1}{2} \hat R_2[\hat \rho] \right\}
\nonumber \\
&=& -\frac{1}{2} \left[\begin{array}{ccc}
-2(\Gamma_1 \rho^{11} + \Gamma_2 \rho^{22}) & (\Gamma_1 + K_1) \rho^{01}
& (\Gamma_2 + \Gamma_3 + K_2) \rho^{02} \\
(\Gamma_1 + K_1) \rho^{10} & 2(\Gamma_1 \rho^{11} - \Gamma_3 \rho^{22})
& (\Gamma_1 + \Gamma_2 + \Gamma_3 + K_1 + K_2) \rho^{12} \\
(\Gamma_2 + \Gamma_3 + K_2) \rho^{20} & (\Gamma_1 + \Gamma_2 + \Gamma_3 + K_1 + K_2) \rho^{21}
& 2(\Gamma_2 + \Gamma_3) \rho^{22}
\end{array} \right],
\label{A9}
\end{eqnarray}
\end{widetext}
in which the $K_\alpha \rho^{\alpha\alpha} \hat n^\alpha$, $\alpha = 1,2$, terms cancel between the third line of (\ref{A6}) and the first term in the first line of (\ref{A7}). The mean field equation of motion (\ref{2.24}) now takes the form
\begin{equation}
\partial_t \hat \rho = i({\cal C}_D[\hat \rho] + {\cal C}_\Omega[\hat \rho]
+ {\cal C}_\Delta[\hat \rho]) + {\cal L}[\hat \rho],
\label{A10}
\end{equation}
which may be written out explicitly in the form (\ref{2.50}) given in the main text.

\section{Single photon, single Rydberg level critical point high temperature asymptotic forms}
\label{app:1ph1RyhighTasymp}

We first note the derivatives
\begin{eqnarray}
f'(\zeta) &=& -1 - \zeta f(\zeta)
\nonumber \\
f''(\zeta) &=& \zeta + (\zeta^2 - 1) f(\zeta).
\label{B1}
\end{eqnarray}
following immediately from (\ref{3.15}), which leads to
\begin{eqnarray}
\Phi_\nu'(w) &=& -\nu \mathrm{Im}[(w + i\nu) f(w + i\nu)]
\nonumber \\
\Phi_\nu''(w) &=& \nu^2 + \nu \mathrm{Im}\{[(w + i\nu)^2 - 1] f(w + i\nu) \}. \ \ \ \ \ \
\label{B2}
\end{eqnarray}
In the high temperature limit, $\nu \to 0$, one may use the above to Taylor expand
\begin{eqnarray}
f(w + i\nu) &=& f(w) + i\nu f'(w) - \frac{1}{2} \nu^2 f''(w) + O(\nu^3)
\nonumber \\
&=& f(w) - i\nu [1 + w f(w)]
\label{B3} \\
&&-\ \frac{1}{2} \nu^2 \left[w + (w^2 - 1) f(w) \right] + O(\nu^3),
\nonumber
\end{eqnarray}
with
\begin{eqnarray}
\mathrm{Im}[f(w)] &=& -\sqrt{\frac{\pi}{2}} e^{-w^2/2}
\nonumber \\
\mathrm{Re}[f(w)] &=& -\sqrt{\frac{\pi}{2}} e^{-w^2/2} F(w)
\nonumber \\
F(w) &=& \sqrt{\frac{2}{\pi}} \int_0^w e^{u^2/2} du.
\label{B4}
\end{eqnarray}
Taking the second derivative of this series, the critical line second derivative condition is
\begin{eqnarray}
0 &=& \mathrm{Im}[f''(w + i\nu)]
\nonumber \\
&=& (w^2 - 1) \mathrm{Im}[f(w)]
\label{B5} \\
&&+\ \nu \left\{2 - w^2 + w(3-w^2) \mathrm{Re}[f(w)] \right\} + O(\nu^2)
\nonumber
\end{eqnarray}
which may be put in the form
\begin{eqnarray}
w^2 &=& 1 + \nu \left[\sqrt{\frac{2}{\pi}} (2-w^2) e^{w^2/2} +  w(w^2 - 3) F(w) \right].
\nonumber \\
&&+\ O(\nu^2).
\label{B6}
\end{eqnarray}
For $\nu = 0$ one obtains $w_c = \pm 1$, and substituting this into the right hand side one obtains
\begin{eqnarray}
|w_c(\nu)| &=& 1 + \nu w_{c,1} + O(\nu^2)
\nonumber \\
w_{c,1} &=& F(1) - \sqrt{\frac{e}{2\pi}} \simeq 0.2956936.
\label{B7}
\end{eqnarray}
Using this result, one obtains
\begin{eqnarray}
\Phi_c(\nu) &=& \nu \left[\Phi_{c,1} + \frac{1}{2} \Phi_{c,2} \nu + O(\nu^2) \right]
\nonumber \\
|\Phi_c'(\nu)| &=& \nu \left[\Phi_{c,1} - \nu + O(\nu^2) \right]
\nonumber \\
\Phi_{c,1} &=& \sqrt{\frac{\pi}{2e}}
\nonumber \\
\Phi_{c,2} &=& -2 + \sqrt{\frac{2\pi}{2e}} [F(1) - w_{c,1}] = -1.
\label{B8}
\end{eqnarray}
along with
\begin{eqnarray}
A_3(\nu) &=& \nu^4 \Phi_{c,1} [\Phi_{c,1} - 2 \nu + O(\nu^2)]
\nonumber \\
C(\nu) &=& -2\Phi_{c,1} \nu + \frac{1}{2} C_2 \nu^2 + O(\nu^3)
\nonumber \\
C_2 &=& 2 - \Phi_{c,2} -  2 \Phi_{c,1} w_{c,1}
\nonumber \\
&=& 5 - \sqrt{\frac{2\pi}{e}} F(1) = 3.5504431.
\label{B9}
\end{eqnarray}
These asymptotic forms are verified in the right panel of Fig.\ \ref{fig:critlinedat}.

Finally, using these values, one obtains from (\ref{4.8}) the ``universally scaled'' forms
\begin{eqnarray}
\Delta_U &\equiv& \frac{\bar \Delta}{\sqrt{1+\bar \Omega^2}}
= \mathrm{sgn}(\alpha) \frac{C(\nu)}{\sqrt{A_3(\nu)}}
\nonumber \\
&=& -\mathrm{sgn}(\alpha)
\left[\frac{2}{\nu} + \frac{2 - C_2}{\Phi_{c,1}} + O(\nu) \right]
\nonumber \\
V_U &\equiv& \frac{\bar V_\mathrm{eff}}{a(\bar\Omega) \sqrt{1+\bar \Omega^2}}
= \frac{\mathrm{sgn}(\alpha)}{\sqrt{A_3(\nu)}}
\label{B10} \\
&=& \mathrm{sgn}(\alpha) \left[\frac{1}{\Phi_{c,1} \nu^2}
+ \frac{1}{\Phi_{c,1}^2 \nu} + O(1) \right]
\nonumber
\end{eqnarray}
These asymptotes are verified in the right panel of Fig.\ \ref{fig:clinephys} (black dashed lines).

\section{Dynamic linear response formalism}
\label{app:dynlinrespformal}

\subsection{General formulation}
\label{app:genform}

In order to formulate the general linear response theory, let ${\bf X}(t)$ represent a vector of $D$ dynamical variables, in this case a 1D list of density matrix elements $\rho_k^{lm}$, or the equivalent spin components $S_k^\alpha$ in the spin-$\frac{1}{2}$ formulation (\ref{3.3}), and write the Linblad equation of motion in the vector form
\begin{equation}
\partial_t {\bf X}(t) = {\bf H}[{\bf X}(t);{\bf P}(t)]
\label{C1}
\end{equation}
in which various instantiations of ${\bf H}$ will be discussed below, and ${\bf P}$ is a vector of $d$ controllable time-dependent parameters. We linearize the equations of motion, assuming small perturbations
\begin{equation}
\delta {\bf X}(t) = {\bf X}(t) - {\bf X}_0,\ \
\delta {\bf P}(t) = {\bf P}(t) - {\bf P}_0
\label{C2}
\end{equation}
in which ${\bf X}_0, {\bf P}_0$ satisfy the stationary condition
\begin{equation}
{\bf H}({\bf X}_0;{\bf P}_0) = 0,
\label{C3}
\end{equation}
in the present case corresponding to the steady state mean field equations analyzed in the main body of the paper. Dropping nonlinear terms, one obtains the linear response equation
\begin{equation}
(\partial_t  - {\bf G}_0) \delta {\bf X}(t)
= {\bf F}({\bf X}_0;{\bf P}_0) \delta {\bf P}(t)
\label{C4}
\end{equation}
in which
\begin{eqnarray}
G_{0,lm} &=& \partial_{X_m} H_l({\bf X}_0;{\bf P}_0),\ \ l,m = 1,2,\ldots,D
\nonumber \\
F_{0,l\mu} &=& \partial_{P_\mu} H_l({\bf X}_0;{\bf P}_0),\ \ \mu = 1,2,\ldots,d
\label{C5}
\end{eqnarray}
are $D \times D$ and $D \times d$ response and forcing matrices, respectively.

The formal solution takes the form
\begin{eqnarray}
\delta {\bf X}(t) &=& \int_{-\infty}^t dt'
{\bm \Gamma}(t-t') {\bf F}_0 \delta {\bf P}(t')
\nonumber \\
{\bm \Gamma}(t) &=& \theta(t) e^{{\bf G}_0 t}.
\label{C6}
\end{eqnarray}
The Heaviside function, $\theta(s) = 1$ for $s \geq 0$, vanishing otherwise, explicitly enforces causality of the $D \times D$ Green tensor ${\bm \Gamma}$. If ${\bf G}_0$ is assumed to be diagonalizable,
\begin{equation}
{\bf G}_0 {\bf u}^R_l = \lambda_l {\bf u}^R_l,\ \
({\bf u}^L_l)^\dagger {\bf G}_0 = \lambda_l ({\bf u}^L_l)^\dagger
\label{C7}
\end{equation}
with, in general, complex eigenvalues $\lambda_l$, and with left and right eigenvectors ${\bf u}^{L,R}_k$ normalized to obey the orthogonality conditions
\begin{equation}
({\bf u}^L_l)^\dagger {\bf u}^R_m = \delta_{lm},
\label{C8}
\end{equation}
one obtains the decompositions
\begin{eqnarray}
{\bf G}_0 &=& \sum_{l=1}^D \lambda_l {\bf u}_l^R ({\bf u}_l^L)^\dagger
\nonumber \\
{\bm \Gamma}(t) &=& \theta(t) \sum_{l=1}^D e^{\lambda_l t}
{\bf u}_l^R ({\bf u}_l^L)^\dagger
\label{C9}
\end{eqnarray}
with $\mathrm{Re}(\lambda_l) \leq 0$ for the stable dissipative systems under consideration here.

The Fourier space response, equivalently corresponding to the pure tones
\begin{equation}
\delta {\bf P}(t) = \delta \hat {\bf P}(\omega) e^{-i\omega t},\ \
\delta {\bf X}(t) = \delta \hat {\bf X}(\omega) e^{-i\omega t},
\label{C10}
\end{equation}
takes the form
\begin{equation}
\delta \hat {\bf X}(\omega) = \hat {\bm \Gamma}(\omega) {\bf F}_0  \delta \hat {\bf P}(\omega)
\label{C11}
\end{equation}
with frequency domain Green tensor
\begin{equation}
\hat {\bm \Gamma}(\omega) = -(i\omega + {\bf G}_0)^{-1}
= -\sum_{l=1}^D \frac{{\bf u}_l^R ({\bf u}_l^L)^\dagger}{i\omega + \lambda_l}.
\label{C12}
\end{equation}
The response clearly contains a sharply peaked contribution, centered on $\omega = -\mathrm{Im}(\lambda_l)$, if the dissipation is weak, $|\mathrm{Re}(\lambda_l)| \ll 1$. Note that the operator inverse here could performed without the diagonalization, but the latter serves to highlight the multi-Lorentzian character of the response.

\subsection{Mean field approximation and thermal averages}
\label{app:MFthermave}

In the mean field approximation the equation of motion (\ref{C1}) is reduced to the set of one-body equations
\begin{equation}
\partial_t {\bf X}_k = {\bf H}_k \equiv {\bf h}({\bf X}_k,\bar {\bf X};{\bf P}_k)
\label{C13}
\end{equation}
in which only the degrees of freedom for the same single atom $k$ appear, and
\begin{equation}
\bar {\bf X}(t) = \frac{1}{N_a} \sum_{k=1}^{N_a} {\bf X}_k(t)
\label{C14}
\end{equation}
represents the much smaller set of mean values used to approximate the full interaction terms in (\ref{C1}). In the applications of interest here, ${\bf h}$ actually depends only a further subset of the latter, corresponding to the Rydberg level populations. The linear response matrix elements now take the form
\begin{eqnarray}
{\bf G}_{0,kl} &=& \frac{\partial {\bf H}_k}{\partial {\bf X}_l}
= {\bf G}_{0,k} \delta_{kl} + \frac{1}{N_a} {\bf G}_{\mathrm{MF},k}
\nonumber \\
{\bf G}_{0,k} &=& \frac{\partial {\bf H}_k}{\partial {\bf X}_k},\ \
{\bf G}_{\mathrm{MF},k} = \frac{\partial {\bf H}_k}{\partial \bar {\bf X}},
\label{C15}
\end{eqnarray}
with forcing tensor
\begin{equation}
{\bf F}_{0,kl} = {\bf F}_{0,k} \delta_{kl},\ \
{\bf F}_{0,k} = \frac{\partial {\bf H}_k}{\partial {\bf P}_k}.
\label{C16}
\end{equation}
The individual atom linear response equation is
\begin{equation}
(\partial_t - {\bf G}_{0,k}) \delta {\bf X}_k
- {\bf G}_{\mathrm{MF},k} \delta \bar {\bf X}
= {\bf F}_{0,k} \delta {\bf P}_k,
\label{C17}
\end{equation}
and its frequency domain solution is
\begin{eqnarray}
\delta \hat {\bf X}_k(\omega) &=& \hat {\bm \Gamma}_k(\omega)
[{\bf F}_{0,k} \delta \hat {\bf P}_k(\omega) + {\bf G}_{\mathrm{MF},k} \delta \hat {\bf X}(\omega)]
\nonumber \\
\hat {\bm \Gamma}_k(\omega) &=& -(i\omega + \hat {\bf G}_{0,k})^{-1},
\label{C18}
\end{eqnarray}
in which $\hat {\bf X}(\omega)$ is the Fourier transform of $\bar {\bf X}(t)$. This result produces $\delta \hat {\bf X}_k(\omega)$ once its average, appearing on the right hand side, is known. The desired closed equation is obtained by averaging both sides. Defining the single atom mean response matrix
\begin{equation}
{\bm \Sigma}(\omega) = \left\langle \hat {\bm \Gamma}_k(\omega)
{\bf G}_{\mathrm{MF},k} \right\rangle
\label{C19}
\end{equation}
and mean forcing vector
\begin{equation}
\delta \hat {\bf f}(\omega) = \left\langle \hat {\bm \Gamma}_k(\omega)
{\bf F}_{0,k} \delta \hat {\bf P}_k(\omega) \right\rangle
\label{C20}
\end{equation}
one obtains the solution
\begin{equation}
\delta \hat {\bf X}(\omega)
= [\openone - {\bm \Sigma}(\omega)]^{-1} \delta \hat {\bf f}(\omega).
\label{C21}
\end{equation}
For spatially uniform driving $\delta {\bf P}_k = \delta {\bf P}$ one obtains the simpler form
\begin{equation}
\delta \hat {\bf f}(\omega) = \left\langle \hat {\bm \Gamma}_k(\omega)
{\bf F}_{0,k} \right\rangle \delta \hat {\bf P}(\omega).
\label{C22}
\end{equation}

\subsection{Specialization to multi-photon models}
\label{app:genmultiph}

For the models of central interest in this paper, the remaining $k$-dependence in (\ref{C22}) is produced by the atom-dependent Doppler shifts $\Delta_{\sigma,k} = \Delta_\sigma + {\bf k}_\sigma \cdot {\bf v}_k$ of the detuning $\Delta_\sigma$ of laser beam $\sigma$, in general extending (\ref{2.2}) and (\ref{2.5}) to cases of $N_L$ atomic levels and a ``tree'' of $\sigma = 1, 2, \ldots, N_L-1$ illuminators. This leads to nontrivially varying ${\bf X}_{0,k}$ and to nontrivial mean (Gaussian average) products in (\ref{C19}), (\ref{C20}), and (\ref{C22}). For some applications, the time dependent forcing may include the perturbation $\Delta_\sigma(t) = \Delta_\sigma + \delta \Delta_\sigma(t)$ (AC Stark effect), in which $\delta \Delta_\sigma$ is included as an element of the driving vector $\delta {\bf P}$.

We specialize the results of the previous section to models of this type, making much more explicit the types of statistical averages that need to be performed. We consider equations of motion (\ref{C13}) of the form
\begin{equation}
{\bf h}({\bf X}_k,\bar {\bf X},{\bf P}) = [{\bf A}(\bar {\bf X},{\bf P})
+ {\bf v}_k \cdot {\bf B}]{\bf X}_k + {\bf b}(\bar {\bf X},{\bf P}),
\label{C23}
\end{equation}
hence linear in both ${\bf X}_k$ and the random velocities components ${\bf v}_k$. The inhomogeneous term ${\bf b}$ arises from the probability constraint ${\bf p} \cdot {\bf X}_k \equiv \sum_{\alpha = 1}^{N_L} \rho_k^{\alpha \alpha} = 1$ used to eliminate the ground state occupancy $X_{k,0} = \rho_k^{00}$, hence reducing the original $N_L^2$ homogeneous equations to $D = N_L^2-1$ inhomogeneous equations. For example, looking at (\ref{2.50}), one sees that $\rho^{00}$ appears in only one place, producing the very sparse form $b^{(\alpha \beta)} = \frac{i}{2} \Omega (\delta_{\alpha 0} \delta_{\beta 1} - \delta_{\alpha 1} \delta_{\beta 0})$, which could be time-dependent if $\Omega$ is. Here and below, the notation $(\alpha \beta)$ indicates a combined index for the $D$ dynamical variables in the density matrix application. In all of the applications treated in this paper ${\bf b}$ is therefore independent of $\bar {\bf X}$, but we keep it here for generality.

The dependence of the $D \times D$ matrix ${\bf A}$ on the mean values $\bar {\bf X}$ in our applications will generally be restricted to the one or two Rydberg level occupancies, and these also appear only linearly in the interaction parameters (\ref{2.26}), but we do not specialize to this case yet.

Each element of the $3 \times D \times D$ array ${\bf B}$ is assumed here to be independent of $\bar {\bf X}$, ${\bf P}$, given by a fixed linear combination illumination wavevector components---extracted from the matrix ${\cal C}_\Delta$ in (\ref{A5}) and (\ref{A10}) for the $N_L = 3$ case, and explicitly written out in (\ref{2.50}). In the latter case, the vector ${\bf B}^{(\alpha\beta),(\mu\nu)} = i {\bf k}^{(\alpha\beta)} \delta_{(\alpha \beta),(\mu \nu)}$ is actually diagonal in the dynamical variable indices, with remaining vector structure defined by some physical wavevector linear combination ${\bf k}^{(\alpha\beta)}$ associated with each dynamical variable.

\subsection{Steady state solution and mean field equations}
\label{app:steadymfeq}

The steady state solution (\ref{C3}) is now given explicitly by
\begin{equation}
{\bf X}_{0,k} = -[{\bf A}(\bar {\bf X}_0,{\bf P}_0) + {\bf v}_k \cdot {\bf B}]^{-1} {\bf b}_0,
\label{C24}
\end{equation}
resulting in the mean field equation
\begin{equation}
\bar {\bf X}_0 = -\left \langle[{\bf A}(\bar {\bf X}_0,{\bf P}_0)
+ {\bf v} \cdot {\bf B}]^{-1} \right \rangle_{\bf v} {\bf b}_0,
\label{C25}
\end{equation}
with Maxwell distribution average $\langle\, \cdot\, \rangle_{\bf v}$. The matrix inverse here is a ratio of determinants of $D \times D$ matrices, hence of polynomials in the components of ${\bf v}$ of degree at most $D$. With some further simplifications, this will form the basis below for the exact evaluation of the Gaussian averages.

\subsection{Dynamic linear response}
\label{app:dynlinresp}

The frequency domain linear response equation, obtained by linearizing (\ref{C23}) about the steady state mean field solution, takes the form
\begin{eqnarray}
(\partial_t - {\bf A}_0 - {\bf v}_k \cdot {\bf B}) \delta {\bf X}_k
&=& (\delta \bar {\bf X} \cdot \nabla_{\bf \bar X}
+ \delta {\bf P} \cdot \nabla_{\bf P}) {\bf A}_0 {\bf X}_{0,k}
\nonumber \\
&&+\ (\delta \bar {\bf X} \cdot \nabla_{\bf \bar X}
+ \delta {\bf P} \cdot \nabla_{\bf P}) {\bf b}_0
\nonumber \\
\label{C26}
\end{eqnarray}
in which the subscript 0 indicates evaluation of at the mean field solution following any derivative operation acting on ${\bf A}$ or ${\bf b}$ (with ${\bf X}_{0,k}$ unaffected here). The frequency domain solution is
\begin{widetext}
\begin{eqnarray}
\delta \hat {\bf X}_k(\omega)
&=& -(i\omega \openone + {\bf A}_0 + {\bf v}_k \cdot {\bf B})^{-1}
[\delta \bar {\bf X}(\omega) \cdot \nabla_{\bf \bar X}
+ \delta \hat {\bf P}(\omega) \cdot \nabla_{\bf P}]\left({\bf A}_0 {\bf X}_{0,k} + {\bf b}_0 \right)
\nonumber \\
&=& \left(i\omega \openone + {\bf A}_0 + {\bf v}_k \cdot {\bf B} \right)^{-1}
[\delta \bar {\bf X}(\omega) \cdot \nabla_{\bf \bar X}
+ \delta \hat {\bf P}(\omega) \cdot \nabla_{\bf P}]{\bf A}_0
({\bf A}_0 + {\bf v}_k \cdot {\bf B})^{-1} {\bf b}_0
\nonumber \\
&&-\  \left(i\omega \openone + {\bf A}_0 + {\bf v}_k \cdot {\bf B} \right)^{-1}
[\delta \bar {\bf X}(\omega) \cdot \nabla_{\bf \bar X}
+ \delta \hat {\bf P}(\omega) \cdot \nabla_{\bf P}]{\bf b}_0
\label{C27}
\end{eqnarray}
in which in the second line the gradients act only on the ${\bf A}$ immediately adjacent to the square bracket. Defining the averaged two and four index tensors
\begin{equation}
g^{pq} = -\left\langle \left[(i\omega \openone + {\bf A}_0 + {\bf v} \cdot {\bf B})^{-1}
\right]^{pq} \right\rangle_{\bf v},\ \ \
G^{pq;rs} = \left\langle \left[(i\omega \openone + {\bf A}_0 + {\bf v} \cdot {\bf B})^{-1} \right]^{pq}
\left[({\bf A}_0 + {\bf v} \cdot {\bf B})^{-1} \right]^{rs} \right\rangle_{\bf v},
\label{C28}
\end{equation}
\end{widetext}
with all indices running from 1 to $D$, and the derived arrays
\begin{eqnarray}
\Gamma_A^{pu} &=& \sum_{q,r,s} G^{pq;rs}
[\partial_{\bar X_u} {\bf A}]_0^{qr} b_0^s
\nonumber \\
F_A^{p \lambda} &=& \sum_{q,r,s} G^{pq;rs}
[\partial_{P_\lambda} {\bf A}]_0^{qr} b_0^s
\nonumber \\
\Gamma_b^{pu} &=& \sum_q g^{pq}
[\partial_{\bar X_u} {\bf b}]_0^q
\nonumber \\
F_b^{p \lambda} &=& \sum_q g^{pq}
[\partial_{P_\lambda} {\bf b}]_0^q,
\label{C29}
\end{eqnarray}
in which the distinguished index $\lambda$ runs over the number of driving parameters, one obtains the averaged equation
\begin{eqnarray}
(\openone - {\bm \Gamma}_A - {\bm \Gamma}_b) \delta \bar {\bf X}(\omega)
= ({\bf F}_A + {\bf F}_b) \delta \hat {\bf P}(\omega),
\label{C30}
\end{eqnarray}
with solution
\begin{equation}
\delta \bar {\bf X}(\omega) = (\openone - {\bm \Gamma}_A - {\bm \Gamma}_b)^{-1}
({\bf F}_A + {\bf F}_b)  \delta \hat {\bf P}(\omega).
\label{C31}
\end{equation}
The ${\bm \Gamma}$ arrays are nonzero only for column indices $u$ corresponding to the subset of averaged parameters $\bar {\bf X}$ that actually appear inside ${\bf A}$, hence appearing on the right hand side of the mean field equation (\ref{C25}). Distinguishing this subspace by superscript `$(1)$', hence writing $\delta \bar {\bf X} = (\delta \bar {\bf X}^{(1)}, \delta \bar {\bf X}^{(2)})$, the resulting block form of (\ref{C31}) simplifies the solution to
\begin{eqnarray}
\delta \bar {\bf X}^{(1)}
&=& (\openone - {\bm \Gamma}^{(11)}_A - {\bm \Gamma}^{(11)}_b)^{-1}
({\bf F}^{(1)}_A + {\bf F}^{(1)}_b)  \delta \hat {\bf P}.
\nonumber \\
\delta \bar {\bf X}^{(2)}
&=& ({\bm \Gamma}^{(21)}_A + {\bm \Gamma}^{(21)}_b) \delta \bar {\bf X}^{(1)}
+ ({\bf F}^{(2)}_A + {\bf F}^{(2)}_b)  \delta \hat {\bf P}.
\nonumber \\
\label{C32}
\end{eqnarray}
in which ${\bm \Gamma}^{(11)}$ and ${\bm \Gamma}^{(21)}$ are the corresponding nonvanishing blocks of ${\bm \Gamma}$, and the ${\bf F}$ matrix superscripts designate the corresponding sets of rows. The first line is substituted into the right hand side of the second.

\subsection{Exact partial fraction evaluation of Gaussian averages}
\label{app:partfracave}

In order to obtain explicit forms for the averages we specialize to the case of collinear laser beams, all parallel or antiparallel to some direction $\hat {\bf k}$, so that only a single component $u_k = \hat {\bf k} \cdot {\bf v}_k$ appears in ${\bf v}_k \cdot {\bf B} = v_k \hat {\bf k} \cdot {\bf B}$. The averages defining the mean field equation (\ref{C25}) and dynamical linear response equation (\ref{C28}) then reduce to Gaussian integrals of ratios of polynomials in $u$ of degree at most $D$ and $2D$, respectively.

To define the general approach, consider the following Gaussian integrals
\begin{equation}
F_{12} = \int \frac{du}{\sqrt{2\pi}} e^{-u^2/2} \frac{P_1(u)}{P_2(u)},
\label{C33}
\end{equation}
written for notational convenience in terms of $u = v/v_\mathrm{th}$. By polynomial division, and subtraction of a simple polynomial if necessary, we may assume that the degree of $P_1$ is less than that of $P_2$ and that they share no common polynomial factor. Factorizing
\begin{equation}
P_2(u) = P_0 \prod_{j=1}^L (u - u_j)^{m_j},
\label{C34}
\end{equation}
in which $m_j$ is the multiplicity of root $u_j$, the method of partial fractions then allows one to express
\begin{equation}
\frac{P_1(u)}{P_2(u)} = \sum_{j=1}^L \sum_{\alpha = 1}^{m_j} \frac{A_{j\alpha}}{(u-u_j)^\alpha}
\label{C35}
\end{equation}
in which $A_{j\alpha}$ are constant coefficients given explicitly by
\begin{eqnarray}
A_{j\alpha} &=& \frac{1}{(m_j-\alpha)!}\partial_{u}^{m_j-\alpha}
\left[\frac{P_1(u)}{P^{(j)}_2(u)} \right]_{u=u_j}
\nonumber \\
P_2^{(j)}(u) &\equiv& \frac{P_2(u)}{(u-u_j)^{m_j}} = P_0 \prod_{k \neq j} (u-u_k)^{m_k},\ \ \ \ \ \
\label{C36}
\end{eqnarray}
corresponding to a local Taylor expansion of the nonsingular terms about each root $u_j$. It follows that $F_{12}$ may be expanded in the form
\begin{equation}
F_{12} = \sum_{j=1}^L \sum_{\alpha=1}^{m_j} \frac{A_{j\alpha}}{(\alpha-1)!}
\partial_\zeta^{\alpha-1} f(\zeta)|_{\zeta = u_j}
\label{C37}
\end{equation}
with basic integral
\begin{equation}
f(\zeta) = \int \frac{du}{\sqrt{2\pi}} \frac{e^{-u^2/2}}{u - \zeta},
\label{C38}
\end{equation}
identical to (\ref{3.12}) and with explicit error function form evaluation (\ref{3.14}). The $\zeta$-derivatives, if needed, follow via iteration of the obvious identity
\begin{equation}
\partial_\zeta f(\zeta) = -\zeta f(\zeta) - 1.
\label{C39}
\end{equation}

The roots of the denominator in (\ref{C33}) are not in general known analytically, hence numerical solution is required (the single photon special case described in App.\ \ref{app:1ph1Rydynexact} being an exception). However, factorization algorithms are extremely efficient, while subsequent exact evaluation of the Gaussian integral hugely speeds up the evaluations required for subsequent numerical solution of the mean field equations (Secs.\ \ref{sec:2phmfsoln}, \ref{sec:2Ryresonant}).

\section{Exact dynamical linear response forms for the single photon, single Rydberg level case}
\label{app:1ph1Rydynexact}

We now specialize that general results derived in Apps.\ \ref{app:genmultiph}--\ref{app:partfracave} to the single photon, single Rydberg level case [Fig.\ \ref{fig:atomsetup}(a)] and which forms the basis for results shown in Sec.\ \ref{sec:dynlinresp1}. Using the spin representation ${\bf X}_k = {\bf S}_k$, defined by (\ref{3.3}) along with $S_k^z = 2n_k - 1$, and ${\bf P}_k = (\Delta_k, \Omega_k^x, \Omega_k^y)$, comparing (\ref{3.1}) to (\ref{C23}) one obtains
\begin{eqnarray}
h^x({\bf X},\bar S^z;{\bf P}) &=& -(\Delta + \bar n V_\mathrm{eff}) S^y
- \frac{\Gamma + K}{2} S^x + \Omega^y S^z
\nonumber \\
h^y({\bf X},\bar S^z;{\bf P}) &=& (\Delta + \bar n V_\mathrm{eff}) S^x
- \frac{\Gamma + K}{2} S^y - \Omega^x S^z
\nonumber \\
h^z({\bf X},\bar S^z;{\bf P}) &=& -\Omega^y S^x + \Omega^x S^y - \Gamma (S^z + 1),
\label{D1}
\end{eqnarray}
with $\bar n = \frac{1}{2}(1 + \bar S^z)$, which leads to arrays
\begin{eqnarray}
{\bf A}(\bar {\bf X},{\bf P}) &=& \left(\begin{array}{ccc}
- \frac{\Gamma + K}{2} & -(\Delta + \bar n V_\mathrm{eff}) & \Omega^y \\
\Delta + \bar n V_\mathrm{eff} & - \frac{\Gamma + K}{2} & -\Omega^x \\
-\Omega^y & \Omega^x & -\Gamma
\end{array} \right)
\nonumber \\
\hat {\bf k} \cdot {\bf B} &=& \left(\begin{array}{ccc}
0 & -k_P & 0 \\
k_P & 0 & 0 \\
0 & 0 & 0
\end{array} \right),\ \
{\bf b} = \left(\begin{array}{c}
0 \\ 0 \\ -\Gamma
\end{array} \right).
\label{D2}
\end{eqnarray}
From (\ref{C26}) one obtains
\begin{eqnarray}
\delta \bar {\bf X} \cdot \nabla_{\bar {\bf X}} {\bf A}
&=& \frac{1}{2} \delta \bar S^z
\left(\begin{array}{ccc}
0 & -V_\mathrm{eff} & 0 \\
V_\mathrm{eff} & 0 & 0 \\
0 & 0 & 0
\end{array} \right)
\nonumber \\
\delta {\bf P} \cdot \nabla_{\bf P} {\bf A}
&=& \left(\begin{array}{ccc}
0 & -\delta \Delta & \delta \Omega^y \\
\delta \Delta & 0 & -\delta \Omega^x  \\
-\delta \Omega^y  & \delta \Omega^x  & 0
\end{array} \right).
\label{D3}
\end{eqnarray}

The steady state solution (\ref{C24}) is obtained from the last column (due to the very simple form of ${\bf b}$) of the inverse of the matrix ${\bf A} - v_k \hat {\bf k} \cdot {\bf B}$ and reproduces the solution (\ref{3.6}) in the form
\begin{eqnarray}
S^z(v) &=& \frac{2}{a} \frac{1}{1 + (u_{\bar n} + u)^2/\nu^2} - 1
\nonumber \\
\hat {\bf \Omega}^\perp \cdot {\bf S}(v) &=& \frac{2\Gamma}{a |\Omega|} \frac{1}{1 + (u_{\bar n} + u)^2/\nu^2}
\nonumber \\
\hat {\bf \Omega} \cdot {\bf S}(v) &=& -\frac{4\Gamma k_Pv_\mathrm{th}}{a|\Omega|(\Gamma + K)}
\frac{u_{\bar n} + u}{1 + (u_{\bar n} + u)^2/\nu^2}, \ \ \ \ \ \
\label{D4}
\end{eqnarray}
in which $u = v/v_\mathrm{th}$, and recalling also the definitions (\ref{3.7}), (\ref{3.9}), and (\ref{3.10}).

Moving on now to the dynamical forms, it is useful to extend the dimensionless combinations (\ref{4.7}) to the frequency-dependent forms
\begin{eqnarray}
&&\{\tilde \Delta_k(\omega), \delta \tilde \Delta_k(\omega),
\tilde V_\mathrm{eff}(\omega) \}
= \frac{\{\Delta_k, \delta \Delta_k, V_\mathrm{eff}\}}{-i\omega + \frac{\Gamma + K}{2}}
\nonumber \\
&&\{\tilde \Omega_k(\omega), \delta \tilde \Omega_k(\omega) \}
= \frac{\{|\Omega_k|, \delta \Omega_k\}}
{\sqrt{\left(-i\omega + \frac{\Gamma + K}{2} \right)(-i\omega + \Gamma)}}
\nonumber \\
&&\{\tilde S_k^{x,y}(\omega), \delta \tilde S_k^{x,y}(\omega) \}
= \tilde \gamma(\omega) \{S_k^{x,y}(\omega), \delta S_k^{x,y}(\omega) \}
\nonumber \\
&&\{\tilde S_k^z, \delta \tilde S_k^z\} = \{S_k^z, \delta S_k^z\},\ \
\tilde \gamma(\omega) = \sqrt{\frac{-i\omega + \frac{\Gamma + K}{2}}{-i\omega + \Gamma}}.
\nonumber \\
\label{D5}
\end{eqnarray}
In terms of these, the frequency domain solution (\ref{C27}) may be put in the compact form
\begin{equation}
\delta \tilde {\bf S}_k = \tilde {\bm \Gamma}_k \left[\tilde {\bf F}_{0,k}
\left(\begin{array}{c}
\delta \tilde \Delta_k \\
\delta \tilde \Omega^x_k \\
\delta \tilde \Omega^y_k
\end{array} \right)
+ \frac{\tilde V_\mathrm{eff}}{2}
\left(\begin{array}{c} -\tilde S_k^y \\ \tilde S_k^x \\ 0 \end{array} \right)
\delta \tilde S^z \right]
\label{D6}
\end{equation}
in which the second term in the square brackets originates from the first line of (\ref{D3}) and the first term, defined by the scaled array
\begin{equation}
\tilde {\bf F}_{0,k} = \left(\begin{array}{ccc}
-\tilde S_k^y & 0  & \tilde S_k^z \\
\tilde S_k^x & -\tilde S_k^z & 0\\
0 & \tilde S_k^y & -\tilde S_k^x
\end{array} \right),
\label{D7}
\end{equation}
originates from the second line of (\ref{D3}), both multiplying the steady state solution $\tilde {\bf S}_k$ obtained from (\ref{D4}). The scaling (\ref{D5}) produces the mapping
\begin{equation}
-(i\omega \openone + {\bf A}^0 + {\bf v}_k \cdot {\bf B})^{-1}\ \to\ \ \tilde {\bm \Gamma}_k
= (\openone - \tilde {\bf G}_{0,k})^{-1}.
\label{D8}
\end{equation}
At this point, we simplify the notation by defining the natural coordinate system for each atom in which the $x$-axis is along the steady state direction $\hat {\bm \Omega}_k$, hence
\begin{eqnarray}
\Omega_k^x = |\Omega_k|,&& \Omega_k^y = 0
\nonumber \\
\hat {\bf \Omega}_k \cdot {\bf S}_k = S^x_k, &&
\hat {\bf \Omega}_k^\perp \cdot {\bf S}_k = S^y_k.
\label{D9}
\end{eqnarray}
The perturbation $\delta {\bm \Omega}_k$, however, remains unconstrained. With this choice one obtains
\begin{equation}
\tilde {\bf G}_{0,k} = \left(\begin{array}{ccc}
0 & -\tilde \Delta_k & 0 \\
\tilde \Delta_k & 0 & -|\tilde \Omega_k| \\
0 & |\tilde \Omega_k| & 0
\end{array} \right)
\label{D10}
\end{equation}
is the scaled version of ${\bf A}^0 + {\bf v}_k \cdot {\bf B}$, and the inverse takes the explicit form
\begin{equation}
\tilde {\bm \Gamma}_k = \frac{1}{\tilde P_k} \left(\begin{array}{ccc}
1 + |\tilde \Omega_k|^2 & -\tilde \Delta_k & \tilde \Delta_k |\tilde \Omega_k| \\
\tilde \Delta_k & 1 & -|\tilde \Omega_k| \\
\tilde \Delta_k |\tilde \Omega_k|
& |\tilde \Omega_k| & 1 + \tilde \Delta_k^2
\end{array} \right),
\label{D11}
\end{equation}
with determinant factor
\begin{equation}
\tilde P_k = 1 + \tilde \Delta_k^2 + |\tilde \Omega_k|^2.
\label{D12}
\end{equation}

Assuming now that the only $k$-dependence enters through the velocity $\Delta_k = \Delta(v_k) = \Delta + k_P v_k$ one obtains
\begin{eqnarray}
\tilde \Delta(v) &=& \tilde \Delta_\mathrm{Dopp}(u_{\bar n} + u)
\nonumber \\
\tilde P(v) &=& (1 + \tilde \Omega^2) [1 + (u_{\bar n} + u)^2/\tilde \nu^2]
\nonumber \\
\tilde \Delta_\mathrm{Dopp} &=& \frac{k_P v_\mathrm{th}}{-i\omega + \frac{\Gamma + K}{2}}
\nonumber \\
\tilde \nu &=& \frac{\sqrt{1 + \tilde \Omega^2}}{\tilde \Delta_\mathrm{Dopp}}.
\label{D13}
\end{eqnarray}

Examining the various products appearing in (\ref{D6}), based on all of the previous results, the thermal average of the linear response equation (\ref{D6}) reduces to evaluation of averages of the form
\begin{eqnarray}
{\bf F}_0 &=& \left\langle \frac{\tilde {\bf S}(v)}{\tilde P(v)} \right\rangle
\nonumber \\
{\bf F}_1 &=& \left\langle \frac{\tilde \Delta(v) \tilde {\bf S}(v)}{\tilde P(v)} \right\rangle
\nonumber \\
{\bf F}_2 &=& \left\langle \frac{\tilde \Delta(v)^2 \tilde {\bf S}(v)}{\tilde P(v)} \right\rangle
= \tilde {\bf S} - (1 + |\tilde \Omega|^2){\bf F}_0. \ \ \ \ \ \
\label{D14}
\end{eqnarray}
whose arguments are ratios of polynomials in $u$, with denominators of degree 4 and numerator at most of degree 3. One obtains
\begin{eqnarray}
\left(\openone - \frac{1}{2} \tilde V_\mathrm{eff}
\tilde {\bm \Sigma} \hat {\bf z}^T \right) \delta \tilde {\bf S}
= \tilde {\bf f} \left(\begin{array}{c}
\delta \tilde \Delta \\ \delta \tilde \Omega^x \\ \delta \tilde \Omega^y
\end{array} \right)
\label{D15}
\end{eqnarray}
with explicit solution
\begin{eqnarray}
\delta \tilde {\bf S} &=& \left[\openone
- \frac{1}{2} \tilde V_\mathrm{eff} \tilde {\bm \Sigma} \hat {\bf z}^T \right]^{-1}
\tilde {\bf f}
\left(\begin{array}{c}
\delta \tilde \Delta \\ \delta \tilde \Omega^x \\ \delta \tilde \Omega^y
\end{array} \right)
\nonumber \\
&=& \left[\openone + \frac{\frac{1}{2} \tilde V_\mathrm{eff}}
{1 - \frac{1}{2} \tilde V_\mathrm{eff} \tilde \Sigma^z}
\tilde {\bm \Sigma} \hat {\bf z}^T \right] \tilde {\bf f}
\left(\begin{array}{c}
\delta \tilde \Delta \\ \delta \tilde \Omega^x \\ \delta \tilde \Omega^y
\end{array} \right).
\label{D16}
\end{eqnarray}
The thermally averaged response and forcing arrays are given by
\begin{widetext}
\begin{eqnarray}
\tilde {\bm \Sigma} &=& \left\langle \tilde {\bm \Gamma}(v)
\left[\begin{array}{c}
-\tilde S^y(v) \\ \tilde S^x(v) \\ 0
\end{array} \right] \right\rangle
= \left[\begin{array}{ccc}
- F_1^x - (1+|\tilde \Omega|^2) F_0^y \\
F_0^x - F_1^y \\
|\tilde \Omega| (F_0^x - F_1^y)
\end{array} \right]
\nonumber \\
\tilde {\bf f} &=& \langle \tilde {\bm \Gamma}(v) \tilde {\bf F}_0(v) \rangle
= \left[\begin{array}{ccc}
-(1 + |\tilde \Omega|^2) F_0^y - F_1^x & F_1^z + |\tilde \Omega| F_1^y
& (1 + |\tilde \Omega|^2) F_0^z - |\tilde \Omega| F_1^x \\
F_0^x - F_1^y & -F_0^z  - |\tilde \Omega| F_0^y
& F_1^z + |\tilde \Omega| F_0^x \\
|\tilde \Omega| (F_0^x - F_1^y) & - |\tilde \Omega| F_0^z + F_0^y + F_2^y
& |\tilde \Omega| F_1^z - F_0^x - F_2^x
\end{array} \right].
\label{D17}
\end{eqnarray}
\end{widetext}

Finally, obtaining closed forms for (\ref{D9}) reduce to Gaussian integrals of the form
\begin{equation}
\Phi^{(l)}_{\nu \tilde\nu}(w) = \int \frac{du}{\sqrt{2\pi}}
\frac{e^{-u^2/2} (w-u)^l}{[1 + (w-u)^2/\nu^2][1 + (w-u)^2/\tilde \nu^2]}
\label{D18}
\end{equation}
for $l = 0,1,2$. As described in App.\ \ref{app:partfracave}, one obtains via partial fraction expansion (in this case, fully analytically factorable)
\begin{eqnarray}
\Phi^{(0)}_{\nu\tilde\nu}(w)
&=& \frac{\tilde \nu^2 \phi^{(0)}_\nu(w) - \nu^2 \phi^{(0)}_{\tilde \nu}(w)}
{\tilde \nu^2 - \nu^2}
\nonumber \\
\Phi^{(1)}_{\nu\tilde\nu}(w)
&=& -\frac{\tilde \nu^2 \phi^{(1)}_\nu(w)
- \nu^2 \phi^{(1)}_{\tilde \nu}(w)}{\tilde \nu^2 - \nu^2}
\nonumber \\
\Phi^{(2)}_{\nu\tilde\nu}(w)
 &=& -(\nu \tilde \nu)^2 \frac{\phi^{(0)}_\nu(w)
- \phi^{(0)}_{\tilde \nu}(w)}{\tilde \nu^2 - \nu^2}.
\label{D19}
\end{eqnarray}
in which
\begin{eqnarray}
\phi^{(0)}_{\tilde \nu}(w) &=& \frac{\tilde \nu}{2i}
[f(w + i\tilde \nu) - f(w - i\tilde\nu)]
\nonumber \\
\phi^{(1)}_{\tilde \nu}(w) &=& \frac{\tilde \nu^2}{2}
[f(w + i\tilde \nu) + f(w - i\tilde\nu)]
\nonumber \\
\phi^{(0)}_\nu(w) &=& \phi^{(0)}_{\tilde \nu}(w)|_{\omega = 0}
= \nu \mathrm{Im}[f(w + i\nu)]
\nonumber \\
\phi^{(1)}_\nu(w) &=& \phi^{(1)}_{\tilde \nu}(w)|_{\omega = 0}
= \nu^2 \mathrm{Re}[f(w + i\nu)].\ \ \ \ \ \
\label{D20}
\end{eqnarray}

The (mean field, steady state) averages of the spin components (\ref{D4}) are given by
\begin{eqnarray}
\bar n &=& \frac{1}{a} \phi^{(0)}_\nu(u_{\bar n})
\nonumber \\
\bar S^z &=& 2 \bar n - 1,\ \
\bar S^y \ =\ \frac{2\Gamma}{\Omega} \bar n
\nonumber \\
\bar S^x &=& \frac{2\Gamma \bar \Delta_\mathrm{Dopp}}{a\Omega} \phi^{(1)}_\nu(u_{\bar n})
\label{D21}
\end{eqnarray}
consistent with (\ref{3.5})--(\ref{3.11}). Of course, since $u_{\bar n}$ depends on $\bar n$ via (\ref{3.9}) the first line here defines the mean field consistency equation that is solved for $\bar n$ given the system parameters---See Secs.\ \ref{sec:finiteTcritline}--\ref{sec:biphasehystsensor}.

Using (\ref{D13}) and (\ref{D4}) one obtains
\begin{eqnarray}
F_0^x &=& -\frac{2\Gamma \bar \Delta_\mathrm{Dopp}}{a\Omega}
\frac{\tilde \gamma}{1 + \tilde \Omega^2} \Phi_{\nu\tilde\nu}^{(1)}(w)
\nonumber \\
F_0^y &=& \frac{2\Gamma}{a\Omega}
\frac{\tilde \gamma}{1 + \tilde \Omega^2} \Phi_{\nu\tilde\nu}^{(0)}(w)
\nonumber \\
F_0^z &=& \frac{1}{1 + \tilde \Omega^2}
\left[\frac{2}{a} \Phi_{\nu\tilde\nu}^{(0)}(w) - \phi^{(0)}_{\tilde \nu}(w) \right]
\nonumber \\
F_1^x &=& -\frac{2\Gamma \bar \Delta_\mathrm{Dopp}}{a\Omega}
\frac{\tilde \gamma \tilde \Delta_\mathrm{Dopp}}{1 + \tilde \Omega^2}
\Phi_{\nu\tilde\nu}^{(2)}(w)
\nonumber \\
F_1^y &=& \frac{2\Gamma}{a\Omega}
\frac{\tilde \gamma \tilde \Delta_\mathrm{Dopp}}{1 + \tilde \Omega^2}
\Phi_{\nu\tilde\nu}^{(1)}(w)
\nonumber \\
F_1^z &=& \frac{\tilde \Delta_\mathrm{Dopp}}{1 + \tilde \Omega^2}
\left[\frac{2}{a} \Phi_{\nu\tilde\nu}^{(1)}(w) + \phi^{(1)}_{\tilde \nu}(w) \right].
\label{D22}
\end{eqnarray}
The result for ${\bf F}_2 = \tilde {\bf S} - (1 + \tilde \Omega^2) {\bf F}_0$ follows from these results, together with (\ref{D4}) [recalling also the scaling $\tilde S^{x,y} = \tilde \gamma(\omega) \bar S^{x,y}$ defined in (\ref{D5})].

It is straightforward to verify that the zero frequency limit of the above results reproduces the static derivatives $\partial \bar S^\alpha/\partial \Delta$, $\partial \bar S^\alpha/\partial \Omega$ derived directly from the mean field equation (\ref{3.11}) together with transverse spin averages (\ref{4.13}).

These results form the basis of the examples presented in Sec.\ \ref{sec:dynlinresp1}.

\section{Exact formulation of the three-state mean field equations}
\label{app:3statemfform}

In this appendix we derive exact forms for the mean field equations for setups (b) and (c) in Fig.\ \ref{fig:atomsetup}: (2-photon, 1 Rydberg level) and (1-photon, RF local oscillator, 2 Rydberg level), respectively. These form the basis for the solutions presented in Secs.\ \ref{sec:2phmfsoln} and \ref{sec:2Ryresonant}, respectively. We begin by deriving results for the most general three atomic state case, based on the steady state forms of the equations of motion (\ref{2.50}), and then specialize to the above two cases. Recall here that these equations depend on the suppressed atom index $k$ through the Doppler shifts
\begin{equation}
\Delta_k = \Delta + {\bf k} \cdot {\bf v}_k,\ \
\Delta_{R,k} = \Delta_R + {\bf k}_R \cdot {\bf v}_k
\label{E1}
\end{equation}
corresponding to (\ref{2.5}) with an obvious notational adjustment.

\subsection{Three-state mean field equations}
\label{app:3statemfeq}

The equations of motion (\ref{2.50}) (plus the complex conjugates of the last three) are of the form defined by (\ref{C13}) and (\ref{C23}), with ${\bf X}_k$ corresponding to the $D = 2 + 6 = 8$ independent real and imaginary parts of the density matrix components $\rho^{\alpha \beta} = \rho^{\beta \alpha *}$, $\beta \geq \alpha$, and different from $\rho^{00}$. Thus, we define the dynamical variable vector (suppressing for now the atom index $k$)
\begin{equation}
{\bm \rho} = \left[n^1, n^2, \rho^{01\,\prime}, \rho^{01\,\prime\prime},
\rho^{02\,\prime}, \rho^{02\,\prime\prime} , \rho^{12\,\prime},
\rho^{12\,\prime\prime} \right]
\label{E2}
\end{equation}
with real level occupancies $n^\alpha = \rho^{\alpha\alpha}$ and with real and imaginary parts $\rho^{\alpha\beta} = \rho^{\alpha\beta\,\prime} + i \rho^{\alpha\beta\,\prime\prime}$ otherwise. Defining the resonant denominators
\begin{eqnarray}
D_1 &=& \Delta + V_1 + \frac{i}{2}(\Gamma_1 + K_1)
\nonumber \\
D_2 &=& \Delta + \Delta_R + V_2 + \frac{i}{2}(\Gamma_2 + \Gamma_3 + K_2)
\label{E3} \\
D_3 &=& \Delta_R + V_2 - V_1 + \frac{i}{2}(\Gamma_1 +\Gamma_2 + \Gamma_3 + K_1 + K_2),
\nonumber
\end{eqnarray}
the arrays
\begin{widetext}
\begin{equation}
{\bf A} + {\bf v} \cdot {\bf B} = \left(\begin{array}{cccccccc}
-\Gamma_1 & \Gamma_3 & -\Omega'' & \Omega' & 0 & 0 & \Omega_R'' & -\Omega_R' \\
0 & -(\Gamma_2 + \Gamma_3) & 0 & 0 & 0 & 0 & -\Omega_R'' & \Omega_R' \\
\Omega'' & \frac{1}{2} \Omega'' & -D_1'' & -D_1' & \frac{1}{2} \Omega_R'' & -\frac{1}{2} \Omega_R' & 0 & 0 \\
-\Omega' & -\frac{1}{2} \Omega' & D_1' & -D_1'' & \frac{1}{2} \Omega_R' & \frac{1}{2} \Omega_R'' & 0 & 0 \\
0 & 0 & -\frac{1}{2} \Omega_R'' & -\frac{1}{2} \Omega_R' & -D_2'' & -D_2' & \frac{1}{2} \Omega'' & \frac{1}{2} \Omega' \\
0 & 0 & \frac{1}{2} \Omega_R' & -\frac{1}{2} \Omega_R'' & D_2' & -D_2'' & -\frac{1}{2} \Omega' & \frac{1}{2} \Omega'' \\
-\frac{1}{2} \Omega_R'' & \frac{1}{2} \Omega_R'' & 0 & 0 & -\frac{1}{2}\Omega'' & \frac{1}{2} \Omega' & -D_3'' & -D_3' \\
\frac{1}{2} \Omega_R' & -\frac{1}{2} \Omega_R' & 0 & 0 & -\frac{1}{2}\Omega' & -\frac{1}{2} \Omega'' & D_3' & -D_3''
\end{array} \right),\ \
{\bf b} = \left(\begin{array}{c}
0 \\ 0 \\ -\frac{1}{2} \Omega'' \\ \frac{1}{2} \Omega' \\ 0 \\ 0 \\ 0 \\ 0
\end{array} \right)
\label{E4}
\end{equation}
\end{widetext}
may then be read off (\ref{2.50}), following the substitution $n^0 = 1 - n^1 - n^2$. The ${\bf v}$-dependence is implicit in the detuning parameters (\ref{E1}), from which the matrix ${\bf v} \cdot {\bf B}$, limited to the second diagonals of the last six rows and columns, may be separately read off. The $\bar n^1, \bar n^2$ dependence is implicit in the interaction parameters defined by (\ref{2.26}), both contained in the real parts $D_\alpha'$.

\subsection{Three-state thermal averages}
\label{app:3statethermave}

The first two elements of the matrix inverse relation (\ref{C24}) produces solutions for $n^1({\bf v})$ and $n^2({\bf v})$, and the mean field closure equation derived from the thermal average (\ref{C25}) is obtained in the form
\begin{eqnarray}
\bar n^1 &=& \int d{\bf v} P_\mathrm{th}({\bf v}) n^1(\bar n^1, \bar n^2; {\bf v})
\nonumber \\
\bar n^2 &=& \int d{\bf v} P_\mathrm{th}({\bf v}) n^2(\bar n^1, \bar n^2; {\bf v})
\label{E5}
\end{eqnarray}
with thermal (Gaussian) Maxwell distribution $P_\mathrm{th}$ given by (\ref{2.52}). The right hand side depends on ${\bf v}$ through the Doppler shifts (\ref{E1}), and on the mean values $\bar n^1$, $\bar n^2$ through the interaction parameters (\ref{2.26}) appearing in (\ref{E3}).

Despite the apparent complexity, analytic expressions for the right hand side of (\ref{E5}) may be derived using the partial fraction technique described in App.\ \ref{app:partfracave}. This reduces the results to sums of fundamental (error function) averages (\ref{3.12}) which are then used to analyze the resulting mean field equations. As described in App.\ \ref{app:partfracave}, for the setup in Fig.\ \ref{fig:atomsetup}(b) we need to make the simplifying assumption that ${\bf k}$ and ${\bf k}_R$ are parallel (or antiparallel) so that only a single velocity component enters (\ref{E1}):
\begin{equation}
\Delta(u_l) = \Delta + k u_l,\ \ \Delta_R(u_l) = \Delta_R + k_R u_l,
\label{E6}
\end{equation}
in which $u_l = \hat {\bf k} \cdot {\bf v}_l$ and $k_R$ may take either sign. For the setup in Fig.\ \ref{fig:atomsetup}(c), as described below (\ref{2.5}), we approximate ${\bf k}_R = 0$. In either case, the Gaussian averages reduce to the required one-dimensional integrals.

An inspection of the determinants involved in the adjoint matrix solution for the matrix inverse components yields the following conclusions. For setup (b) the numerators for $n^1$ and $n^2$ have degrees 4 and 2, respectively, while the denominators [both given by the determinant of (\ref{E4})] have degree 6. For setup (c), in which $D_3$ is now velocity independent, the numerators both have degree 2, while the denominators have degree 4. Of course, for setup (b) one has $V_1 = 0$ and $V_2 = V_\mathrm{eff} \bar n^2$ so that only $\bar n^2$ appears on the right, and the thermally averaged equation for $n^2$ produces the single mean field equation. The resulting solution $\bar n^2$ is then substituted into the equation for $\bar n^1$ (as well as those for the thermal averages of any other desired density matrix components). For setup (c) although the polynomials are simpler, the two equations must now be solved simultaneously.

\subsection{Three-state dynamical linear response}
\label{app:3statedlr}

Finally, we consider the thermal averages defined in App.\ \ref{app:dynlinresp}, specifically the average of the frequency domain relation (\ref{C27}), in turn requiring evaluation of the two- and four-index tensors (\ref{C28}). Recall that, via (\ref{C24}), the vector $({\bf A}^0 + {\bf v}_k \cdot {\bf B})^{-1} {\bf b}$ corresponds to the previously described steady state solutions for the eight independent real and imaginary parts of $\rho^{\alpha \beta}$. The array $i\omega \openone + {\bf A}^0 + {\bf v}_k \cdot {\bf B}$ includes a simple diagonal perturbation of the of the matrix (\ref{E4}). Its inverse hence follows immediately from the general results above, and the thermal averages are of products of polynomial ratios of the same degrees determined above.

Note here that the vector elements (\ref{E2}) were used to define a convenient set of real time-domain equations. However, the Fourier transform of these involves the complex vector ${\bm \rho}(\omega) = {\bm \rho}(-\omega)^*$, consistent with the now complex contribution to the matrix diagonal.

The perturbation arrays $(\nabla_{\bar{\bf X}} {\bf A})_0$, $(\nabla_{\bf P} {\bf A})_0$, $\nabla_{\bar{\bf X}} {\bf b}_0$, $\nabla_{\bf P} {\bf b}$ in (\ref{C27}) may be read off (\ref{A4}). For the latter two, only
\begin{equation}
\nabla_{\Omega'} b^\alpha = \frac{1}{2} \delta_{\alpha 4},\ \
\nabla_{\Omega''} b^\alpha = -\frac{1}{2} \delta_{\alpha 3}
\label{E7}
\end{equation}
are nonzero. In the examples treated here, the probe amplitude $\Omega$ will not be perturbed. Thus, ${\bf b}$ will be taken as fixed and the arrays ${\bm \Gamma}_b$ and ${\bf F}_b$ in (\ref{C29})--(\ref{C31}) vanish. Defining the sparse antisymmetric forms
\begin{equation}
[{\bf C}_{PQ}]^{rs} = \delta_{rP} \delta_{sQ} - \delta_{rQ} \delta_{sP},
\label{E8}
\end{equation}
one obtains
\begin{eqnarray}
[\partial_\Delta {\bf A}]_0 &=& [\partial_{V_2} {\bf A}]_0 = {\bf C}_{43} + {\bf C}_{65}
\nonumber \\
{[\partial_{\Delta_R} {\bf A}]}_0 &=& {\bf C}_{65} + {\bf C}_{87}
\nonumber \\
{[\partial_{V_1} {\bf A}]}_0 &=& {\bf C}_{43} - {\bf C}_{87}
\nonumber \\
{[\partial_{\bar n^\alpha} {\bf A}]}_0 &=& V_\mathrm{eff}^{1\alpha} {[\partial_{V_1} {\bf A}]}_0
+ V_\mathrm{eff}^{2\alpha} {[\partial_{V_2} {\bf A}]}_0,
\label{E9}
\end{eqnarray}
along with
\begin{equation}
{\bf v} \cdot {\bf B} = ({\bf k} \cdot {\bf v}) [\partial_\Delta {\bf A}]_0
+ ({\bf k}_R \cdot {\bf v}) [\partial_{\Delta_R} {\bf A}]_0.
\label{E10}
\end{equation}
The $\Omega$ and $\Omega_R$ gradients are somewhat more complicated due to the more complex structure of the first two rows and columns of ${\bf A}$.

For setup (b) we consider the Stark-induced coupling beam detuning perturbation $\Delta_R$, and only the mean value $\bar n^2$ appears in ${\bf A}$. Example results are shown in Sec.\ \ref{sec:2ph1Rydlr}. For setup (c) we consider the local oscillator perturbation $\delta \Omega_R$. Both $\bar n^1, \bar n^2$ appear in ${\bf A}^0$, while ${\bf k}_R = 0$ in (\ref{E10}). Example results are shown in Sec.\ \ref{sec:1ph2Rydlr}.


\begin{thebibliography}{}
\bibitem{foot:mF} To be complete, each state $|l \rangle$ actually represents a rotationally degenerate azimuthal quantum number subset $-F_l \leq m_{Fl} \leq F_l$. In the single state description the effective transition dipole moments, hence Rabi frequencies $\Omega_{lk}$, specified in the model then represent a rotational average defined by the laser and LO field polarizations. In the simplest case, the latter two are parallel (and orthogonal to the laser beam direction) and $m_F$ defined relative to this axis is conserved.
\bibitem{foot:corotate} The Hamiltonian (\ref{2.1}) is written here in the ``co-rotating'' frame of reference in which a unitary transformation $|l \rangle \to e^{i(\varepsilon_l + \Delta_l) t} |l \rangle$, $l > 0$, with detunings defined by (\ref{2.5}), has been performed to absorb the laser and local oscillator frequency factors into the state definitions. This simplifies the time dependence appearing in the model to that of the frequency-differenced RF signal (\ref{2.7}). In particular, the model is time-independent for perfect tuning, $\omega_\mathrm{LO} = \omega_\mathrm{RF}$.
\bibitem{foot:Rydecay} There are cases in which the second Rydberg level lies lower in energy than the first, in which case the last line of (\ref{2.17}) would replace $\hat \sigma^{12}$ by $\hat \sigma^{21}$. The analysis for this case differs only in minor details, and for simplicity we do not consider it further here.
\end{thebibliography}
\end{document}